\documentclass[12pt]{report}
\usepackage{utdiss1}
\usepackage{amsfonts, amscd}
\usepackage{amssymb}
\usepackage{eucal,eufrak}
\usepackage{verbatim}
\usepackage{latexsym}
\usepackage{epsfig}

\def\IH{{\mathbb H}}
\def\IR{{\mathbb R}}
\def\IC{{\mathbb C}}
\font\teneufm=eufm10
\font\seveneufm=eufm7
\font\fiveeufm=eufm5
\newfam\eufmfam
\textfont\eufmfam=\teneufm
\scriptfont\eufmfam=\seveneufm
\scriptscriptfont\eufmfam=\fiveeufm

\newcommand{\infinity}{\ensuremath{\infty}}
\newcommand{\dd}{\ensuremath{\partial}}
\newcommand{\ww}{\omega}
\newcommand{\ee}{\varepsilon}
\newcommand{\kk}{\kappa}
\newcommand{\Sc}{\scriptstyle}
\newcommand{\te}{\theta}
\newcommand{\fdot}{\dot{f}}
\newcommand{\fOdot}{\dot{f}_0}

\newcommand{\fddot}{\ddot{f}}
\newcommand{\delr}{\triangle r}
\newcommand{\delt}{\triangle t}
\newcommand{\df}{\delta f}
\newcommand{\old}{\ensuremath{\mbox{old}}}
\newcommand{\newk}{\ensuremath{\mbox{new}_1}}
\newcommand{\neww}{\ensuremath{\mbox{new}_2}}
\newcommand{\grad}{\raisebox{.5 ex}{\ensuremath{\bigtriangledown}}}
\newcommand{\lin}{\ensuremath{\langle}}
\newcommand{\rin}{\ensuremath{\rangle}}
\newcommand{\infnty}{\infty}
\title{Numerical Investigations of Singularity Formation in Non-Linear Wave Equations in the Adiabatic Limit} 
\author{Jean-Marie Linhart}
\address{2207 S. Fifth Street \#111 \\ Austin, TX 78704}
\supervisor{Lorenzo  Sadun}
\previousdegrees{B.S., M.A.}
\degree{DOCTOR OF PHILOSOPHY}
\degreeabbr{Ph.D.}
\graduationmonth{May}
\graduationyear{1999}
\committeesize{5}
\begin{document}

\copyrightpage
\titlepage
\signaturepage
\begin{dedication}
To the unadulterated, unmitigated, unforgiven and unforgiving
stubbornness that will carry you through the pits of hell (or a
dissertation) if necessary.
\end{dedication}

\begin{acknowledgments}
Lorenzo Sadun, my dissertation supervisor, managed to drag me through
this (kicking and screaming, sometimes), and he deserves a great deal
of admiration for that difficult feat as well as gratitude from me.

Ivanna Albertin has been my mentor, supervisor and friend while I've
been working for Schlumberger Well Services while completing my
dissertation.  She's helped me to learn about problems in the real
world and also that there is a good life beyond the ivory tower.  She
made it possible for me to work on my dissertation at my own pace.
She often reminded me that I better just hang on and get done, and so
I have.  Every day I am grateful for her guidance and friendship.

My father, George Linhart, has repeatedly come to my aid morally and
financially while I have been working on my doctorate; sometimes he
gave me help that I needed even when I was insisting that I didn't
need it.  Dad taught me that it is only important to get it right in
the end, and I know he loves me even when I am (considerably) less
than perfect.  Even when I doubted, he always knew that I had what it
takes to get a Ph.D.

My teacher, John Blankenship, reminded me many times that my best is
always good enough and that to obtain success all I have to do is to
chop the wood and carry the water and chop the wood and carry the
water and so forth and so on, ad nauseum.  His school provided a place
of sanity for me when graduate school was anything but sane.  I
appreciate his support and the support of all my brothers and
sisters (also known as instructors, friends, comrades and flying
femmes) at Austin Cha Yon Ryu.

There are many friends who were there for me while I've been working
on this degree.  Marcia Branstetter, Joanna McDaniel, Heather
Caldwell and Eric Hollas all deserve a special mention.

\end{acknowledgments}

\utabstract 
\index{Abstract} 

This dissertation deals with singularity formation in spherically
symmetric solutions of the hyperbolic Yang Mills equations in (4+1)
dimensions and in spherically symmetric solutions of $\IC P^1$ wave
maps in (2+1) dimensions.  These equations have known moduli spaces of
time-independent (static) solutions.  Evolution occurs close to the
moduli space of static solutions.  The evolution is modeled
numerically using an iterative finite differencing scheme, and
modeling is done close to the adiabatic limit, {\it i.e.,\ } with small
velocities.  The stability of the numerical scheme is analyzed and
growth is shown to be bounded, yielding a convergence estimate for the
numerical scheme.  The trajectory of the approach is characterized, as
well as the shape of the profile at any given time during the
evolution.

\tableofcontents
\listoftables
\listoffigures

\chapter{Introduction}

\section{Introduction to the equations}
One of the classic problems of physics is that of the motion of a
particle.

In finite dimensions, {\it i.e.,\ } in $\IR^{n}$, we have a particle described by
its position $\vec{x}$ and velocity $\dot{\vec{x}}$, and the particle
is under the influence of a potential $V(\vec{x})$ (corresponding
perhaps to gravity or an electrical potential).  The Lagrangian is
the kinetic (energy of motion) minus the potential and is given by
\[ L = \frac{1}{2} m|\dot{\vec{x}}|^2 - V(\vec{x}).\]
The Euler-Lagrange equations give the equations of motion which are
\[ m\ddot{\vec{x}} = - \grad V(\vec{x}).\]

In field theory, the position of the particle is not given by a vector
$\vec{x} \in \IR^{n}$ but rather by a function which is a point in an
infinite dimensional vector space.  Call this position $f(\vec{x})$
and instead of the inner product on $\IR^{n}$, the inner product is
based on a function space, such as the ${\mathcal{L}^2}$ inner
product.  Using the $\mathcal{L}^2$ norm as an example, if there is a
potential $V(f)$, and the Lagrangian action is given by:
\[ L = \int |\dot{f}(\vec{x})|^2 - V\left(f(\vec{x})\right) d\vec{x}, \]
the calculus of variations gives the equations of motion
\[ \ddot f = - \frac{\delta V}{\delta f}.\]

We will study two situations similar to this in this dissertation.  

First, we address the situation of the Yang Mills Lagrangian in 4
dimensions.  The Yang Mills equation is a generalization of Maxwell's
equations in a vacuum.  We wish our particles to have certain internal
and external symmetries, which give rise to the various geometrical
objects in the problem.  The state of our particles are given by gauge
potentials or connections $A$ on $\IR^{4}$, where we identify
$\IR^{4}$ as $\IH$, the quarternions. The gauge potentials have values
in the Lie algebra of $SU(2)$ which can be viewed as pure imaginary
quarternions, $\mbox{Im}(\IH)$.  The curvature $F_{ij} = \dd_iA_j -
\dd_jA_i + [A_i,A_j]$ where $[A_i,A_j] = A_iA_j - A_jA_i$ is the
bracket in the Lie Algebra, gives rise to the potential $V(A) = \lin
F, F \rin $ which is a nonlinear function of $A$.  The action is:
\begin{equation} L = \frac{1}{2}
\int \ \lin F_{ij},F_{ij}\rin d\vec{x}. \label{YMPLagr}
\end{equation}
The local minima of [\ref{YMPLagr}] are the instantons on 4
dimensional space.  These correspond to solutions of Maxwell's
equations in the vacuum, if we consider the Yang Mills Lagrangian to
be a generalization of Maxwell's equations.  We now consider the wave
equation generated by this potential with Lagrangian:
\begin{equation} L = \frac{1}{2}\int \lin\dot{A_i},\dot{A_i}\rin -\frac{1}{2} \lin F_{ij},F_{ij}\rin d\vec{x} \label{YMLagr}
\end{equation}
Here we use the summation convention; however in taking inner products
of two forms, we only sum over $i < j$, and not all $i, j$, which has
here been expressed by dividing by 2 in computing $\lin F_{ij},
F_{ij}\rin$.  Now use the calculus of variations by taking $A
\rightarrow A + \delta A$, to obtain
\begin{eqnarray*} \lefteqn{L(A + \delta A) =}\\
& & \int \frac{1}{2}\lin\dot{A_i}, \dot{A_i}\rin + \lin \dot{A_i}, \dot{\delta A_i}\rin  + \frac{1}{2} \lin\dot{\delta A_i}, \dot{\delta A_i}\rin - \\
& &
\frac{1}{4}\lin\dd_iA_j + \dd_i\delta A_j - \dd_j A_i - \dd_j \delta A_i + [A_i + \delta A_i,A_j + \delta A_j],\\
& &
\dd_iA_j + \dd_i\delta A_j - \dd_j A_i - \dd_j \delta A_i + [A_i + \delta A_i,A_j + \delta A_j]\rin d\vec{x} \end{eqnarray*}
Using the definition of $F$, one obtains
\begin{eqnarray*} L(A + \delta A) &=& \int \frac{1}{2}\lin\dot{A_i}, \dot{A_i}\rin + \lin\dot{A_i},\dot{\delta A_i}\rin
 +\frac{1}{2}\lin\dot{\delta A_i}, \dot{\delta A_i}\rin - 
\frac{1}{4}\lin F_{ij},F_{ij}\rin \\
& & +\frac{1}{2}\lin F_{ij}, \delta F_{ij} \rin  + \frac{1}{4}\lin \delta F_{ij}, \delta F_{ij}\rin 
d\vec{x}.  \end{eqnarray*}
One now takes the linear part of the variation, 
\[ \int \lin\dot{A_i}, \dot{\delta A_i}\rin + \frac{1}{2}\lin F_{ij}, \delta F_{ij}\rin 
d\vec{x},\]
and analyzes it.  First work on 
\begin{eqnarray*} \int \lin F_{ij},\delta F_{ij}\rin &=& \int \lin F_{ij}, \dd_i\delta A_j\rin
- \lin F_{ij}, \dd_j \delta A_i\rin + \lin F_{ij}, [\delta A_i,
A_j]\rin + \lin F_{ij}, [A_i, \delta A_j]\rin\\ 
&=& \int \lin F_{ij}, \dd_i\delta A_j\rin-\lin 
F_{ij}, \dd_j\delta A_i\rin - \\
& & \mbox{Tr}\bigl(F_{ij}(\delta A_i A_j - A_j
\delta A_i + A_i \delta A_j - \delta A_j A_i)\bigr) \\
&=& \int -2\lin \dd_j F_{ij}, \delta A_i\rin -  \mbox{Tr}\bigl(F_{ij}(\delta A_i A_j - A_j \delta A_i)\bigr) \\
\noalign{\mbox{by integration by parts and the cyclic property of traces:}}\\
&=& \int - 2\lin \dd_j F_{ij}, \delta A_i\rin -  \mbox{Tr}\bigl(F_{ij}A_j - F_{ij} A_j)\bigr)\delta A_i \\
&=& \int -2\lin \dd_j F_{ij} + [F_{ij}, A_j], \delta A_i\rin\\
&=& \int - 2\lin \grad_j F_{ij}, \delta A_i\rin.
\end{eqnarray*}
Here $\grad_j$ represents the covariant derivative in the $j$ direction.
Now integrate the first term by parts to obtain:
\[ \int \lin\dot{A_i}, \dot{\delta A_i}\rin d\vec{x} = - \int \lin \ddot{A_i}, \delta A_i\rin d\vec{x}.\]
Put these two things together to get
\begin{equation} \ddot{A_i} = -\grad_j F_{ij}\label{genPDEa}\end{equation}
This is our evolution equation for the 4+1 dimensional model.

The second situation to be addressed is that of the $\IC P^1$ model in
2+1 dimensions.  

In the finite dimensional analog of this case, we consider
motion on a manifold under the influence of a potential.  The
Lagrangian in this case is
\[ L = \frac{1}{2} g_{ij}(\vec{x})\dot{x_i}\dot{x_j} - V(\vec{x}),\]
where $g_{ij}$ is the metric tensor that allows us to measure length
on the curved space of the manifold.  The equations of motion are of the form
\[ D_t \dot{\vec{x}} = -\grad V,\] where $D_t$ is a covariant
derivative with respect to time. 

In the $\IC P^1$ model we have motion on an infinite dimensional manifold
as we consider maps from $\IR^{2+1} \rightarrow S^2$, to the two sphere,
of a particular degree.  If $\phi:\IR^{2+1} \rightarrow S^2$ the
Lagrangian is
\[ \int_{\IR^2} |\dot{\phi}|^2 - |\grad\phi|^2. \] 
Identifying $S^2$ as $\IC + \{\infinity\},$ one can consider $u:\IR^{2+1}
\rightarrow \IC + \{\infinity\}$.  The Lagrangian is
\begin{equation} \int_{\IR^2} \frac{|\dot{u}|^2}{(1 + |u|^2)^2} - 
\frac{|\grad u|^2}{(1 + |u|^2)^2}.\label{cp1lag}\end{equation}
One can see this is one again the kinetic minus the potential energy.

The calculus of variations on this Lagrangian in conjunction with
integration by parts yields the following equation of motion for the
$\IC P^1$ model:
\begin{equation}
(1 + |u|^2)(\dd_t^2u - \dd_x^2 u - \dd_y^2 u) = 2\bar{u}(|\dd_t u|^2 - |\dd_x u|^2
- |\dd_y u|^2) \label{genPDEb}\end{equation}

\section{Introduction to the adiabatic limit}

The origin of the term ``adiabatic'' seems to be with reversible
processes: processes that do not change the entropy of a system and
consequently that can be undone.  One way of enacting such a
reversible process it to make the changes infinitely slowly.

Another way of looking at making changes infinitely slowly is to break
space-time into two pieces, space and time.  If one then rescales the
spatial piece to be extremely small, or the time piece to be extremely
large, it then takes correspondingly more time to move from point to
point.  In the limit as we take the spatial piece to be infinitely
small or the time piece to be infinitely large, we are moving
infinitely slowly.

Naturally, this sort of rescaling can occur in any problem that can be
broken into two pieces.  Any space that is a Cartesian product, say an
infinite cylinder, $S^1 \times \IR$ or a torus $S^1\times S^1$.  Or
even any space that locally looks like a product, such as our old
friend the M\oe bius band, or, if one is even more adventurous and
brave, a twisted bundle.  The main requirement is that we have a
metric on the space that splits into two pieces,
\[ g(x,y)\lin x, x\rin + h(x,y)\lin y,y\rin. \]
Then one can introduce a parameter $\lambda$ without changing the
topology of the space:
\[ g(x,y)\lin x, x\rin + \lambda h(x,y)\lin y,y\rin. \]
An adiabatic limit is simply taking the parameter $\lambda\rightarrow
0$ or $\lambda \rightarrow \infty$, which makes the space $Y$ look
either infinitely large or infinitely small.  While changing from one
finite and nonzero $\lambda$ to another causes no changes in the
topological properties of the space, the limit has no such immunity.
Conclusions based on an adiabatic limit must be examined with care,
for what happens in the limit {\it might not\ } be close to what
happens near the limit.

Adiabatic limits take on special meaning in the case of space-time.
Taking an adiabatic limit results in moving either infinitely slowly
or infinitely quickly, usually infinitely slowly.  When we have a
partial differential equation with known moduli spaces of static
solutions, the adiabatic limit tells us that as velocity tends towards
zero, we should get motion along the geodesics of these moduli spaces.
We approximate solutions with small velocity with these geodesics, and
hence this is called the Geodesic Approximation.  Investigating this
phenomenon in the two cases mentioned in the introduction is the
concern of this dissertation.

\chapter{The 4+1 dimensional model}

The first things to identify in this problem are the static
solutions to equation [\ref{genPDEa}].  These are simply the 4
dimensional instantons investigated in \cite{Atiyah}.  In \cite{Atiyah} the form
of all such instantons in the degree one sector is shown to be:
\[ A(x) = \frac{1}{2}\left\{\frac{(\bar{x} - \bar{a})dx - d\bar{x} (x -a)}{\lambda^2 + |x-a|^2}\right\}
\qquad x = x_1 + x_2i + x_3 j + x_4 k\in \IH. \] The curvature $F$ of
this potential is computed by $F = dA + [A,A]$ and is:
\[ F = \frac{d\bar{x}\wedge dx}{(\lambda^2 + |x-a|^2)^2}.\]
One notices the denominators are radially symmetric about $x= a$.  

An instanton of unit size centered at the origin would be
\[ A(x) = \frac{1}{2}\left\{\frac{\bar{x} dx - d\bar{x} x}{1 + |x|^2}.\right\}
\]
One motion one can study from these static
instantons would be to consider connections the form
\[A(r, t) = \frac{1}{2}\left\{ \frac{\bar{x}dx - d\bar{x} x}{f(r,t) + r^2}\right\} \qquad r = \sqrt{x_1^2 + x_2^2 + x_3^2 + x_4^2}
\]
and derive an equation of motion for $f(r,t)$ from
[\ref{genPDEa}]. This is actually quite computationally extensive.

To get at this, start with connections of the form
\[A(r, t) = \frac{1}{2}g(r,t)\left\{\bar{x}dx - d\bar{x} x\right\}.\]
Now we have a formula for $F$ given by $F = dA + [A,A]$.  But $F$
cannot easily be expressed as before.  The components can be computed
using Maple, with quarternions defined as
\[
i = \left[ \begin{array}{*{2}{c}} 0 & i \\ i & 0 \end{array}\right], \qquad
j = \left[ \begin{array}{*{2}{c}} 0 & -1 \\ 1 & 0 \end{array}\right], \qquad
k = \left[ \begin{array}{*{2}{c}} i & 0 \\ 0 & -i \end{array}\right] \]
the usual derivative, and the bracket is that of two matrices: $[M,N] = MN - NM$.  First, computing $F$ one obtains components such as
\begin{eqnarray*} F_{12} &=& i\left[\frac{x_1^2 g'(r,t) + 
2r g(r,t) + x_2^2 g'(r,t)}{r} - 2 x_4^2 g(r,t)^2  - 
2 x_3^2  g(r,t)^2\right] + \\
& & j\left[\left(\frac{g'(r,t)}{r} + 2 g(r,t)^2\right)\left(x_2x_3 - 
x_1x_4\right)\right] + \\
& &  k\left[\left(\frac{g'(r,t)}{r} + 2 g(r,t)^2\right)\left(x_1x_3 + 
x_2 x_4\right)\right]
\end{eqnarray*}
Once $F$ has been obtained one may apply [\ref{genPDEa}] to obtain
an equation for $g(r,t)$.  After much labor, one obtains:
\begin{equation} \ddot{g} = 12 g^2 + \frac{5g'}{r} + g'' - 8 g^3 r^2.\label{protPDEa}\end{equation}

It is much less difficult to now compute a differential equation for $f(r,t)$
if \[ g(r,t) = \frac{1}{f(r,t) + r^2}.\]  It is
\begin{equation} \ddot{f} = f'' + \frac{5f'}{r} - \frac{8 f'r}{f + r^2} 
+ \frac{2}{f + r^2}\left((\fdot)^2 - (f')^2\right) .\label{PDEa}\end{equation}

The static solutions for $f(r,t)$ are simply horizontal lines,
$f(r,t) = c$.  The adiabatic limit expects that motion under small
velocities should progress from line to line, {\it i.e.,\ } $f(r,t) = c(t)$.
$f(r,t) = 0$ is a singularity of the system, where the instantons are
not well defined.  We can use this numerical approximation to the
adiabatic limit to observe progression from $f(r,t) = c_0 > 0$ towards
this singularity.  Our initial assumption is that $f(\cdot, t) =
c(t)$.

\section{Numerics for the 4+1 dimensional model}

A finite difference method is used to compute the evolution of
[\ref{PDEa}] numerically.  Unless otherwise noted, centered
differences are used consistently, so that
\begin{eqnarray*} g'(x) &\approx& \frac{g(x + \delta) - g(x-\delta)}{2\delta}\\
g''(x) &\approx& \frac{g(x+\delta) + g(x-\delta) - 2 g(x)}{\delta^2}. 
\end{eqnarray*}

In order to avoid serious instabilities in [\ref{PDEa}], the terms
\begin{equation} f'' + \frac{5f'}{r}\label{linpart}\end{equation} 
must be modeled in a special way.  Allow
\[ f'' + \frac{5f'}{r} = \mathcal{L}f\]
where \[ \mathcal{L} = r^{-5}\partial_r r^5 \partial_r.\] In Appendix
A, we will see that this operator has negative real spectrum, hence it
is stable.  However the naive central differencing scheme on
[\ref{linpart}] always results in uncontrolled growth near the origin.
General wisdom holds that when one has difficulties with the numerics in
one part of a problem one should find a differencing scheme for that
specific part in the natural to that specific part.  Applying
this allowed for the removal of the problem near the origin.
Instead of using centered differences on $f''$ and on $f'$, we
difference the operator:
\[ \mathcal{L}f = r^{-5} \partial_r r^5 \partial_r. \]
The ``natural differencing scheme'' is \[ \mathcal{L}f \approx
r^{-5}\left[\frac{\left(r + \displaystyle{\frac{\delta}{2}}\right)^{5} \left(\displaystyle{\frac{f(r + \delta) -
f(r)}{\delta}}\right) - \left(r - \displaystyle{\frac{\delta}{2}}\right)^5 \left(\displaystyle{\frac{f(r) - f(r-
\delta)}{\delta}}\right)}{\delta}\right].\]

With the differencing explained, to derive $f(r,t + \delt)$, one
always has a guess for $f(r,t+\delt)$ given by either the initial velocity,
e.g. $f(r,t+\delt) = f(r,t) + v_0\delt$, or by $f(r,t+\delt) = 2f(r,t)
- f(r,t-\delt)$.  Use this to compute $\dot{f}(r,t)$ on the right hand
side of [\ref{PDEa}].  Then solve for $f(r,t +\delt)$ in the
difference for $\ddot{f}(r,t)$, and iterate this procedure to get a
more precise answer.  So, one iterates
\begin{eqnarray*} f(r,t+\delt) &=& 2f(r,t) - f(r,t-\delt) + (\delt)^2\left[
f''(r,t) - \frac{5f'(r,t)}{r}\right.\\ & & - \left.
\frac{2\dot{f}(r,t)^2}{f(r,t) + r^2} - \frac{2f'(r,t)^2}{f(r,t) + r^2}
- \frac{8 f'(r,t)r}{f(r,t) + r^2}\right],\end{eqnarray*} where all
derivatives on the right hand side are represented by the appropriate
differences.

There remains the question of the boundary conditions.  The function
$f$ is only modeled out to a value $r = R \gg 0$.  Initial data for
$f(r,0)$ was originally a horizontal line.  The corresponding boundary
conditions are that $f(R,t) = f(R-\delr, t)$, and that \[ f(0,t) =
\frac{4}{3} f(\delr, t) - \frac{1}{3} f(2\delr,t)\] {\it i.e.,\ } that $f$ is
an even function.

Subsequent investigation of the model indicated that the appropriate
form for $f(r,t)$ was a parabola instead of a line, and the $f(R,t)$
boundary condition was changed to reflect this. For the runs with
parabolic initial data, we set the boundary condition at $R$ to be:
\[f'(R,t) = f'(R-\delr, t)\frac{R}{R-\delr}.\]

\section{Predictions}

Equation [\ref{YMLagr}] gives us the Lagrangian for the general
version of this problem.  We are using
\[ A = \frac{1}{2} \left\{ \frac{\bar{x} dx - d\bar{x} x}{f + r^2}\right\}
\qquad r = \sqrt{x_1^2 + x_2^2 + x_3^2 + x_4^2},\]
 and the adiabatic limit says we will move on the moduli space of
these solutions.  This will give us an effective Lagrangian.  The
portion of the integral given by
\[  -\frac{1}{4} \int_{\IR^4} \lin F_{ij},F_{ij}\rin d\vec{x} \]
represents the potential energy and integrates to a topological constant,
hence it may be ignored.  We need to calculate
\[ \frac{1}{2} \int_{\IR^4} \lin\dot{A_i},\dot{A_i}\rin d\vec{x}. \]
First calculate
\[ \lin \dot{A_i},\dot{A_i}\rin = \frac{3r^2 \fdot^2}{(f + r^2)^4}.\]
So the effective Lagrangian is
\[ \int_{\IR^4} \frac{3r^2 \fdot^2}{(f + r^2)^4} d\vec{x}, \]
Letting $\vec{y} = \vec{x}/\sqrt{f}$ we can rewrite this integral as
\[ \frac{3\fdot^2}{f} \int_{\IR^4} \frac{|y|^2}{(1 + |y|^2)^4} d\vec{y}.\]
The integral with respect to $\vec{y}$ converges, hence we have
\[ L = c\frac{\fdot^2}{f}. \]
This is purely kinetic energy.  Since the potential energy is
constant, so is the kinetic energy.  We have
\[ \frac{\fdot^2}{f} = k. \]
Integrating this we get
\[ f = (c_1 t + c_2)^2.\]
If $f = 0$ occurs at time $T$, we find
\[ T = -\frac{c_2}{c_1} \]
hence we rewrite this as
\[ f = a(t - T)^2.\]
This is how we predict that $f(0,t)$ will evolve.

\section{Results}

The computer model was  run under the condition that $f(r,0)
= f_0$ with various small velocities.  The initial velocity is
$\dot{f}(r,0) = v_0$, other input parameters are $R = r_{max}$,
$\delr$ and $\delt$.  

\subsection{Evolution of $f(0,t)$}

The first question to ask is how does the evolution of the origin
occur.  We note that equation [\ref{PDEa}] becomes the regular linear
wave equation 
\[ \fddot = f''\]
as $r\rightarrow \infinity$, and so the interesting nonlinear behavior
is at the origin.  Consequently we track the evolution of $f(0,t)$.

The evolution of $f(0,t)$ is a parabola of the form: 
\[f(0,t) = a(t-T)^2,\] where
\[ a  = \frac{v_0^2}{4f_0}\] and 
\[ T = \frac{2f_0}{|v_0|}.\]

\begin{figure}[H]
\begin{center}
\epsfig{file=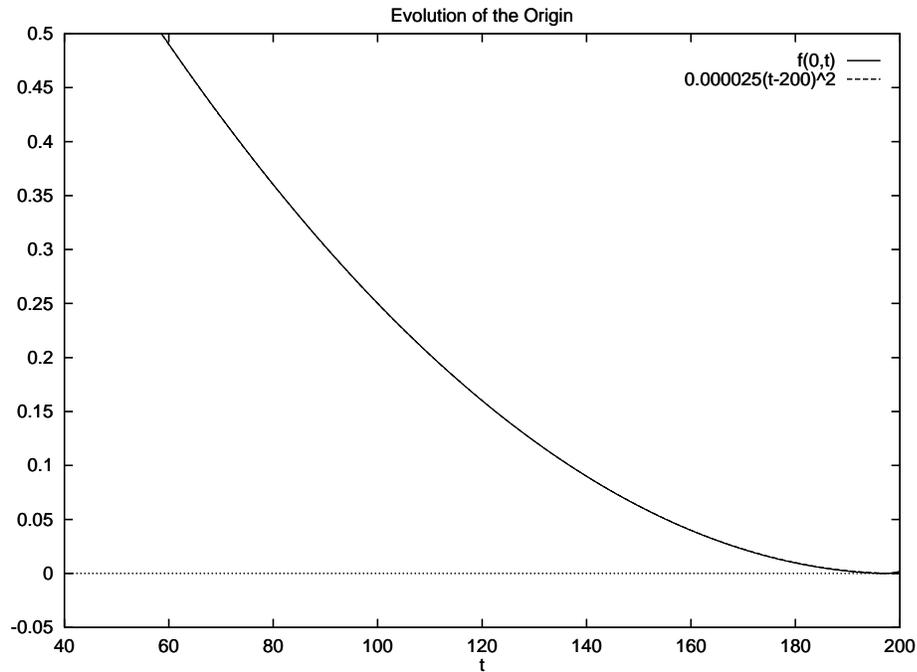}
\caption{4+1 dimensional model, evolution $f(0,t)$.}
\label{o4p1}
\end{center}
\end{figure}

A typical evolution of $f(0,t)$ is given in Figure \ref{o4p1}.  In
this figure, the equation $0.000025(t-200)^2$ neatly overlays the
graph of $f(0,t)$.  This picture represents the evolution where $f_0 =
1.0$ and $v_0 = -0.01$.  Hence $c = \frac{(0.01)^2}{4(1.0)} =
0.000025$ and $T = \frac{2(1.0)}{0.01} = 200,$ as predicted.

The time to ``blow up'' is the parameter $T$ in this equation.  Recall
that $f_0 = f(r,0)$ is the initial height and $\dot{f}_0(r,0) = v_0$
is the initial velocity.  Using a least squares parabolic fit to the
origin data obtained after $f(0,t) \leq 0.5f_0$, one obtains the
parameters $a$ and $T$ for a given origin curve.  Table \ref{op4p1}
shows the behavior.

\begin{table}[H]
\begin{center}
\caption{4+1 Dimensional Model, parabolic fit to $f(0,t)$ vs. Initial conditions $f_0$ and $v_0$}
\label{op4p1}
\[ \begin{array}{*{4}{r}} f_0 & v_0\  & a\ \ \ \ \  & T\  \\ 
1.0&-0.010&0.00002501&200.1\\
2.0& -0.010&0.00001257&399.4\\
0.5&-0.010&0.00005166&99.0\\
4.0&-0.010&0.00000626&799.7\\
4.0&-0.020&0.00002503&400.1\\
4.0&-0.005&0.00000157&1599.3 \end{array}\]
\end{center}
\end{table}

\clearpage

\subsection{Characterization of time slices $f(r,T)$: evolution of a horizontal line}

\begin{figure}[H]
\begin{center}
\epsfig{file=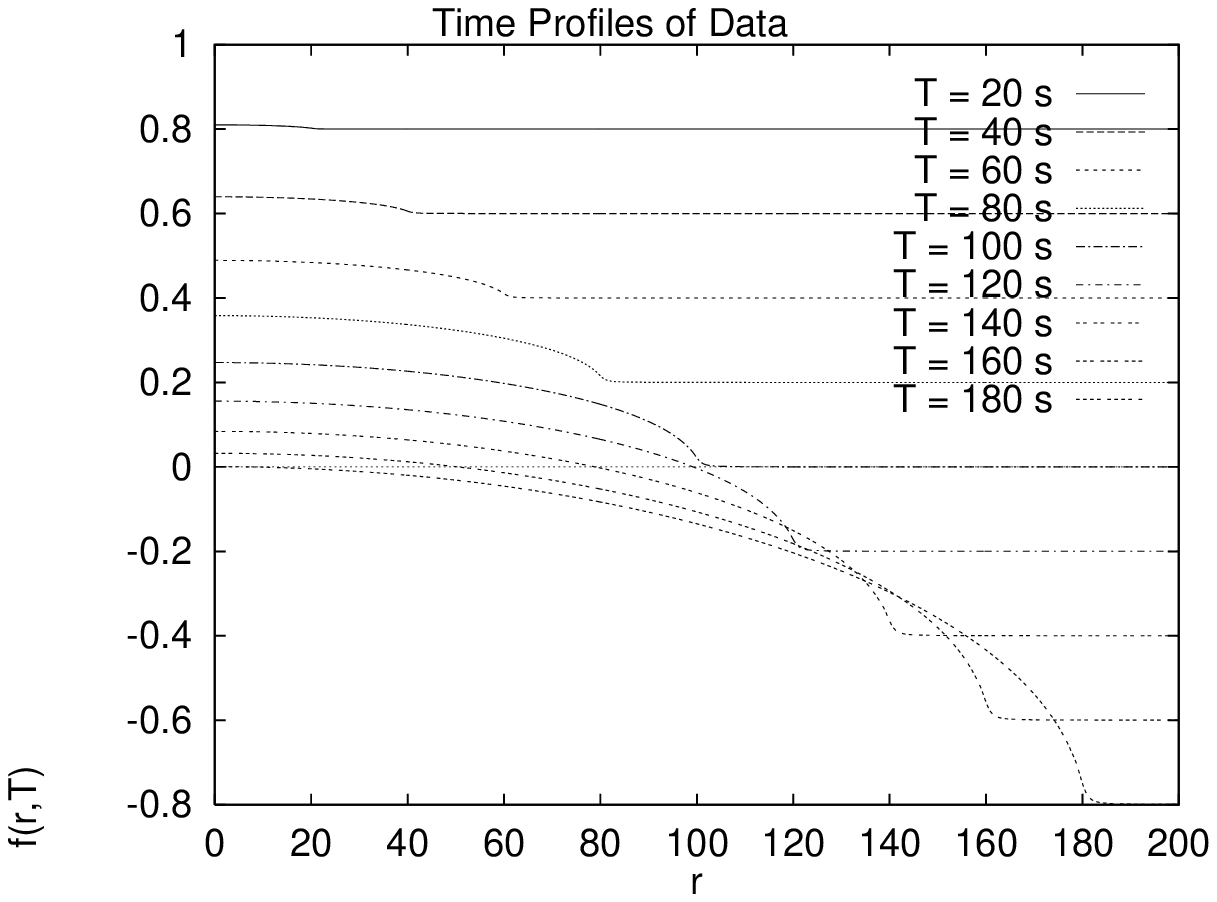}
\caption{$4+1$ Dimensional model, Time Slices $f(r,T)$ evolve an elliptical bump at the origin}
\label{tsl4p1}
\end{center}
\end{figure}

The most striking immediate result is that the initial line, $f(r,0) =
f_0,$ evolved an elliptical bump at the origin that grew as time
passed.  Figure \ref{tsl4p1} shows this behavior.

The elliptical bumps can be modeled as 
\begin{equation} \frac{x^2}{a^2} + \frac{(y-k)^2}{b^2} = 1.\label{ellform}\end{equation}
The question naturally arises as to how the parameters $a$, $b$ and
$k$ evolve.  This is straightforward:  
\[ a = t \]
\[ b = \frac{v_0^2}{4f_0} t^2\]
\[ k = f_0 + v_0t \]
Table \ref{ell4p1} shows these values as calculated using least
squares fitting (to either a line or parabola).  Since the elliptical
fit only works before the right end of the ellipse hits the boundary
at $r = R$ the fit sometimes needed to be restricted to the portion of
the data before this occurred.  While the ellipse is small, there is a
great deal of noise in finding the elliptical parameters, and to get a
good data fit, this noise must often be removed.  Here $m_a$ and $b_a$
is the slope and intercept of the line $a(t)$, and likewise $m_k$ and
$b_k$ are the slope and intercept of the line $k(t)$.  The parameter
$c$ is that in $b(t) = ct^2$

\begin{table}[H]
\begin{center}
\caption{4+1 Dimensional Model, Elliptical Parameters vs. Initial Conditions $f_0$ and $v_0$}
\label{ell4p1}
\[ \begin{array}{*{7}{r}} f_0 & v_0\  & m_a\ \ \  &b_a\ \ & c\ \  & m_k\ \  & b_k\ \ \\
1.0&-0.010&0.999& -0.11&0.00002510&0.0100&-1.00\\
2.0& -0.010&0.999& 0.01&0.00001270&0.0108&-2.00\\
0.5&-0.010&1.007& -0.420&0.00005653&0.0100&-0.51\\
4.0&-0.010&1.001&0.764&0.00000626&0.0100&-4.001\\
4.0&-0.020&0.998&0.177&0.00002527&0.0200&-4.003\\
4.0&-0.005&1.001&-0.834&0.00000157&0.0050&-4.000\end{array}
\]
\end{center}
\end{table}

\begin{figure}[H]
\begin{center}
\epsfig{file=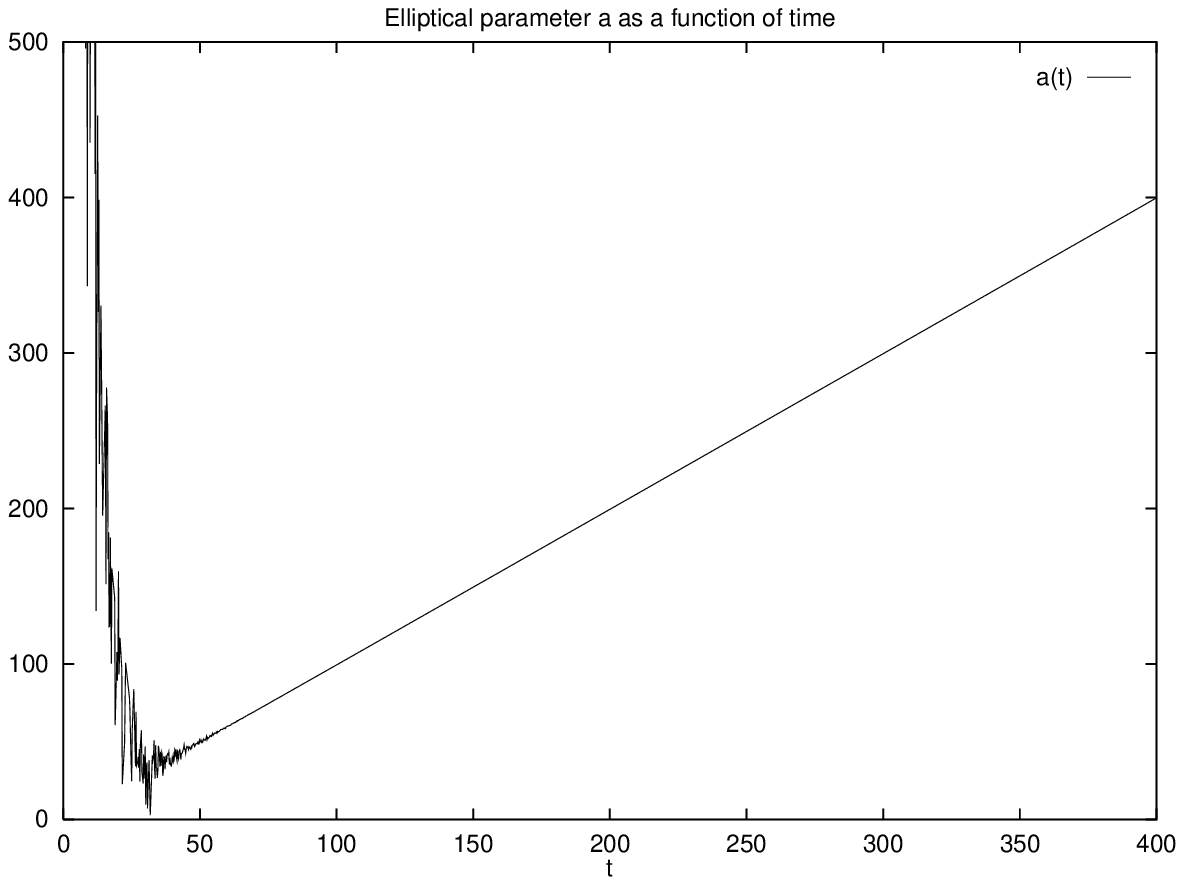}
\caption{4+1 dimensional model: elliptical parameter $a$ as a function of time when $f_0 = 4.0$ and $v_0 = -0.01$.}
\label{a4p1}
\end{center}
\end{figure}

A typical evolution for $a$ with $f_0 = 4.0$ and $v_0 = -0.01$ is in
Figure \ref{a4p1}.  Note the noise when $t < 100$.

\begin{figure}[H]
\begin{center}
\epsfig{file=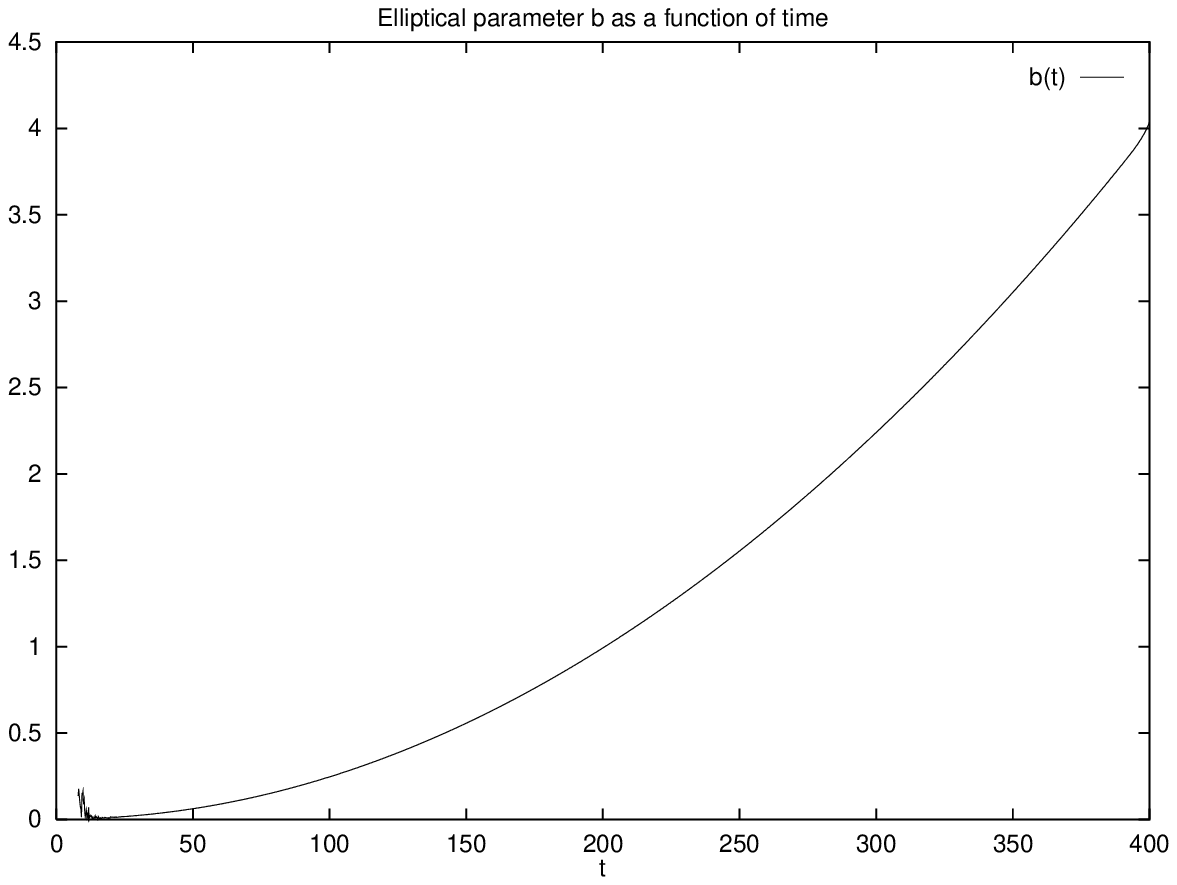}
\caption{4+1 dimensional model: elliptical parameter $b$ as a function of time when $f_0 = 4.0$ and $v_0 = -0.02$.}
\label{b4p1}
\end{center}
\end{figure}

Figure \ref{b4p1} is a typical evolution for $b$ with $f_0 = 4.0$ and $v_0 = -0.02$.

\begin{figure}[H]
\begin{center}
\epsfig{file=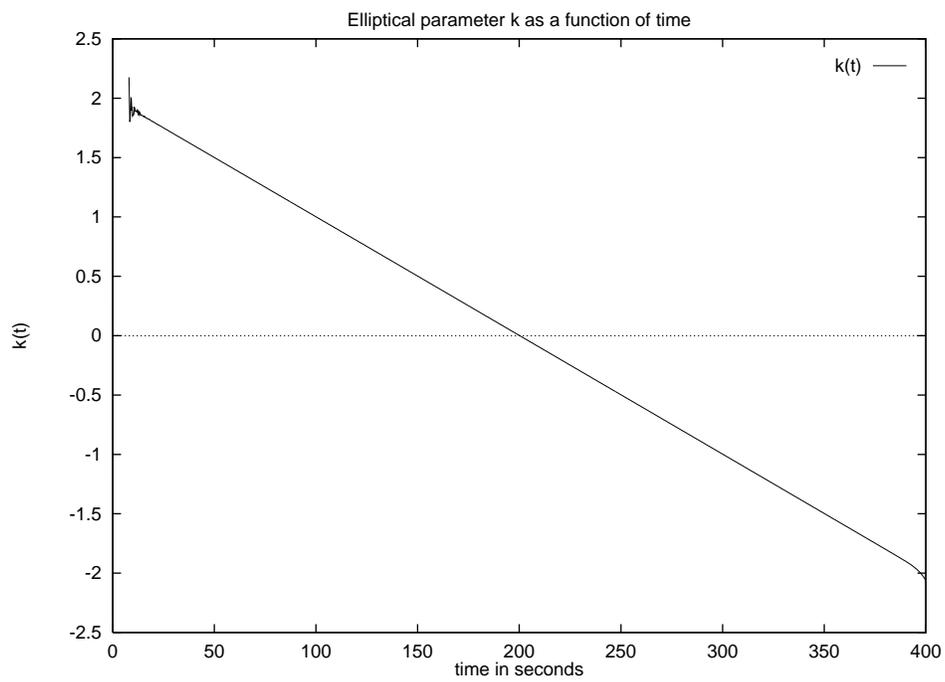}
\caption{4+1 dimensional model: elliptical parameter $k$ as a function of time when $f_0 = 2.0$ and $v_0 = -0.01$.} 
\label{k4p1}
\end{center}
\end{figure}

Figure \ref{k4p1} is a typical evolution for $k$ with $f_0 = 2.0$ and
$v_0 = -0.01$.

\clearpage

\subsection{Characterization of time slices $f(r,T)$: evolution of a parabola}

\medskip
The elliptical bump that formed in the evolution of a horizontal line
and the various configurations that ensued after it bounced off the
$r=0$ and $r = R$ boundary suggested that perhaps the curve was trying
to obtain the shape of a parabola.  After all, near $r = 0$, ellipses
are excellent approximations for parabolas of the form

\begin{equation} f(r,t) = pr^2 + h.\label{parabform}\end{equation}

To get the parabola, calculate from the general form of our ellipse
in [\ref{ellform}] 
\[ \frac{dy}{dx} = -\frac{x^2b^2}{(y-k)a^2}\]
so
\[ \frac{d^2y}{dx} = \frac{-b^2}{(y-k)a^2} - \frac{xb^2}{(y-k)^2a^2}\frac{dy}{dx}\]
At $x = 0$, $y-k = b$ and this gives
\[ \frac{d^2y}{dx} = \frac{-b}{a^2}.\]
Recall from the previous section that $b = ct^2$ and $a = t$, so this
gives \[ \frac{d^2y}{dx^2} = -c.\] The identification of $c$ gives \[
\frac{d^2y}{dx^2} = -\frac{v_0^2}{4f_0}.\] So \[p =
-\frac{1}{2}\frac{d^2y}{dx^2} = -\frac{v_0^2}{8f_0}\] 

\begin{figure}[H]
\begin{center}
\epsfig{file=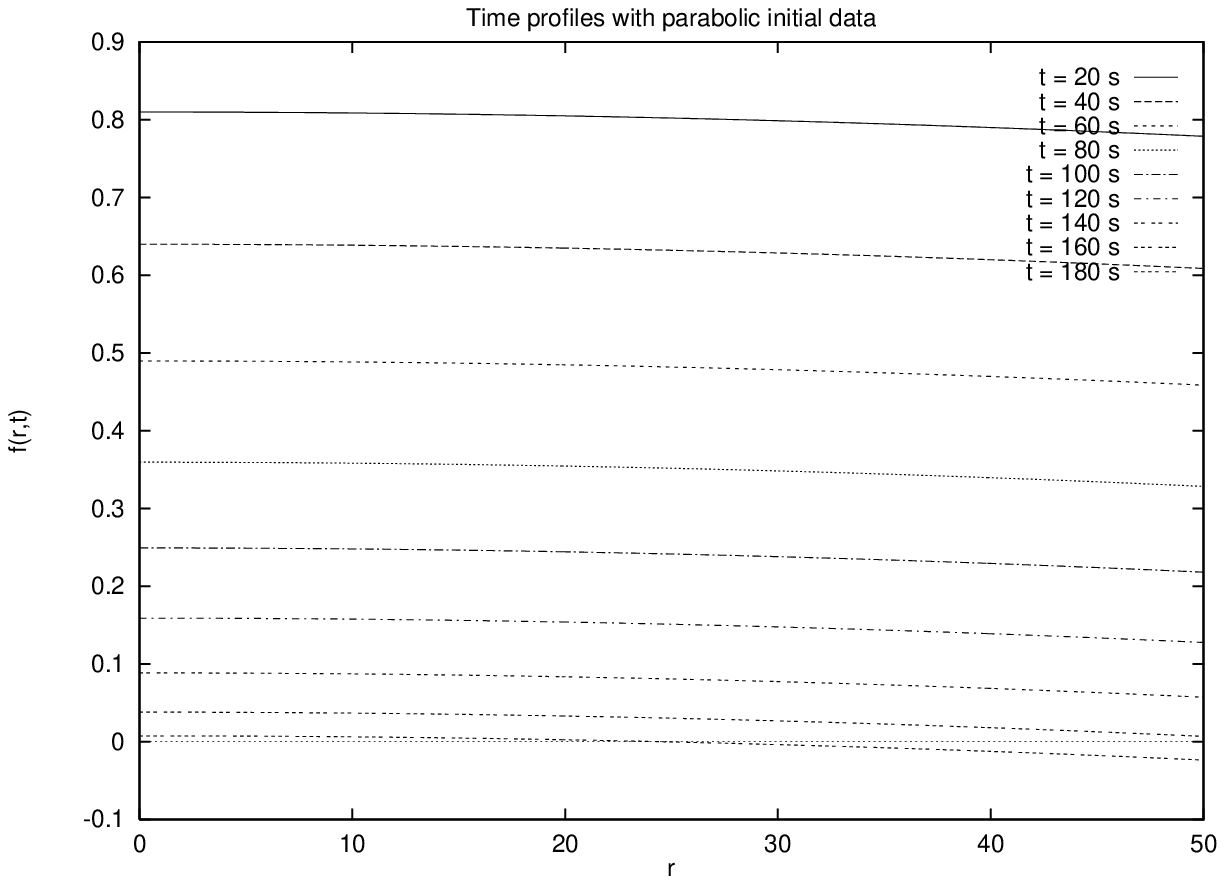}
\caption{$4+1$ dimensional model: Time slices of the evolution of a parabola
are parabolas.}
\label{p4p1}
\end{center}
\end{figure}

When a run is started with this initial data, $\dot{f_0} = v_0 =
-0.01$, $f_0 = f(0,0) = 1.0$ and $p = \frac{v_0^2}{8f_0} =
-0.0000125$, the time slices of the data have this same profile.  This
is shown in figure \ref{p4p1}.

\begin{figure}[H]
\begin{center}
\epsfig{file=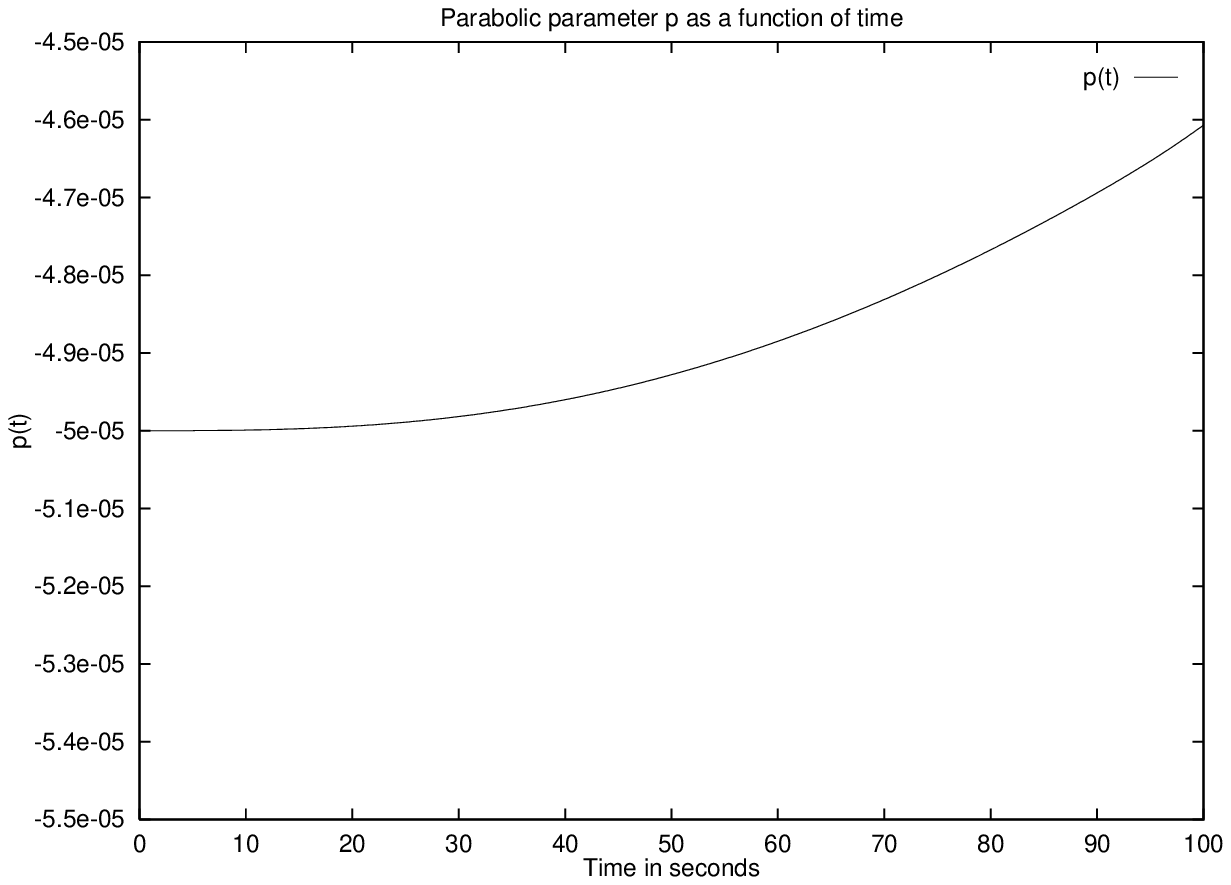}
\caption{$4+1$ dimensional model: evolution of parabolic parameter $p$ with time.}
\label{pp4p1}
\end{center}
\end{figure}

The curvature of the parabola at the origin, as measured by the
parameter $p$ from equation [\ref{parabform}] changes by less than 1
part in 100 during the course of this evolution, a graph of $p$ over
time can be seen in figure \ref{pp4p1}.

\begin{figure}[H]
\begin{center}
\epsfig{file=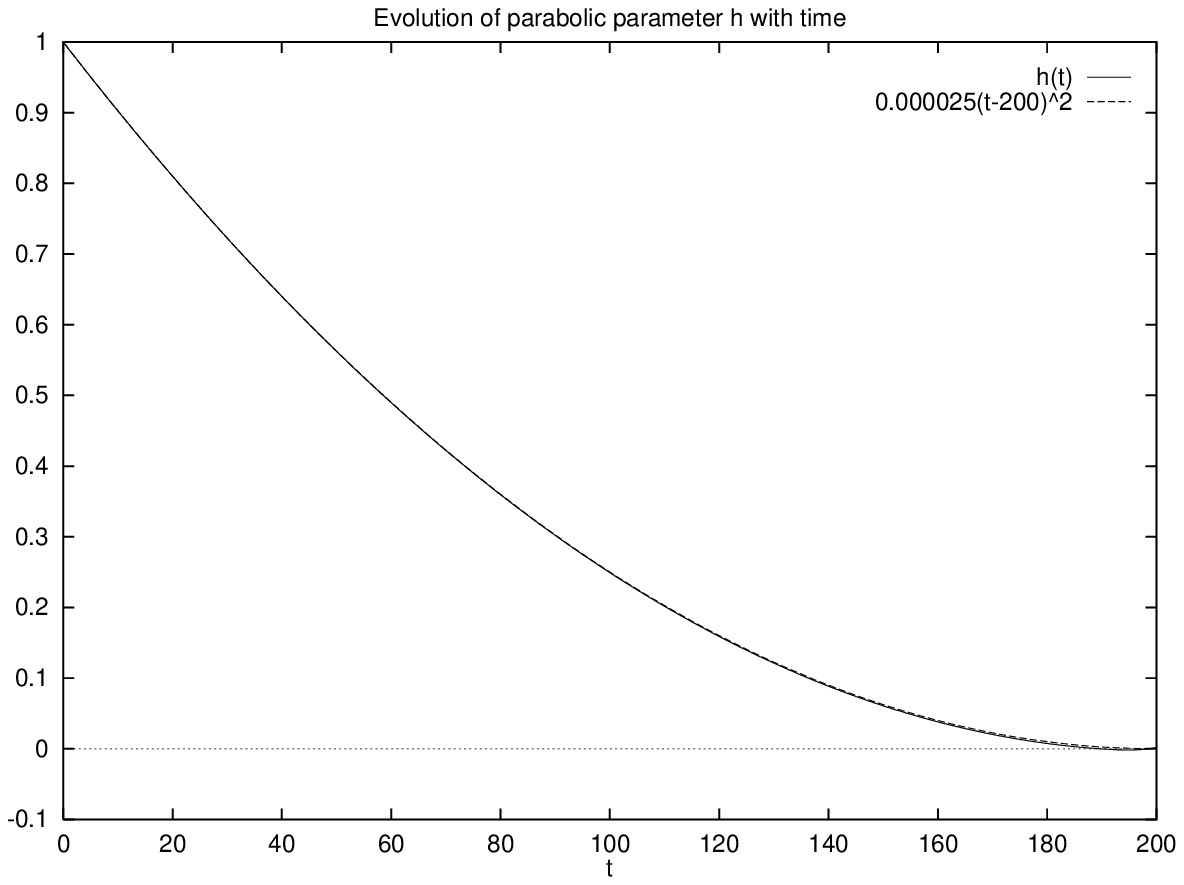}
\caption{$4+1$ dimensional model: Parabolic parameter $h$ as a function of time evolves as $f(0,t)$.}
\label{hp4p1}
\end{center}
\end{figure}

The other parameter in equation [\ref{parabform}] for the parabola,
$h$, should be given by the height of the origin, but this was
calculated in the previous section to be $a(t-T)^2$, substituting the
expressions for $c$ and $T$ we obtain: \[ h(t) =
\frac{v_0^2}{4f_0}\left(t - \frac{2f_0}{|v_0|}\right)^2.\] This is
indeed the correct form, as shown in figure \ref{hp4p1}.  The
initial conditions were $v_0 = -0.01$ and $f_0 = f(0,0) = 1.0$, hence
$h(t) = 0.000025(t - 200)^2$.  The plot of the function overlays the
data.

Now, using both expressions for $p$ and $h$, one can get the general
form of a parabolic $f(r,t)$, which is \begin{equation} f(r,t) =
\frac{v_0^2}{8f_0}r^2 + \frac{v_0^2}{4f_0}\left(t
-\frac{2f_0}{|v_0|}\right)^2\label{pf}\end{equation}
Substitute this into the partial differential equation [\ref{PDEa}],
get a common denominator and simplify to obtain:

\begin{eqnarray*}\lefteqn{ \frac{v_0^2}{4f_0}\left[-\frac{v_0^2}{4f_0}r^2 + 2\frac{v_0^2}{4f_0}\left(t - \frac{2f_0}{|v_0|}\right)^2 + 2r^2\right] \stackrel{?}{=} }\\
& & \frac{v_0^2}{4f_0}\left[\frac{v_0^2}{4f_0}r^2 + 2\frac{v_0^2}{4f_0}\left(t - \frac{2f_0}{|v_0|}\right)^2 + 2r^2\right].\end{eqnarray*}
Clearly these two sides are not the same, and the error between them is
\[ 2\left(\frac{v_0^2}{4f_0}\right)^2r^2.\]
Since our concern is the geodesic approximation, $v_0^2/(f_0)$ is
always chosen to be less than $1/200$.  This term is always much
smaller than the $2r^2$ term, and since we only have accurate numerics
away from the neighborhood of the singularity, the term with $t -
2f_0/|v_0|$ is on the order of magnitude $f_0/|v_0|$, which is large
compared to the correction.

\chapter{The $\IC P^1$ model, charge 1 sector}

The first thing to identify in this problem are the static solutions
determined by equation [\ref{genPDEb}].  These are outlined in
\cite{Ward} among others.  The entire space of static solutions can be
broken into finite dimensional manifolds $\mathcal{M}_n$ consisting of
the harmonic maps of degree $n$.  If $n$ is a positive integer, then
$\mathcal{M}_n$ consists of the set of all rational functions of $z =
x + iy$ of degree $n$.  For this chapter, we restrict our attention to
$\mathcal{M}_1$, the charge one sector, on which all static solutions have the
form
\begin{equation} u = \alpha + \beta(z + \gamma)^{-1}.\label{genc1soln}\end{equation}

In order to simplify, consider only solutions of the form
\[ \beta z^{-1}\]
or look for a real radially symmetric function $f(r,t)$ so that this 
evolution occurs as: 
\[ \frac{f(r,t)}{z}.\]

It is straightforward to calculate the evolution equation for $f(r,t)$.  It is:

\begin{equation} \fddot = f'' + \frac{3f'}{r} - \frac{4rf'}{f^2 + r^2} + 
\frac{2f}{f^2 + r^2}\left(\fdot^2 - f'^2\right).\label{PDEb}\end{equation}

The static solutions for $f(r,t)$ are the horizontal lines $f(r,t)
= c$.  In the adiabatic limit motion under small velocities
should progress from line to line, i.e $f(r,t) = c(t)$.  $f(r,t) = 0$
is a singularity of this system, where the instantons are not well
defined.  We use this to form a numerical approximation to the
adiabatic limit to observe progression from $f(r,0) = c_0 > 0$ towards
this singularity. 
  
\section{Numerics for the $\IC P^1$ charge 1 sector model}

A finite difference method is used to compute the evolution of [\ref{PDEb}]
numerically.  As with the $4+1$ dimensional model, centered differences are used consistently except for
\begin{equation} f'' + \frac{3f'}{r}.\label{instabpart}\end{equation}
In order to avoid serious instabilities in [\ref{PDEb}] this is
modeled in a special way.  Let
\[ \mathcal{L} f = r^{-3} \partial_r r^3 \partial_r f = f'' + \frac{3f'}{r}.\]
In Appendix B it is shown that this operator has negative real
spectrum, hence it is stable.  The naive central differencing scheme
on [\ref{instabpart}] results in unbounded growth at the origin, but
the natural differencing scheme for this operator does not.  It is
\[ \mathcal{L}f \approx r^{-3}\left[ \frac{\left(r + \displaystyle{\frac{\delta}{2}}\right)^3\left(\displaystyle{\frac{f(r+\delta) - f(r)}{\delta}}\right) - \left(r-\displaystyle{\frac{\delta}{2}}\right)^3\left(\displaystyle{\frac{f(r) - f(r-\delta)}{\delta}}\right)}{\delta}\right].\]

With the differencing explained, we want to derive $f(r,t+\delt)$.  We
always have an initial guess at $f(r,t+\delt)$.  In the first time
step it is $f(r,t+\delt) = f(r,t) + v_0\delt$ with $v_0$ the initial
velocity given in the problem.  On subsequent time steps $f(r,t+\delt)
= 2f(r,t) - f(r,t-\delt)$.  This can be used to compute $\fdot(r,t)$
on the right hand side of [\ref{PDEb}].  Then solve for a new and
improved $f(r,t+\delt)$ in the differencing for the second derivative
$\fddot(r,t)$ and iterate this procedure several times to get
increasingly accurate values of $f(r,t)$.

There remains the question of boundary conditions.  At the origin
$f(r,t)$ is presumed to be an even function, and this gives
\[ f(0,t) = \frac{4}{3} f(\delr,t) - \frac{1}{3} f(2\delr,t).\]
At the $r = R$ boundary we presume that the function is horizontal
so $f(R,t) = f(R-\delr,t)$.

\section{Predictions}
Equation [\ref{cp1lag}] gives us the Lagrangian for the general version of this problem.  We are using 
\[ u = \frac{\beta}{z}\]
for our evolution, and via the geodesic approximation, we restrict the Lagrangian integral to this space, to give an effective Lagrangian.
The integral of the spatial derivatives of $u$ gives a constant, the Bogomol'nyi bound, and hence can be ignored.  If one integrates the kinetic term over the entire plane, one sees it diverges logarithmically, so if $\beta$ is a function of time, the soliton has infinite energy.

\begin{figure}[H]
\begin{center}
\epsfig{file=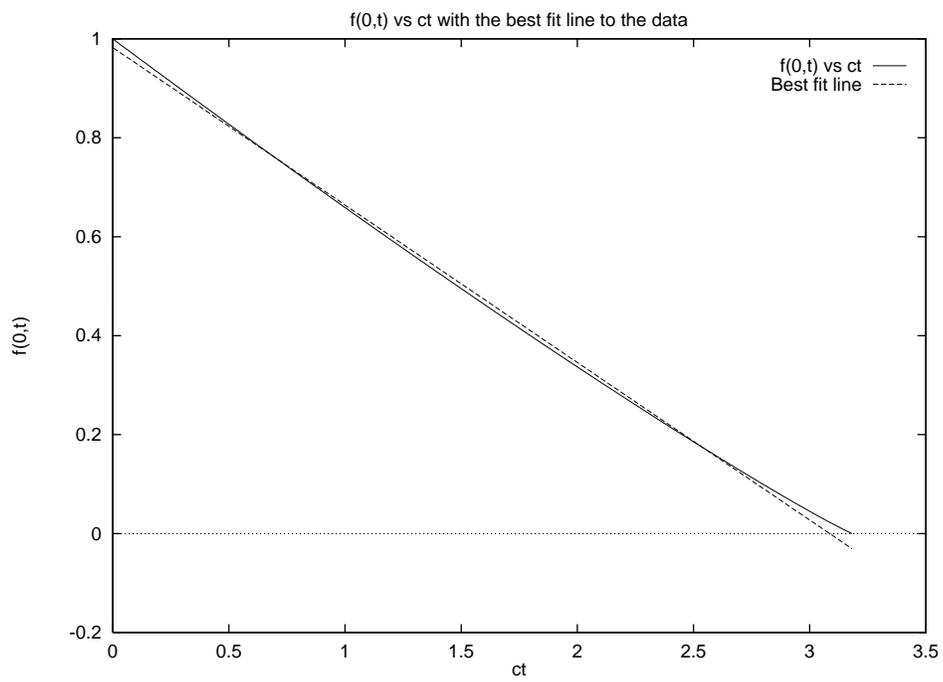}
\caption{$\IC P^1$ model, charge 1 sector: Example plot of $f(0,t)$ vs $ct$ as predicted by the effective Lagrangian.}
\label{exampc1}
\end{center}
\end{figure}

Nonetheless, this is what we wish to investigate.  We cannot address
the entire plane in our numerical procedure either, hence we presume
that the evolution takes place in a ball around the origin of size
$R$.  If $\beta = f(r,t)$ shrinks to $0$ in time $T$, we need $R > T$.
Under these assumptions, up to a multiplicative constant, the effective
Lagrangian becomes

\[ L = \int_0^R r dr \frac{r^2 \dot{f}^2}{(r^2 + f^2)^2} \]
which integrates to
\[ L = \frac{\dot{f}^2}{2} \left[ \ln\left(1 + \frac{R^2}{f^2}\right) -
\frac{R^2}{f^2 + R^2}\right]\] Since the potential energy is constant,
so is the purely kinetic Lagrangian, and
\[ \frac{\dot{f}^2}{2} \left[ \ln\left(1 + \frac{R^2}{f^2}\right) -
\frac{R^2}{f^2 + R^2}\right] = \frac{c^2}{2}, \]
with $c$ (and hence $c^2/2$) a constant.
Solving for $\dot{f}$ we obtain
\begin{equation} \dot{f} = \frac{c}{\sqrt{\left[\displaystyle{\ln\left(1 + 
\frac{R^2}{f^2}\right) - \frac{R^2}{f^2 + R^2}}\right]}}.
\label{fdoteqn} \end{equation}
Since we are starting at some value $f_0$ and evolving toward the singularity at $f = 0$ this gives:
\begin{equation} \int_{f_0}^{f(0,t)} d\, f 
\sqrt{\ln\left(1 + \frac{R^2}{f^2}\right) - \frac{R^2}{f^2 + R^2}} =
\int_0^t c dt.\label{origevo}\end{equation} The integral on the right
gives $ct$.  The integral on the left can be evaluated numerically
for given values of $R$, $f_0$ and $f(0,t)$.  A plot can then be
generated for $ct$ vs. $f(0,t)$.  What we really are concerned with is
$f(0,t)$ vs. $t$, but once the value of $c$ is determined this can be
easily obtained.  One such plot with $f_0 = 1.0$, $R = 100$ of
$f(0,t)$ vs $ct$ is given in Figure \ref{exampc1}.  This curve is
{\it almost,\ } but not quite, linear, as seen by comparison with the
best fit line to this data which is also
plotted in Figure \ref{exampc1}.  The best fit line is obtained by a
least squares method.

\clearpage

\section{Results}

The computer model was run under the condition that $f(r,0)
= f_0$ with various small velocities.  The initial velocity is
$\dot{f}(r,0) = v_0$, other input parameters are $R = r_{max}$,
$\delr$ and $\delt$.  

\subsection{Evolution of $f(0,t)$}
The primary concern with the evolution of the horizontal line is the
way in which the singularity at $f(0,t) = 0$ is approached, because once again as $r \rightarrow \infinity$ equation [\ref{PDEb}] reduces to the linear
wave equation 
\[ \fddot = f'',\]
and so we expect the interesting behavior to occur near $r=0$.  The
model is run with initial conditions that $f(r,0) = f_0$, $\fdot(r,0)
= v_0$.  Other input parameters are $R= r_{\rm max}$, $\delr$ and
$\delt$.

\begin{figure}[H]
\begin{center}
\epsfig{file=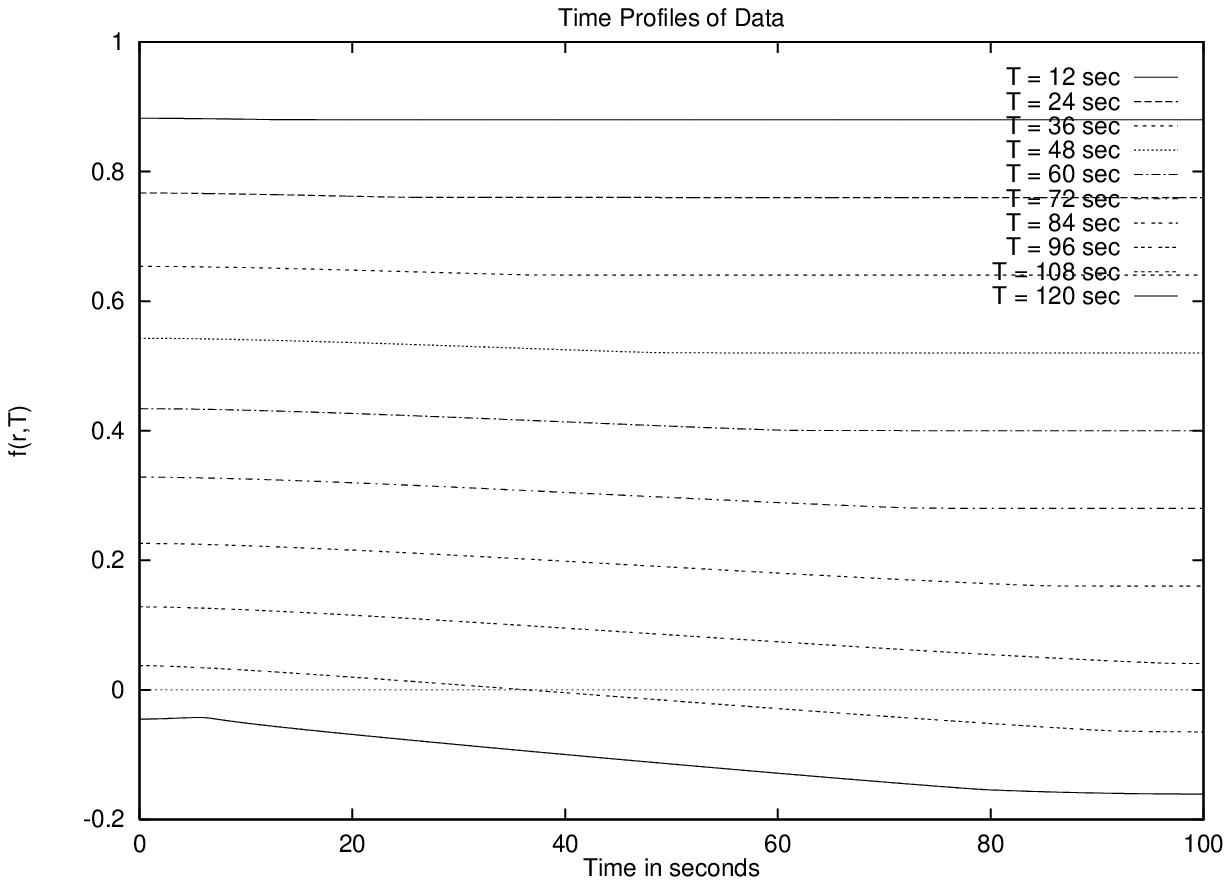}
\caption{$\IC P^1$ model, charge 1 sector: Time slices $f(r,T)$}
\label{tscp1c1}
\end{center}
\end{figure}

The evolution of the initial horizontal line seems to remain largely
flat and horizontal, although there is some slope downward as time
increases.   This is shown in figure \ref{tscp1c1}.  

\begin{figure}[H]
\begin{center}
\epsfig{file=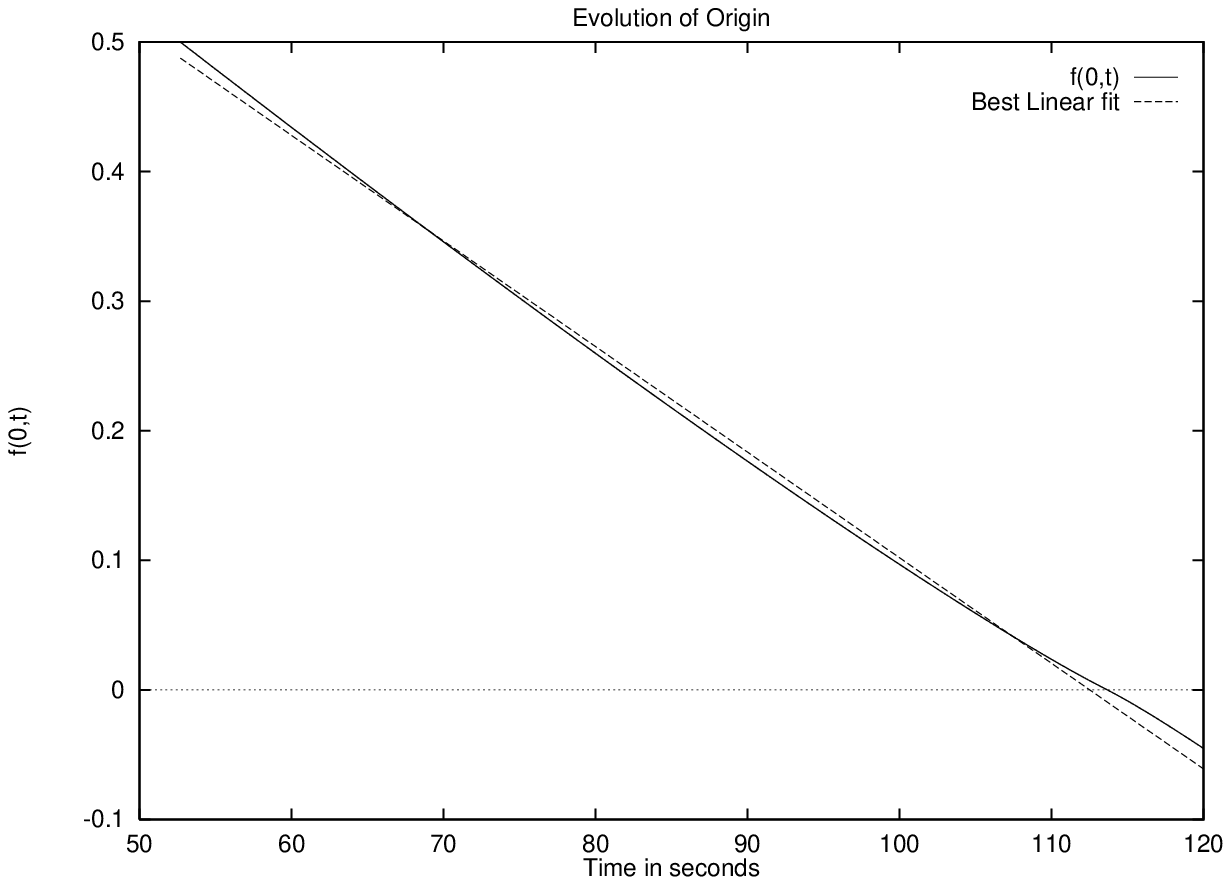}
\caption{$\IC P^1$ model, charge 1 sector: Evolution of $f(0,t)$ is not quite linear.}
\label{ocp1c1}
\end{center}
\end{figure}

We track $f(0,t)$ as it heads toward this singularity, and
find that its trajectory is not quite linear, as seen in figure
\ref{ocp1c1}.  This is suggestive of the result obtained in the
predictions for this model.

A simple characterization this evolution can be obtained via the slope
$m$ of the best fit line (via least squares) and the time $T$ at which
$f(0,T) = 0$.  Step sizes of $\delr = 0.01$ and $\delt = 0.001$ were
used with $R = 100.0$.  Table \ref{ch1f0} displays the initial data
$f_0 = f(r,0)$ and $v_0 = \dot{f}(r,0)$ along with the consequent $T$
and $m$ values.

\begin{table}[H]
\begin{center}
\caption{$\IC P^1$ model, charge 1 sector: Parameters for best fit line to $f(0,t)$ vs. initial data $f_0$ and $v_0$}
\label{ch1f0}
\[ \begin{array}{*{4}{r}} f_0 & v_0\  &\ \ \ T\ \  & m\  \\
1.0&-0.010 & 113 & -0.00815\\
2.0&-0.010 & 228 & -0.00811\\
3.0&-0.010 & 346 & -0.00793\\
4.0&-0.010 & 466 & -0.00836\\
5.0&-0.010 & 588 & -0.00828\\
1.0&-0.020 &  58 & -0.01586\\
1.0&-0.030 &  39 & -0.02336\\
1.0&-0.040 &  30 & -0.03069\\
\end{array}\]
\end{center}
\end{table}

We have 
\[ T \approx \frac{1.2 f_0}{|v_0|} \]
and 
\[ m \approx \frac{3|v_0|}{4}. \]

A better method would be to try to use the result from equation
[\ref{origevo}] in chapter 3.2.  This requires a determination of the parameters $R$
and $c$.  We already have $f_0$ and $f(0,t)$. To determine $R$ and
$c$, observe from equation [\ref{fdoteqn}] that:
\begin{equation} \frac{1}{\dot{f}^2} = \frac{\left[ 
\displaystyle{\ln\left(1 + \frac{R^2}{f^2}\right) - 
\frac{R^2}{f^2+R^2}}\right]}{c^2} \label{fdoteqnb}\end{equation}
Since $R$ is large and $f$ is small 
\[ \ln\left(1 + \frac{R^2}{f^2}\right) \approx \ln\left({\frac{R^2}{f^2}}\right)\]
and
\[ \frac{R^2}{f^2 + R^2} \approx 1.\]
Consequently we can rewrite equation [\ref{fdoteqnb}] as
\[ \frac{1}{\dot{f}^2} \approx \frac{\left[ 
\ln(R^2) - \ln(f^2) - 1\right]}{c^2}. \] 
The plot of
$\ln(f)= \ln(f(0,t))$ vs $1/{\dot{f}^2} = 1/{\dot{f}^2(0,t)}$
should be linear with the slope $m = 2/c^2$ and the intercept
$b = (2\ln(R) - 1)/c^2$.  Such a plot is easily obtained from the
model, and given slope and intercept, the parameters $c$ and $R$ are
easily obtained.

\begin{figure}[H]
\begin{center}
\epsfig{file=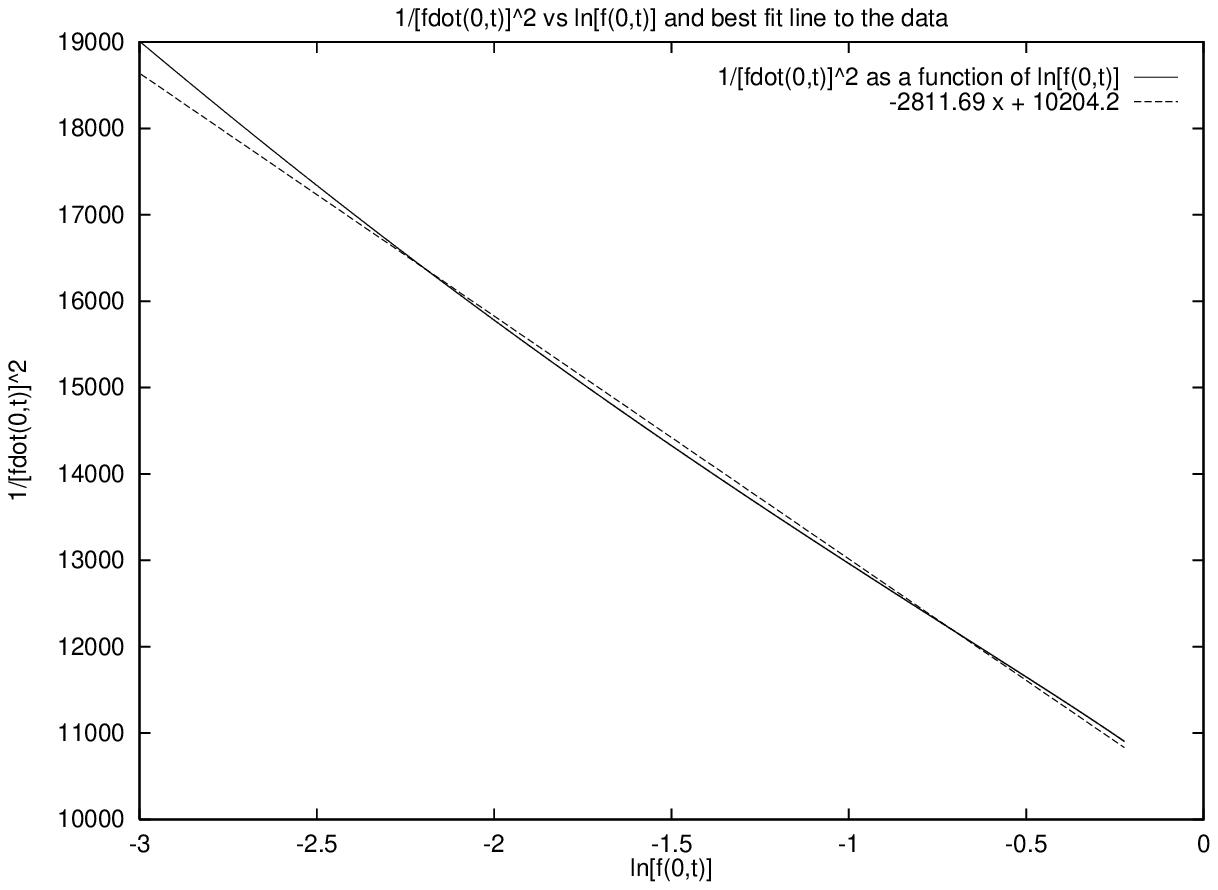}
\caption{$\IC P^1$ model, charge 1 sector: Plot of $1/(\dot{f})^2$ vs $\ln(f)$ and the best fit line to this data.}
\label{lnffdot}
\end{center}
\end{figure}

Figure \ref{lnffdot} is a plot of $\ln(f(0,t))$
vs. $1/\dot{f}^2(0,t)$, with initial conditions $\delr = 0.01$, $\delt
= 0.001$, $f_0 = 1.0$ and $v_0 = -0.01$.  It is easily seen that
although the plot of $\ln(f(0,t))$ vs. $1/\dot{f}^2(0,t)$ is nearly
straight, it is not quite a straight line.  This indicates that the
values of $R$ and $c$ are changing with time.  

\begin{figure}[H]
\begin{center}
\epsfig{file=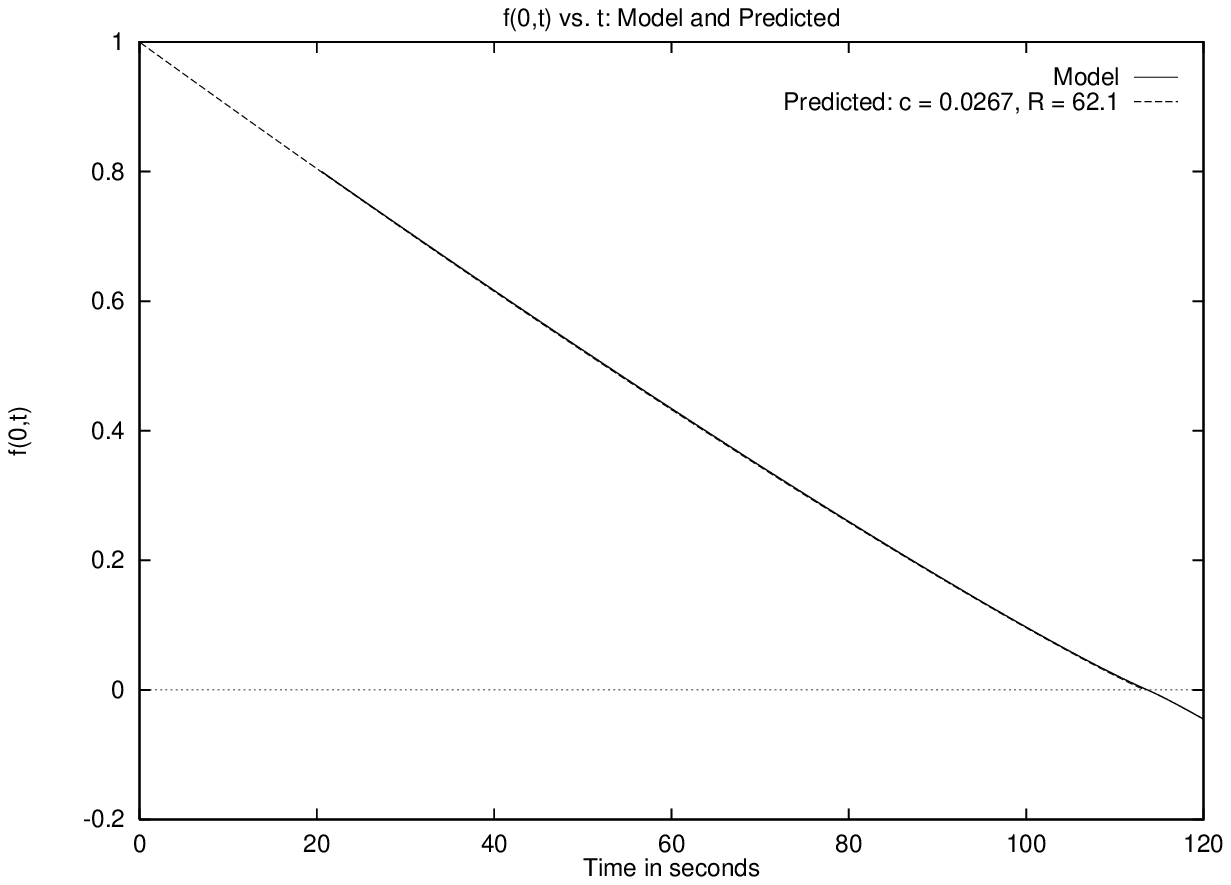}
\caption{$\IC P^1$ model, charge 1 sector: Predicted course of $f(0,t)$ from equation [\ref{origevo}] and actual course of $f(0,t)$ vs. t.}
\label{fevol}
\end{center}
\end{figure}

The best fit line $y = mx + b$ has slope $m = -2810$ and $b = 10200$.
We have
\[c = \sqrt{\frac{-2}{m}}\]
and
\[R = \exp\left(-\frac{b}{m} + \frac{1}{2}\right).\]
This gives values $c = 0.0267$ and $R = 62.1$.  Using these values of
$c$ and $R$ in the calculation of equation [\ref{origevo}], we obtain the plot
of $f(0,t)$ vs $t$ given in Figure \ref{fevol}.  This is overlayed
with the model data for $f(0,t)$ vs. $t$ for comparison.  These two
are virtually identical.

The next question is how do the parameters $c$ and $R$ vary with the
initial conditions.  The first piece of bad news is that they are
strongly dependent on the size of $\delr$ when $\delr > 0.01$,
however, they appear to converge as $\delr \rightarrow 0$.  Table
\ref{crdelrtab} contains data of $c$, $R$ vs. $\delr$ and $\delt$
under initial conditions $f_0 = 1.0$ and $v_0 = -0.01$. 

\begin{figure}[H]
\begin{center}
\epsfig{file=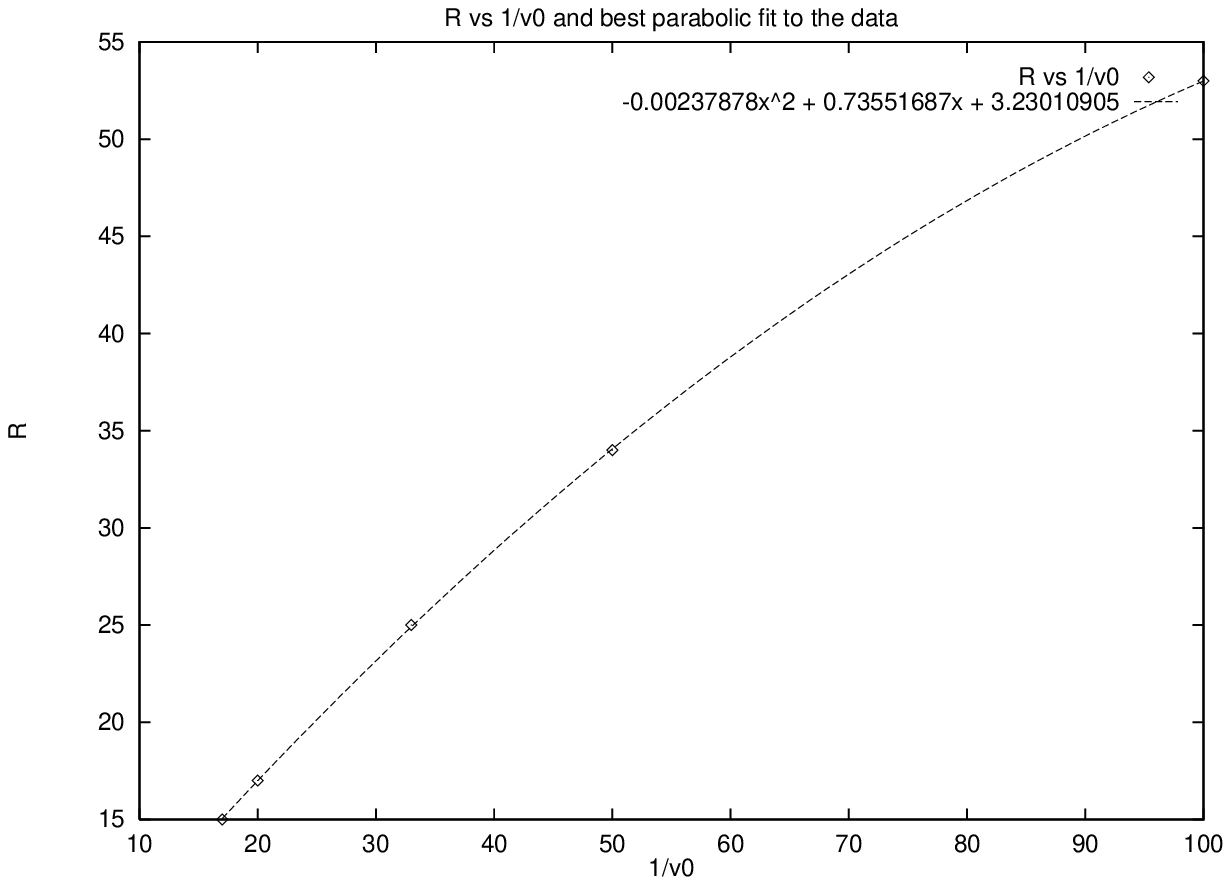}
\caption{$\IC P^1$ model, charge 1 sector: R vs 1/v0 and the 
best fit parabola.}
\label{crvfit}
\end{center}
\end{figure}

Table \ref{crvtab} contains the data for $c$ and $R$ vs. change in
the initial velocity $v_0$, under the initial conditions $f_0 = 1.0$,
$\delr = 0.01$ and $\delt = 0.001$.  The data for $R$ varying with $v_0$ 
fits well to the parabola 
\[ -0.00237878x^2 + 0.73551687x + 3.23010905.\]
This fit is shown in figure {\ref{crvfit}}.  We know this must break
down for small initial velocities, because it would send the value of
$R$ to $-\infinity$.  The values of $R$ do seem to be levelling out.
This suggests a direction for further research.

Table \ref{crftab} containing the data for $c$ and $R$ vs. change in
$f_0$, the initial height.  The parameter $R$ varies close to linearly
with $f_0$, while the parameter $c$ remains nearly constant, changing
by less than $7\%$ over the course of the runs.

\begin{table}[H]
\begin{center}
\caption{$\IC P^1$ model, charge 1 sector: $c$ and $R$ vs. $\delr$ and $\delt$.
$f_0 = 1.0$ and $v_0 = -0.01$.}
\label{crdelrtab}
\[ \begin{array}{*{4}{r}} \delr & \delt & c & R \\
0.100&     0.001 &  0.0281 &   101.\\
0.050&     0.001 &  0.0273 &    76.6\\
0.025&     0.001 &  0.0271 &    71.0\\
0.020&     0.001 &  0.0271 &    70.3\\
0.010&     0.001 &  0.0270 &    69.4\\
0.100&     0.002 &  0.0281 &   101.\\
0.100&     0.004 &  0.0281 &   101.\\
0.100&     0.005 &  0.0281 &   101.\\
\end{array}\]
\end{center}
\end{table}

\begin{table}[H]
\begin{center}
\caption{$\IC P^1$ model, charge 1 sector: $c$ and $R$ vs. $v_0$.
$f_0 = 1.0$.}
\label{crvtab}
\[ \begin{array}{{r}{l}{r}} v_0  & c & R \\
-0.01&  0.0263&  53\\
-0.02&  0.0485&  34\\
-0.03&  0.0683&  25\\
-0.05&  0.104&   17\\
-0.06&  0.121&   15 \\
\end{array}\]
\end{center}
\end{table}

\begin{table}[H]
\begin{center}
\caption{$\IC P^1$ model, charge 1 sector: $c$ and $R$ vs. $f_0$.
$v_0 = -0.01$.}
\label{crftab}
\[ \begin{array}{{r}{l}{r}} f_0  & c & R \\
1.0&   0.0267&    62\\
2.0&   0.0263&   108\\
3.0&   0.0260&   150\\
4.0&   0.0259&   190\\
\end{array}\]
\end{center}
\end{table}

\clearpage

\subsection{Characterization of time slices $f(r,T)$}
\begin{figure}[H]
\begin{center}
\epsfig{file=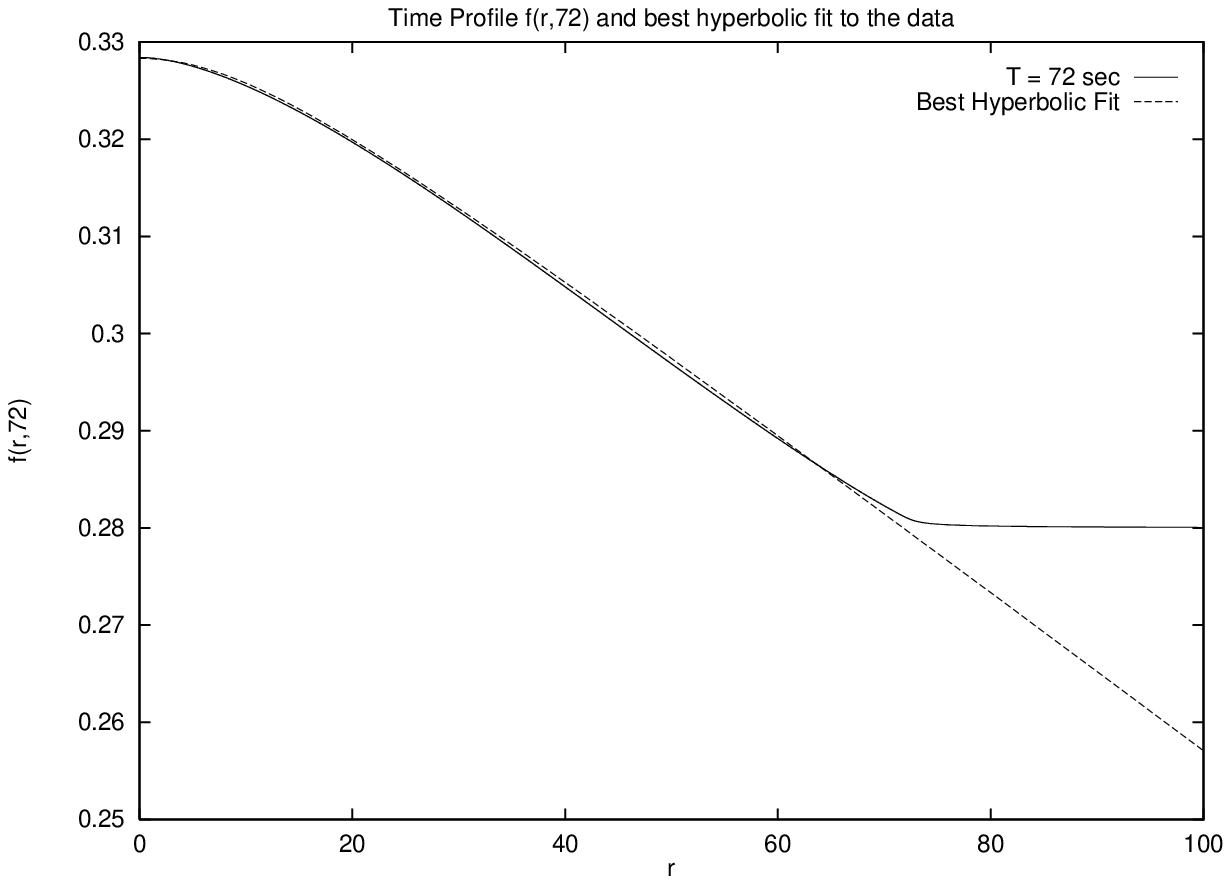}
\caption{$\IC P^1$ model, charge 1 sector: Time slices evolve hyperbolic bump at origin.}
\label{hcp1c1}
\end{center}
\end{figure}

Making a closer inspection of the time profiles $f(r, T)$ with $T$
fixed as in Figure \ref{tscp1c1}, one may observe that the initial part of
the data is close to a hyperbola as seen in Figure \ref{hcp1c1}.
The best hyperbolic fit is determined by a least squares method.

The equation for the hyperbola is 

\[ \frac{(y-k)^2}{b^2} - \frac{x^2}{a^2} = 1.\]

\begin{figure}[H]
\begin{center}
\epsfig{file=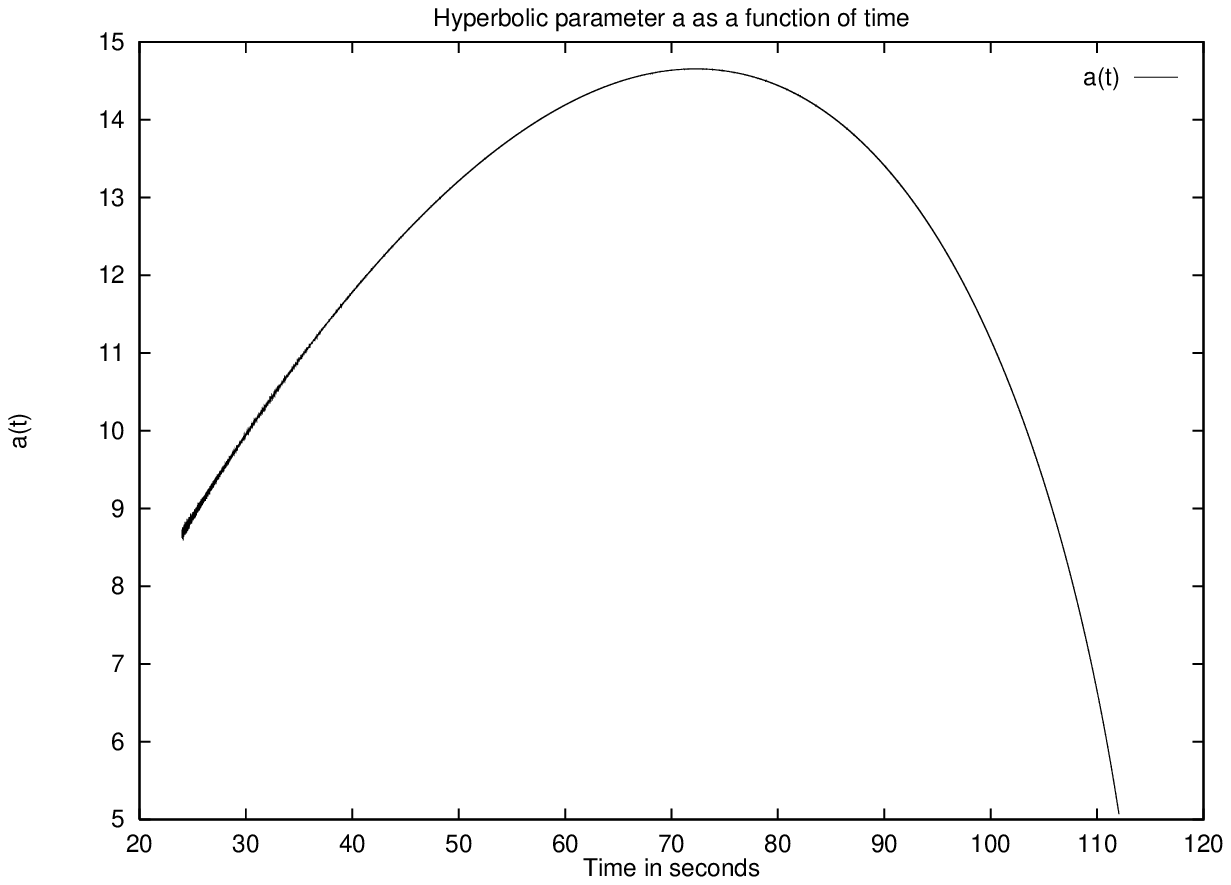}
\caption{$\IC P^1$ model, charge 1 sector: Plot of hyperbolic parameter $a$ vs time, $f_0 = 1.0$, $v_0 = -0.01$.}
\label{ha}
\end{center}
\end{figure}

\begin{figure}[H]
\begin{center}
\epsfig{file=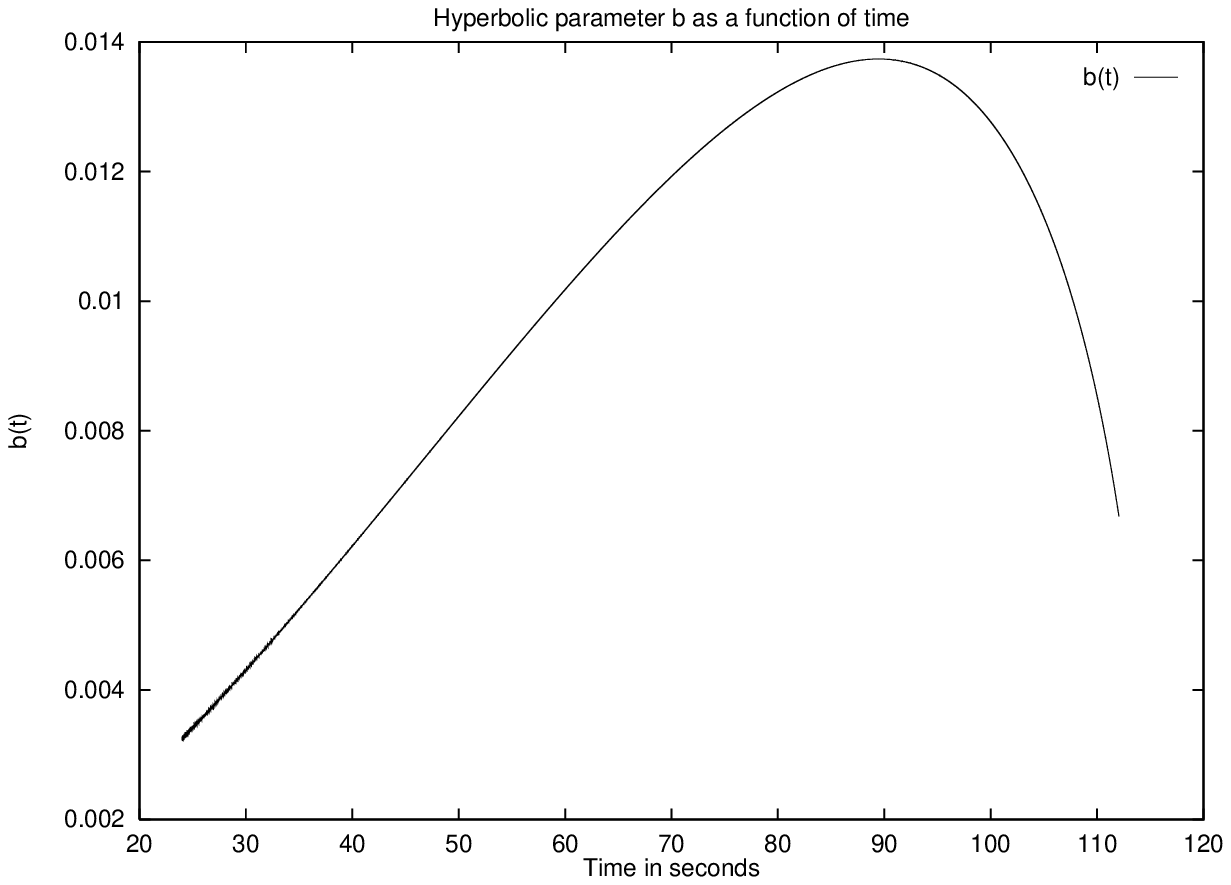}
\caption{$\IC P^1$ model, charge 1 sector: Plot of hyperbolic parameter $b$ vs time, $f_0 = 1.0$, $v_0 = -0.01$.}
\label{hb}
\end{center}
\end{figure}

\begin{figure}[H]
\begin{center}
\epsfig{file=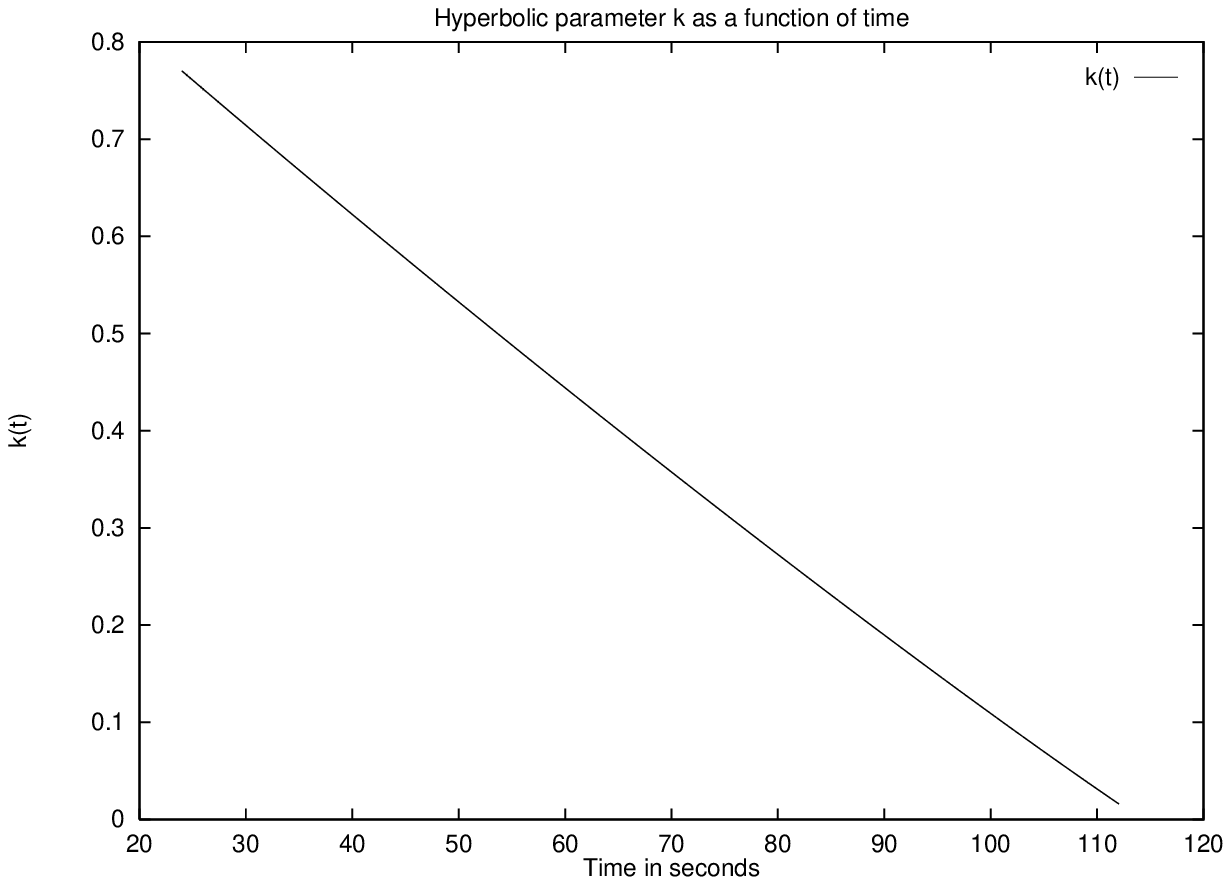}
\caption{$\IC P^1$ model, charge 1 sector: Plot of hyperbolic parameter $k$ vs time, $f_0 = 1.0$, $v_0 = -0.01$.}
\label{hk}
\end{center}
\end{figure}

One would naturally ask about the evolution of the hyperbolic
parameters $a$ and $b$ with time, however, neither of these is
particularly edifying.  A simple calculation shows that $k$ should
follow $f(0,t)$ closely if $b$ is small, as it is.  Figure \ref{ha}
shows an example of $a$ vs. time, Figure \ref{hb} shows an example
of $b$ vs. time, and Figure \ref{hk} shows an example of $k$
vs. time.  In all these figures $f_0 = 1.0$ and $v_0 = -0.01$.  

\begin{figure}[H]
\begin{center}
\epsfig{file=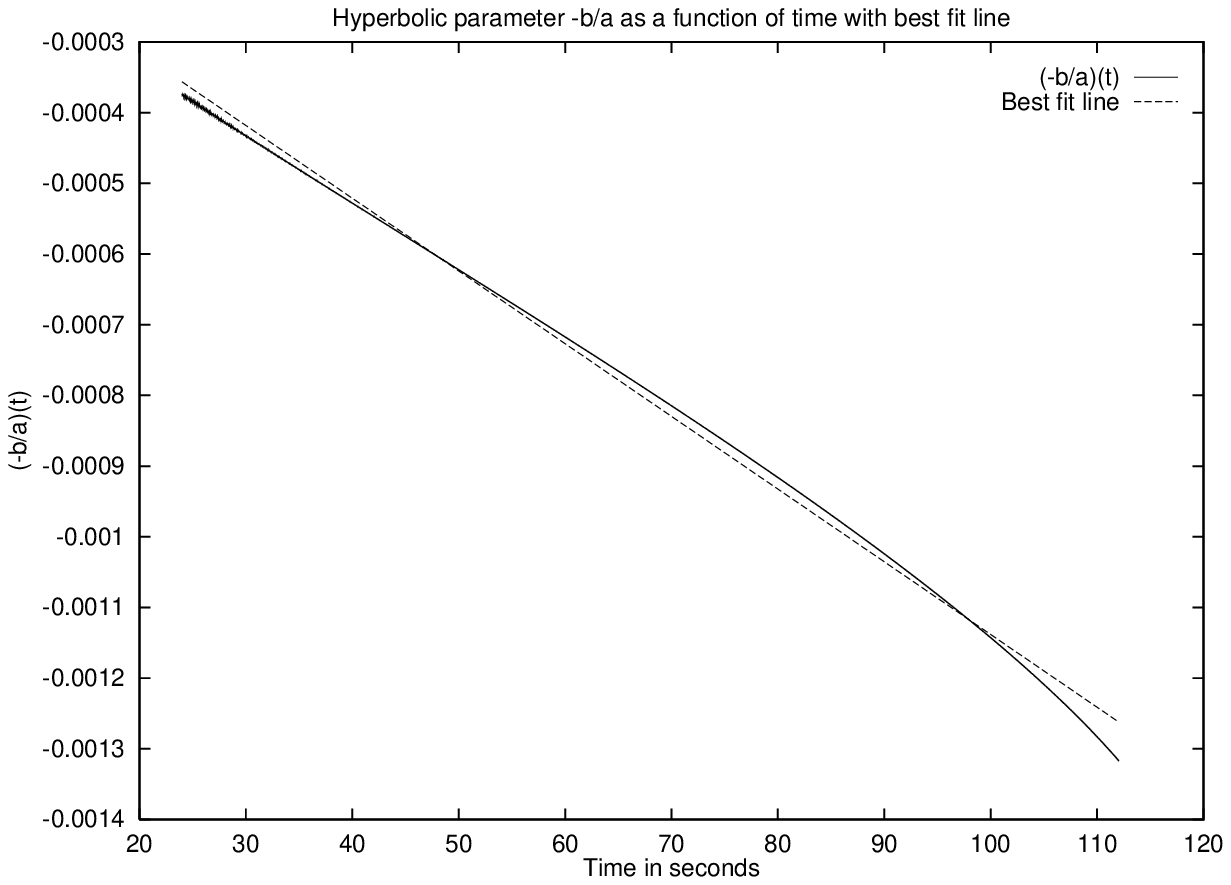}
\caption{$\IC P^1$ model, charge 1 sector: Plot of hyperbolic parameter $-b/a$ vs time, $f_0 = 1.0$, $v_0 = -0.01$.}
\label{hba}
\end{center}
\end{figure}

\begin{figure}[H]
\begin{center}
\epsfig{file=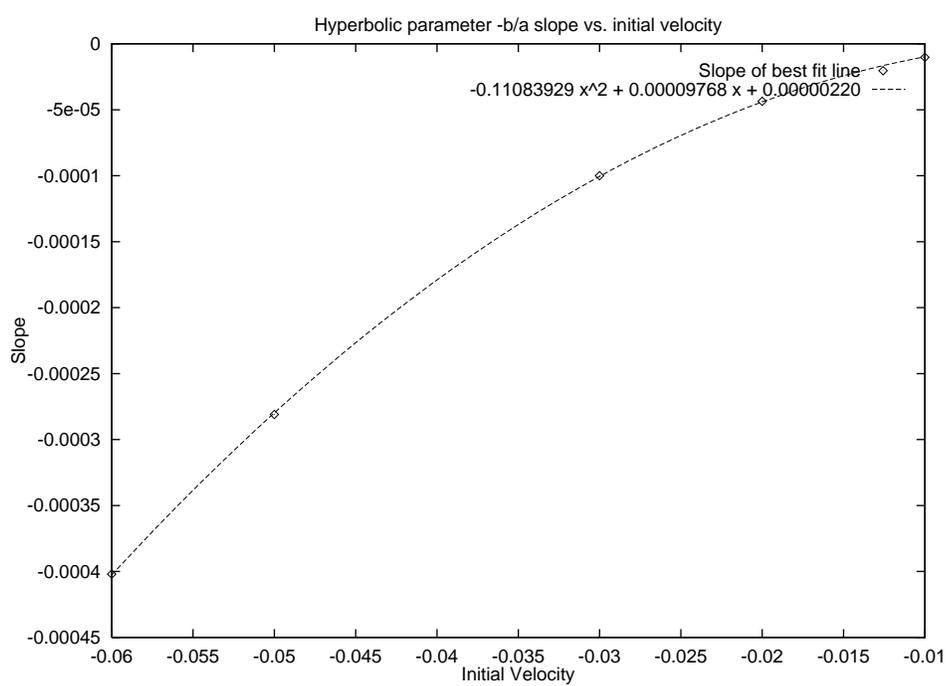}
\caption{$\IC P^1$ model, charge 1 sector: Plot of slope $m$ vs initial velocity $v_0$.}
\label{hvm}
\end{center}
\end{figure}

\begin{figure}[H]
\begin{center}
\epsfig{file=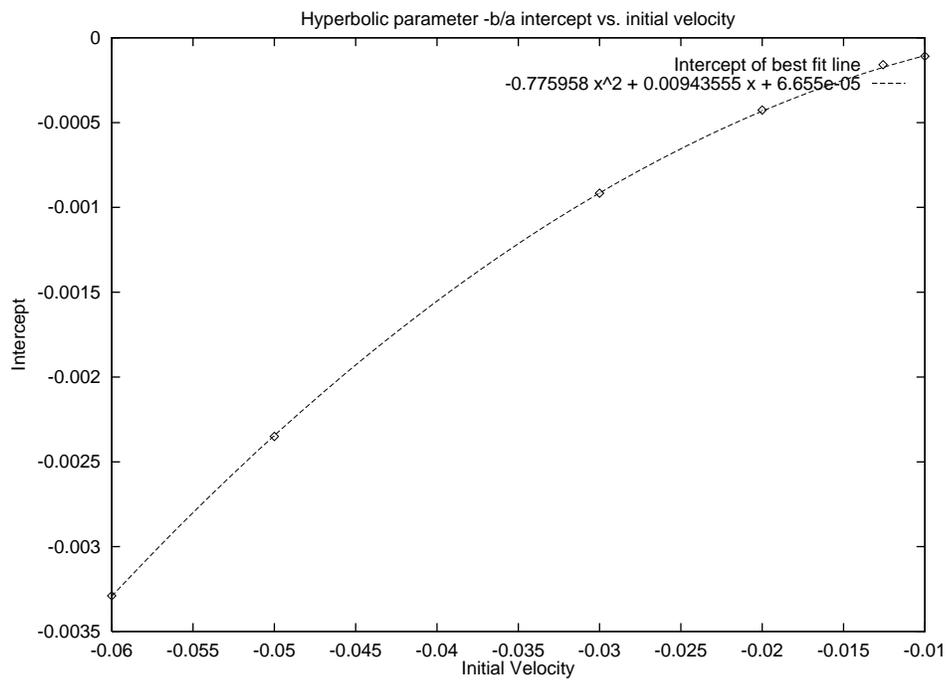}
\caption{$\IC P^1$ model, charge 1 sector: Plot of intercept $b_i$ vs initial velocity $v_0$.}
\label{hvb}
\end{center}
\end{figure}

The evolution of $-b/a$ gives the slope of the asymptotic line to the
hyperbola, and this evolution is close to linear as seen in Figure
\ref{hba}.  One might naturally ask how the slope $m$ and intercept
$b_i$ of the best fit line to the evolution of the hyperbolic $-b/a$
change with various initial conditions on $f_0$ and $v_0$.  The
variation of $m$ and $b_i$ with $f_0$ is shown in Table \ref{hbataba}.
It is easy to see the slope $m$ varies linearly with $1/f_0$, while
the intercept $b_i$ changes by only about $2\%$.  The variation of $m$
and $b_i$ with $v_0$ is shown in Table \ref{hbatabb}.  Here $m$ and $b_i$
vary quadratically with the initial velocity.  In Figure \ref{hvm} we
show the slope $m$ vs. the initial velocity $v_0$ plotted along with
the best parabolic fit $f(x)= -0.11083929 x^2 + 0.00009768 x+
0.00000220$ to this data.  In Figure \ref{hvb} we show the intercept
$b_i$ vs.  the initial velocity $v_0$ plotted along with the best
parabolic fit $f(x) = -0.77595779 x^2 + 0.00943555 x + 0.00006655$ to
this data.

\begin{table}[H]
\begin{center}
\caption{$\IC P^1$ model, charge 1 sector: Best fit line to evolution of $-b/a$
slope $m$ and intercept $b_i$ vs. $f_0$.}
\label{hbataba}
\[ \begin{array}{*{1}{r}*{2}{l}} f_0  & \ \ m &\ \  b_i \\
1.0 & -0.0000103 &  -0.000109\\
2.0 & -0.00000516&  -0.000108\\
3.0 & -0.00000344&  -0.000107\\
4.0 & -0.00000258&  -0.000107\\
\end{array}\]
\end{center}
\end{table}

\begin{table}[H]
\begin{center}
\caption{$\IC P^1$ model, charge 1 sector: Best fit line to evolution of $-b/a$
slope $m$ and intercept $b_i$ vs. $v_0$.}
\label{hbatabb}
\[ \begin{array}{*{1}{r}*{2}{l}} v_0  & \ \ m &\ \  b_i \\
-0.01&  -0.0000103 &  -0.000109\\
-0.02&  -0.0000436 &  -0.000426\\
-0.03&  -0.000100  &  -0.000916\\
-0.05&  -0.000281  &  -0.00235\\
-0.06&  -0.000402 &   -0.00329\\
\end{array}\]
\end{center}
\end{table}

\clearpage

\chapter{The $\IC P^1$ model, charge 2 sector}

It is thought that the charge 2 sector of the $\IC P^1$ model should
behave similarly to the $4+1$ dimensional instanton model.  The charge 2 sector consists of all  rational functions of $z = x+iy$ of degree 2.  A general static solution is of the form 
\[ u  = \alpha + \frac{\beta z + \gamma}{z^2 + \delta z + \epsilon} \]
depending on the five complex parameters $\alpha, \beta, \gamma,
\delta, \epsilon$.  To simplify consider only solutions of the form
\[\frac{\gamma}{z^2}\] 
with $\gamma$ real.  A straightforward calculation determines the
evolution equation for the radially symmetric function $f(r,t) =
\gamma$.  It is:

\begin{equation} \fddot = f'' + \frac{5f'}{r} - \frac{8r^3f'}{f^2 + r^4} +
\frac{2f}{f^2 + r^4}\left(\fdot^2 -
f'^2\right).\label{PDEc}\end{equation} Immediate similarities can be
seen with equation [\ref{PDEa}].  The static solutions are $f(r,T) =
c$ for $T$ fixed and any constant c.  We investigate progression from
$f(r,0) = c_0 > 0$ towards the singularity at $f(r,T) = 0$.

\section{Numerics for the $\IC P^1$ charge 2 sector model}

As with the other models, a finite difference method is used to
compute the evolution of [\ref{PDEc}] numerically.  Centered
differences are used consistently except for
\[ f'' + \frac{5f'}{r}. \]
A naive central differencing scheme here
will yield serious instabilities at the origin.  As before, let
\[ \mathcal{L} f = r^{-5} \partial_r r^5 \partial r f = f'' + \frac{5f'}{r}.\]
This operator has negative real spectrum, hence it is stable.  The
natural differencing scheme for this operator is
\[ \mathcal{L}f \approx r^{-5}\left[ \frac{\left(r + \displaystyle{\frac{\delta}{2}}\right)^5\left(\displaystyle{\frac{f(r+\delta) - f(r)}{\delta}}\right) - \left(r-\displaystyle{\frac{\delta}{2}}\right)^5\left(\displaystyle{\frac{f(r) - f(r-\delta)}{\delta}}\right)}{\delta}\right].\]
This is the differencing scheme used for these terms.

Now with the differencing explained, we derive $f(r,t+\delt)$ in
exactly the same manner as for the 4+1 dimensional model and the
charge 1 sector.  We have an initial guess for $f(r,t+\delt)$, either
given by $f(r,t+\delt) = f(r,t) + v_0\delt$ with $v_0$ the initial
velocity given in the problem, or on subsequent time steps
$f(r,t+\delt) = 2f(r,t) - f(r,t-\delt)$.  We use this guess to compute
$\fdot(r,t)$ on the right hand side of [\ref{PDEc}], and then we can
solve for a new and improved $f(r,t+\delt)$ on the left hand side of
[\ref{PDEc}].  Iterate this procedure to get increasingly accurate
values of $f(r,t)$.

The boundary condition at the origin is found by requiring that
$f(r,t) $ is an even function, hence \[f(0,t) = \frac{4}{3} f(\delr,t)
- \frac{1}{3} f(2\delr, t).\] At the $r = R$ boundary the function
should be horizontal so $f(R,t) = f(R -\delr,t)$.

\section{Predictions}
Equation [\ref{cp1lag}] gives us the Lagrangian for the general version of this problem.  We are using 
\[ u = \frac{\lambda}{z^2}\]
for our evolution, and via the geodesic approximation we restrict the
Lagrangian integral to this space, to give an effective Lagrangian.
The integral of the spatial derivatives of $u$ gives a
constant, and hence can be ignored.  
Under these assumptions, up to a multiplicative constant, the effective Lagrangian
becomes

\[ L = \int_0^{\infinity} r dr \frac{r^4\dot\lambda^2}{\left(r^4 + \lambda^2\right)^2}\]
which integrates to
\[ L = \frac{\dot\lambda^2\pi}{8\lambda}.\]
Since the potential energy is constant, so is the kinetic energy,
hence 
\[ \dot\lambda = k \sqrt{\lambda}. \]
Integrating this one obtains
\[ \lambda = (c_1 t + c_2)^2.\]
If $\lambda = 0$ occurs at time $T$, we find 
\[ T = - \frac{c_2}{c_1},\]
hence we rewrite this as
\[ \lambda(t) = a(t-T)^2.\]
Since equation [\ref{PDEc}] tends towards the linear wave equation
when $r\rightarrow\infinity$, the interesting behavior will occur at
the origin.  This is how we predict that $f(0,t)$ will evolve.

\section{Results}

The computer model was run under the condition that $f(r,0)
= f_0$ with various small velocities.  The initial velocity is
$\dot{f}(r,0) = v_0$, other input parameters are $r_{max} = R$,
$\delr$ and $\delt$.  

\subsection{Evolution of $f(0,t)$}

\begin{figure}[H]
\begin{center}
\epsfig{file=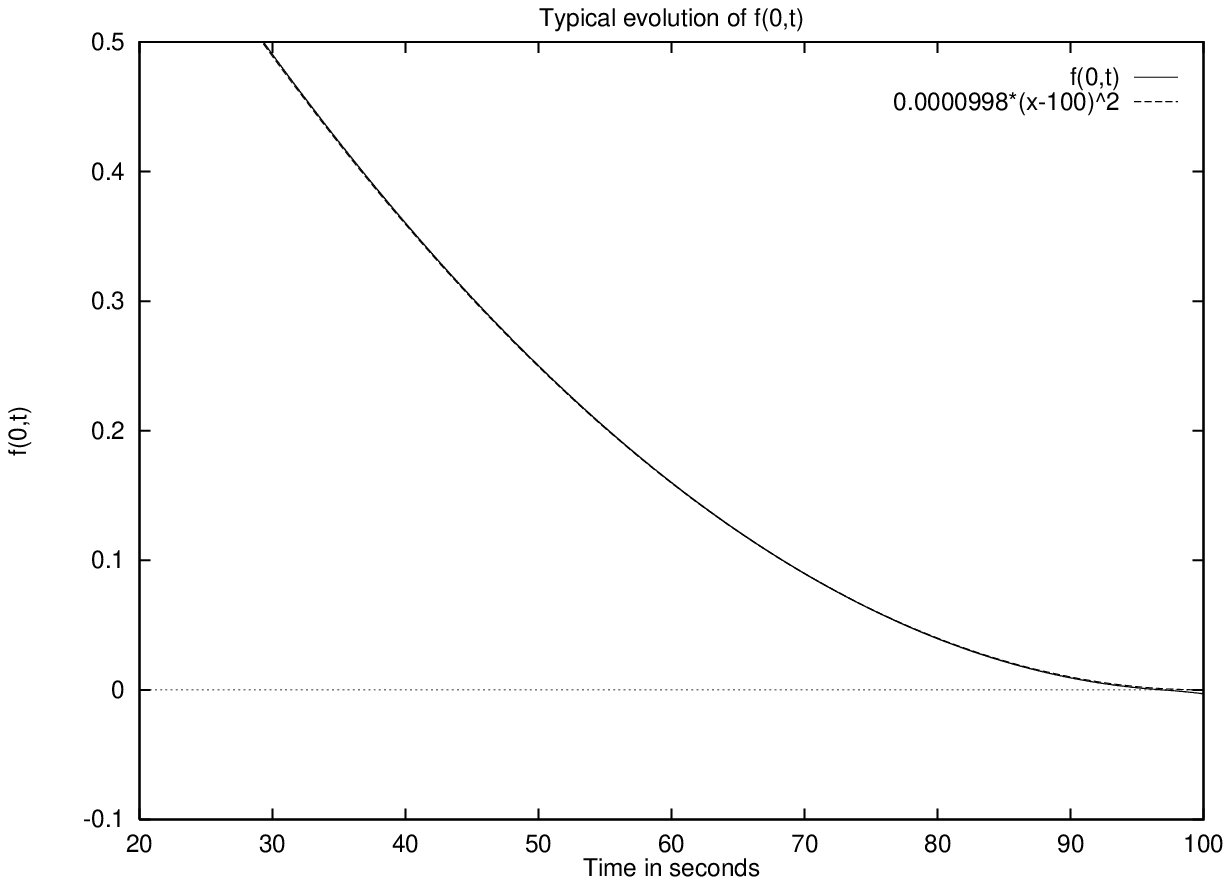}
\caption{$\IC P^1$ model, charge 2 sector: Evolution of $f(0,t)$ and overlaying fit to parabola.}
\label{fc2}
\end{center}
\end{figure}

A typical evolution of $f(0,t)$ is given in Figure \ref{fc2}.
This is best modeled by a parabola of the form $a(t-T)^2 + c$, where c
is small and grows smaller as the grid size decreases.  This curve in
particular is approximated by $0.0000998(t - 100.2)^2 -0.00102$.
This is close to the $a(t-T)^2$ predicted.

We fit $f(0,t)$ to a parabola of the form $a(t-T)^2 + c$ for various
initial conditions.  Table \ref{fch2} gives initial conditions $f_0
= f(r,0)$ and $v_0 = \dot{f}(r,0)$ and the subsequent $T$ and $a$ when
$\delr = 0.1$ and $\delt = 0.005$.

We have
\begin{eqnarray*} T &\approx& \frac{2f_0}{|v_0|}\\
\noalign{\hbox{and}}\\
a &\approx& \frac{|v_0|^2}{4f_0}.
\end{eqnarray*}
\begin{table}[h]
\begin{center}
\caption{$\IC P^1$ model, charge 2 sector: Parameters for best fit parabola to $f(0,t)$ vs. initial data $f_0$ and $v_0$.} 
\[\begin{array}{*{3}{r}{l}}  f_0 &  v_0\ &\ \ \ T\ \ & a\ \\ 
1.0&    -0.01&   204&   0.0000245\\
1.0&    -0.02&   102&   0.0000978\\
1.0&    -0.03&    68&   0.000220\\
1.0&    -0.04&    51&   0.000390\\
1.0&    -0.05&    41&   0.000609\\
0.5&    -0.01&   104&   0.0000479\\
2.0&    -0.01&   404&   0.0000124\\
3.0&    -0.01&   604&   0.00000828\\
\end{array}\]
\label{fch2}
\end{center}
\end{table}

\clearpage

\subsection{Characterization of time slices $f(r,T)$: evolution of a horizontal ine}

\begin{figure}[h]
\begin{center}
 \epsfig{file=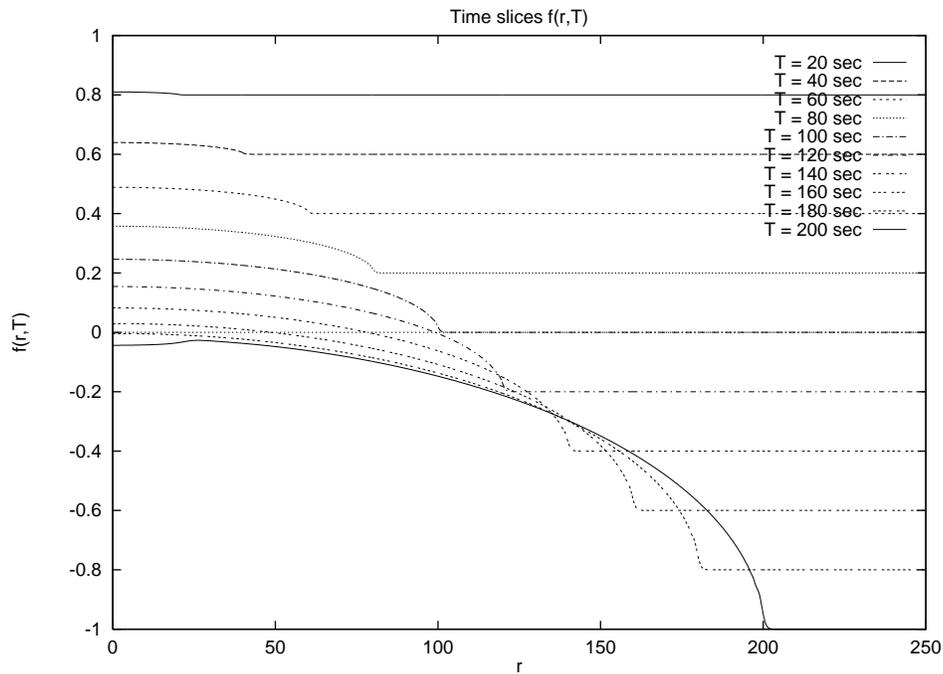}
\caption{$\IC P^1$ model, charge 2 sector: Time slices of $f(r,T)$  evolve an elliptical bump at the origin.}
\label{tsch2}
\end{center}
\end{figure}

With the evolution of $f(0,t)$ taken care of, we consider the shape of
the time slices $f(r,T)$ for a given fixed $T$.  This is rather
striking, as with the 4+1 dimensional model, an elliptical bump forms
at the origin, as seen in figure \ref{tsch2}.

As with the 4+1 dimensional model, this elliptical bump has equation
\begin{equation} \frac{x^2}{a^2} + \frac{(y-k)^2}{b^2} = 1.\label{ellbumpch2}
 \end{equation}
The question naturally arises how the parameters $a$, $b$, and $k$ evolve, and they evolve as they did in the 4+1 dimensional model.
\begin{eqnarray*} a &=& t \\
 b &=& \frac{v_0^2}{4f_0} t^2\\
 k &=& f_0 + v_0t 
\end{eqnarray*}
Table \ref{ellcp1c2} shows these values as calculated using least
squares fitting (to either a line or parabola).  Since the elliptical
fit only works before the right end of the ellipse hits the boundary
at $r = R$ the fit sometimes needed to be restricted to the portion of
the data before this occurred.  While the ellipse is small, there is a
great deal of noise in finding the elliptical parameters, and to get a
good data fit, this noise must often be removed.  Here $m_a$ and $b_a$
is the slope and intercept of the line $a(t)$, and likewise $m_k$ and
$b_k$ are the slope and intercept of the line $k(t)$.  The parameter
$c$ is that in $b(t) = ct^2$

\begin{table}[H]
\begin{center}
\caption{$\IC P^1$ model, charge 2 sector, Elliptical parameters vs. initial conditions $f_0$ and $v_0$}
\label{ellcp1c2}
\[ \begin{array}{*{7}{r}} f_0 & v_0\  & m_a\ \ \  &b_a\ \ & c\ \  & m_k\ \  & b_k\ \ \\
0.5&  -0.01&  0.998&  0.050&  0.0000496&  -0.00998&  0.500\\
1.0&  -0.01&  0.998&  0.023&  0.0000250&  -0.00999&  1.000\\
2.0&  -0.01&  1.000& -0.138&  0.0000125&  -0.01000&  2.000\\
3.0&  -0.01&  0.999& -0.080&  0.00000836& -0.01000&  3.000\\
4.0&  -0.01&  1.000& -0.173&  0.00000628& -0.01000&  4.000\\
1.0&  -0.02&  0.995&  0.206&  0.000100&   -0.01994&  1.000\\
1.0&  -0.03&  0.989&  0.370&  0.000224&   -0.02983&  0.996\\
1.0&  -0.04&  0.983&  0.507&  0.000396&   -0.03965&  0.993\\
\end{array}
\]
\end{center}
\end{table}

\begin{figure}[H]
\begin{center}
\epsfig{file=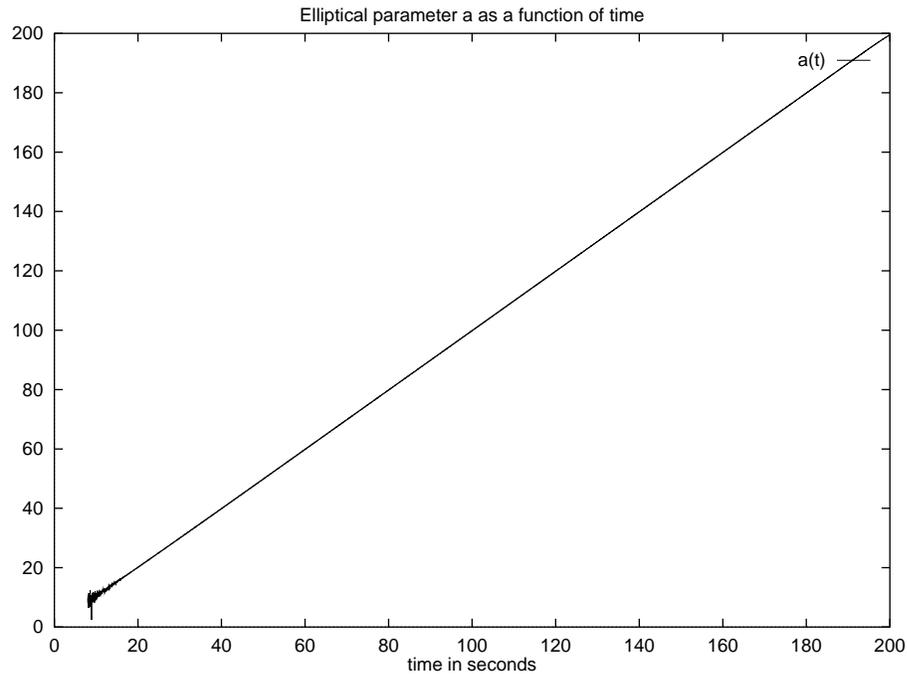}
\caption{$\IC P^1$ model, charge 2 sector: Elliptical parameter $a$ as a function of time.}
\label{acp1c2}
\end{center}
\end{figure}

A typical evolution of $a$ with $f_0 = 1.0$ and $v_0 = -0.01$ is in
Figure \ref{acp1c2}.  Note the initial noise.

\begin{figure}[H]
\begin{center}
\epsfig{file=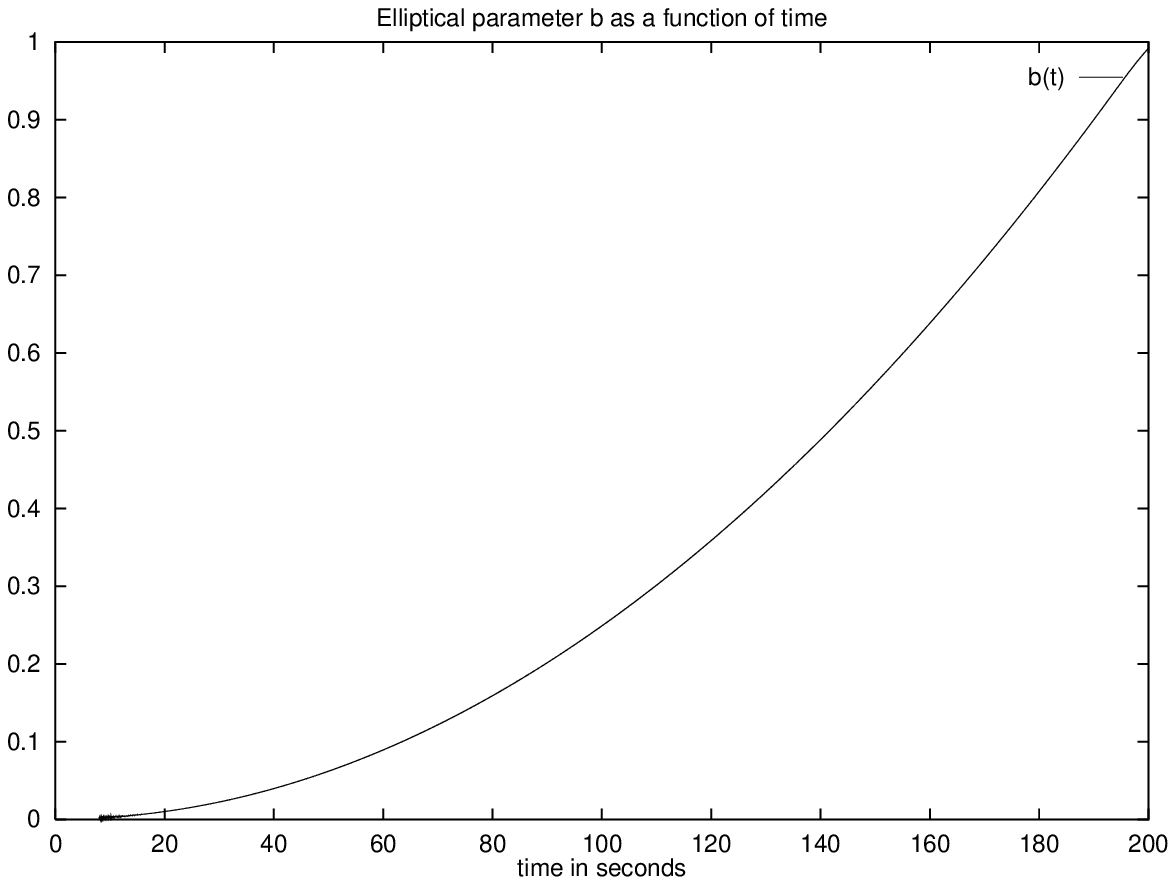}
\caption{$\IC P^1$ model, charge 2 sector: Elliptical parameter $b$ as a function of time.}
\label{bcp1c2}
\end{center}
\end{figure}

Figure \ref{bcp1c2} is a typical evolution for elliptical parameter $b$ as a function of time with $f_0 = 1.0$ and $v_0 =-0.01$.

\begin{figure}[H]
\begin{center}
\epsfig{file=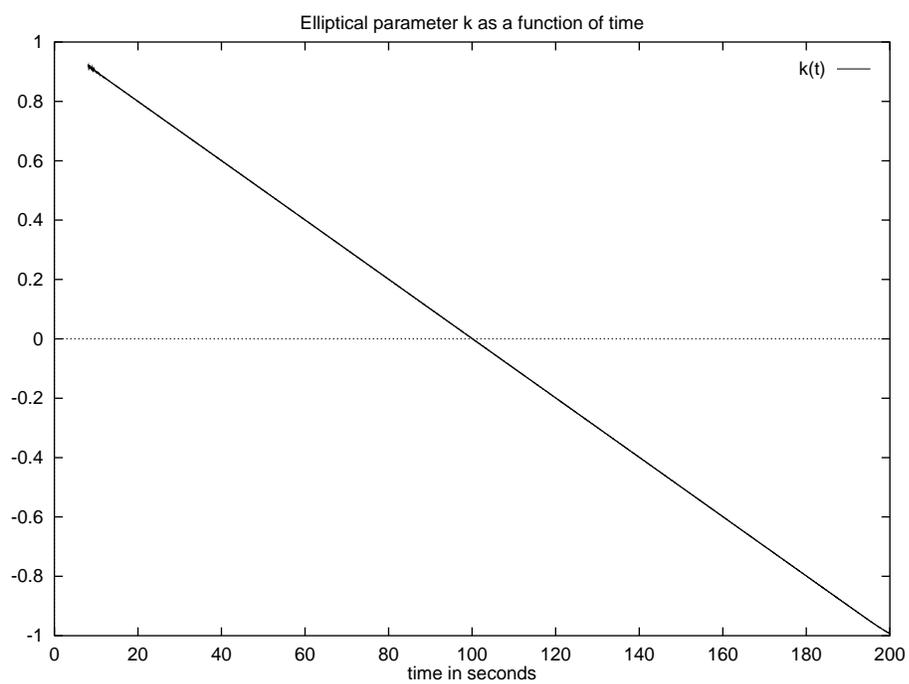}
\caption{$\IC P^1$ model, charge 2 sector: Elliptical parameter $k$ as a function of time.}
\label{kcp1c2}
\end{center}
\end{figure}

Figure \ref{kcp1c2} is a typical evolution for elliptical parameter $k$ as a function of time with $f_0 = 1.0$ and $v_0 =-0.01$.

\clearpage

\subsection{Characterization of time slices $f(r,T)$: evolution of a parabola}

As with the $4+1$ dimensional model, the evolution of the ellipse
suggested the curve was trying to obtain the shape of a parabola of the form

\begin{equation} f(r,t) = p r^2 + h. \label{parabch2}\end{equation}

To get the general form of the parabola, we follow the calculation from the $4+1$ dimensional model. From our ellipse equation
[\ref{ellbumpch2}]:
\[ \frac{dy}{dx} = -\frac{x^2b^2}{(y-k)a^2}\]
so
\[ \frac{d^2y}{dx} = \frac{-b^2}{(y-k)a^2} - \frac{xb^2}{(y-k)^2a^2}\frac{dy}{dx}\]
At $x = 0$, $y-k = b$ and this gives
\[ \frac{d^2y}{dx} = \frac{-b}{a^2}.\]
Recall from the previous section that $b = ct^2$ and $a = t$, so this
gives \[ \frac{d^2y}{dx^2} = -c.\] The identification of $c$ gives \[
\frac{d^2y}{dx^2} = -\frac{v_0^2}{4f_0}.\] So \[p =
-\frac{1}{2}\frac{d^2y}{dx^2} = -\frac{v_0^2}{8f_0}\] 

\begin{figure}[H]
\begin{center}
\epsfig{file=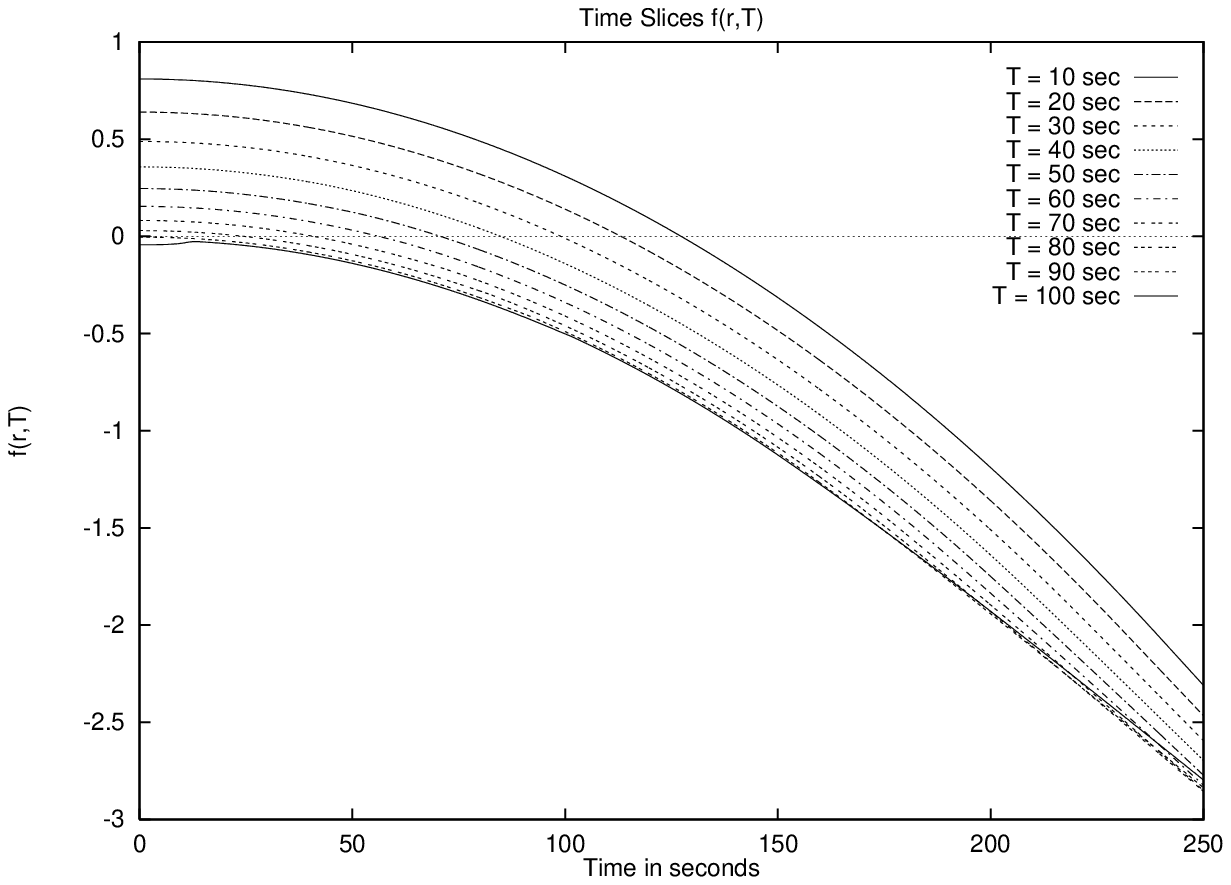}
\caption{$\IC P^1$ model, charge 2 sector: Time slices of the evolution of a 
parabola
are parabolas.}
\label{pch2}
\end{center}
\end{figure}

\begin{figure}[H]
\begin{center}
\epsfig{file=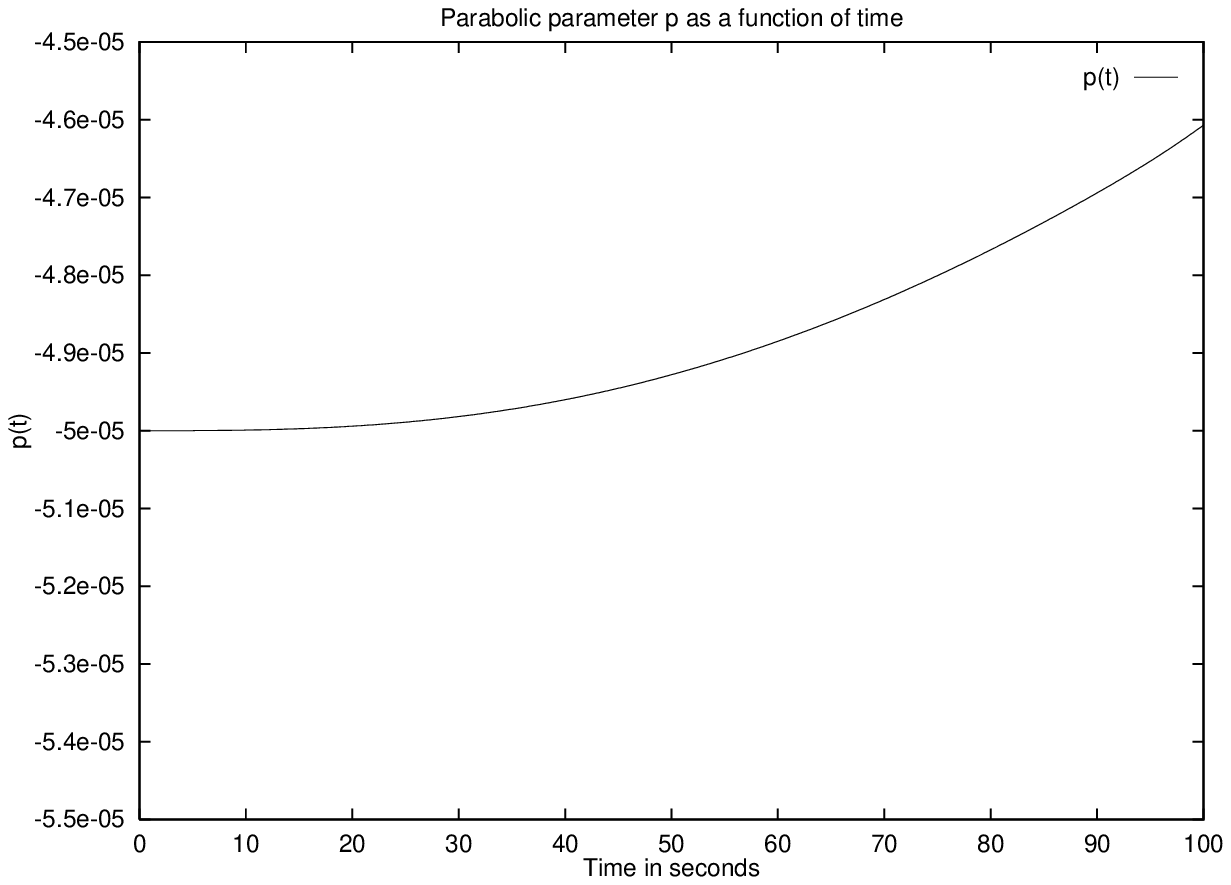}
\caption{$\IC P^1$ model, charge 2 sector: Evolution of parabolic parameter $p$ with time.}
\label{ptch2}
\end{center}
\end{figure}

\begin{figure}[H]
\begin{center}
\epsfig{file=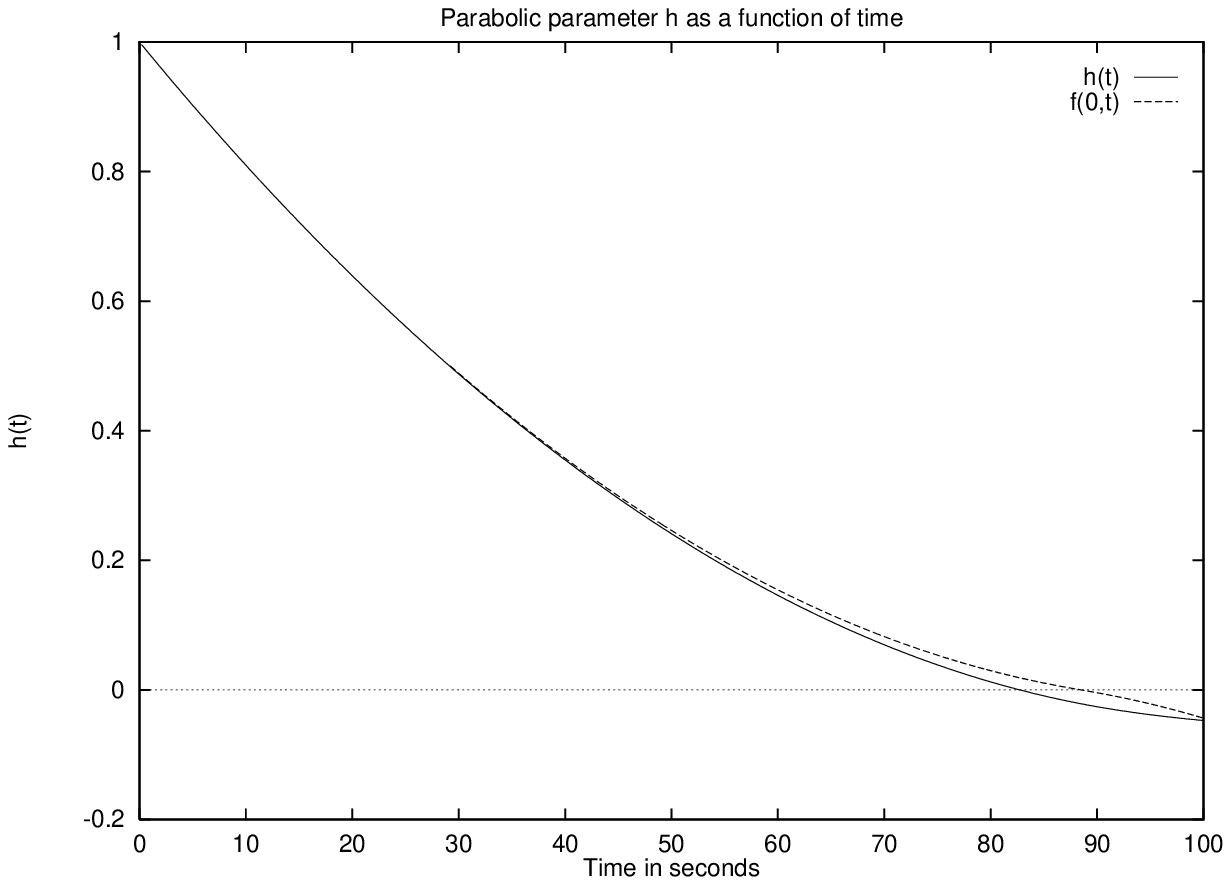}
\caption{$\IC P^1$ model, charge 2 sector: Evolution of parabolic parameter $h$ with time, comparison to $f(0,t)$.}
\label{htch2}
\end{center}
\end{figure}

When a run is started with this initial data, $\dot{f_0} = v_0 =
-0.02$, $f_0 = f(0,0) = 1.0$ and $p = \frac{v_0^2}{8f_0} = -0.00005$,
the time slices of the data have this same profile.  This is shown in
figure [\ref{pch2}].  The parabolic parameter $p$ varies by less than
1 part in 10 during the run as seen in figure [\ref{ptch2}].  The
parabolic parameter $h$ evolves close to $f(0,t)$ as seen in figure
[\ref{htch2}].

For a parabola of the form
\[ f(r,t) = p(t)r^2 + h(t), \]
we have 
\[ p(t) = -\frac{v_0^2}{8f_0},\]
{\it i.e.\ } $p(t)$ is constant, and
\[ h(t) = \frac{v_0^2}{4f_0}\left(t - \frac{2f_0}{|v_0|}\right). \]

Using the identification of $p$ and $h$ in the parabolic form of
$f(r,t)$ and letting
\[ \tau = t - \frac{2f_0}{|v_0|}\]
we have
\[ f(r,t) = -\frac{v_0^2}{8f_0}r^2 + \frac{v_0^2}{4f_0}\tau^2. \]
Substitute this into the partial differential equation [\ref{PDEc}], get a common denominator, and simplify to obtain
\begin{eqnarray*}\lefteqn{ \frac{v_0^6}{32 f_0^3}\left(\frac{r^4}{4} - r^2\tau^2 + \tau^4\right) + \frac{v_0^2}{2f_0}r^4 \stackrel{?}{=}}\\
& & -\frac{v_0^6}{32f_0^3}\frac{r^4}{4} + \frac{v_0^6}{32 f_0^3} \tau^4
+ \frac{v_0^2}{2f_0}r^4.\end{eqnarray*}
The difference between the two sides is
\[ \frac{v_0^6}{64f_0^3}r^4 - \frac{v_0^6}{32f_0^3}r^2\tau^2. \]
As with the 4+1 dimensional mode, our concern is with the geodesic approximation, and so $v_0^2/(f_0)$ is always chosen to be less than $1/200$.  The correction is then much smaller than the term 
\[ \frac{v_0^2}{2f_0}r^4. \]

\clearpage

\chapter{Conclusions}

This dissertation investigates the shrinking of solitons in the $4+1$
dimensional hyperbolic Yang Mills Lagrangian and the charge 1 and
charge 2 sectors of the $\IC P^1$ model.

Some theoretical work on the validity of the adiabatic limit for the
monopole solutions to the Yang--Mills--Higgs theory on Minkowski space
is presented in \cite{Stuart}.

Of these models, only the $\IC P^1$ model, charge 1 sector has been
investigated in any detail in the literature.  Many publications
\cite{Leese}, \cite{Ward}, \cite{Zakr}, \cite{Speight} are concerned with 
the translation and scattering of $\IC P^1$ solitons.  The shrinking
of solitons is investigated in \cite{PZ} and \cite{Speight}.  The
stability of solitons is investigated in \cite{LPZ}.

In \cite{LPZ}, solitons are found to be numerically unstable.  A
solution of the form
\[ \frac{\beta}{z} \]
shrinks spontaneously under their numerical procedure.  This does not
occur in our implementation of any of the three models in this
dissertation.  The static solutions do not evolve in time unless given
an initial rate of shrinking.  Further, stability and convergence
analysis of two of the numerical procedures is provided in the
Appendix.

In \cite{PZ} the time evolution of the shrinking of solitons was
studied.  They arbitrarily cut off the Lagrangian outside of a ball of
radius $R$, to prevent logarithmic divergence of the integral for the
kinetic energy, and then  analyze what happens in the
$R\rightarrow \infinity$ limit.  In \cite{Speight} the problem of the
logarithmic divergence in the kinetic energy integral is solved by
investigating the model on the sphere $S^2$.  The radius of the sphere
determines a parameter for the size analogous to the parameter $R$ for
the size of the ball the Lagrangian is evaluated on in \cite{PZ}.

We use a method analogous to that in \cite{PZ} of cutting off the
Lagrangian outside of a ball of radius $R$, and we find an explicit
integral for the shrinking of the soliton, dependent on two
parameters: $c$ which is a function of the kinetic energy, and $R$
which is the size of the ball on which we evaluate the Lagrangian.
Once these are specified, this integral gives the theoretical
trajectory of the soliton.  The dependence on a cut off $R$ is shown
to be a feature of the dynamics in the 1+1 dimensional partial
differential equation modeled in this dissertation, although the
dependence of $R$ on the initial conditions is unclear.  The shrinking
predicted by the cut off Lagrangian and the shrinking found in the 1+1
dimensional partial differential equation model match.

In \cite{PZ}, via the cut off Lagrangian, an ODE for the evolution is
found, and approximate solutions to this ODE were evaluated.  A 1+1
dimensional partial differential equation was also solved numerically
in \cite{PZ} for the evolution of the shrinking of solitons, although
the initial condition is not a static soliton.  They find the
shrinking rate is given by a power law corrected by a small
logarithmic term, with the corrections vanishing as $R\rightarrow
\infinity$.  In contrast, we find an explicit integral for the
shrinking of a soliton where the Lagrangian is cut off outside of the
ball of radius $R$ and that the corrections between this and the
$R\rightarrow \infinity$ limit are necessary.

Similar analysis of the predicted shrinking is given in the $\IC P^1$
model, charge 2 sector and the 4+1 dimensional model.  These two
models have no divergences.  A prediction is made by evaluating the
effective Lagrangian, and the evolution of the 1+1 dimensional partial
differential equation is shown to be close to that predicted.

In addition to this, in all of the models, the shape of a time slice
$f(r,T)$, with $T$ fixed, is characterized, by elliptical bumps at the
origin in the 4+1 dimensional model and the $\IC P^1$ model, charge 2
sector or by hyperbolic bumps at the origin in the $\IC P^1$ model,
charge 1 sector.

In the 4+1 dimensional model and the $\IC P^1$ model, a better
approximation than the geodesic approximation for solutions to the
partial differential equations are found in the evolution of the
parabolas.

\appendices 
\chapter{Stability and convergence of the 4+1 dimensional model}

In this chapter we analyze the stability of the equation
\begin{equation} \label{gPDE} \fddot = f'' + \frac{5f'}{r} + \frac{2\fdot^2}{f + r^2} 
- \frac{2(f')^2}{f + r^2} - \frac{8rf'}{f + r^2}, 
\end{equation}
and the associated differencing scheme used in finding numerical
solutions.  Here $f = f(r,t)$, and $r$ is a radial variable, hence $r
> 0$. It will be shown that the stability behavior of the differential
equation is qualitatively the same as that of the difference equation,
and that the difference equation converges to the differential
equation as $\Delta r \rightarrow 0$ and $\Delta t \rightarrow 0$.

The usual Von Neumann stability analysis has one substitute $f(r,t) =
f_0(r,t) + \ee e^{i\ww t}e^{i\kk r}$, where $f_0(r,t)$ is presumed to
solve the equation exactly, linearize in $\ee$ and then solve the
resulting equation for $\ww$ in terms of $\kk$.  In this case, if
$\ww$ has a negative imaginary part, then it has a growing mode.  The
equation is stable if it has no growing modes.  This analysis is only
valid when $\displaystyle{\kk \gg \frac{1}{\rm length\ scale}}$ and
$\displaystyle{\kk \gg \frac{1}{r}}$.

\bigskip
\section{Continuum stability: simplified model}

\medskip

The first thing to address is the stability of the linear part of the
partial differential equation [\ref{gPDE}].  This is:
\begin{equation}\label{pPDE} \fddot = f'' + \frac{5f'}{r}. \end{equation}

Set \[ f(r,t) = e^{i\kk r}e^{i\ww t},\] and plug into [\ref{pPDE}]
to obtain the equation:
\[
\omega ^{2} = 
 \kappa^{2} - \,{\displaystyle \frac {5i\,\kappa }{r}} 
\]
Solving this for $\ww$ yields:
\[\ww = \pm{\displaystyle \sqrt{\kappa^2  -\frac{ 5\,i\kk}{r}}} 
\]

So, by this analysis there is always a solution for $\ww$ with a
negative imaginary part, and therefore this equation has a growing
mode and is not stable.

However, if we consider abstractly the equation
\[ \fddot = f'' + \frac{5f'}{r} = \mathcal{L}f\]
where $\mathcal{L}f$ is a linear operator;  it is easy to see that
\[ \mathcal{L} = r^{-5}\partial_r r^5 \partial_r. \]
and hence that 
\[ \mathcal{L} = r^{-5/2}(r^{-5/2} \partial_r r^{5/2})(r^{5/2} \partial_r
r^{-5/2}) r^{5/2} = -B^{-1} A^{\dagger}AB\]
where $B = r^{5/2}$ and 
\[ A = r^{5/2} \partial_r r^{-5/2}. \]
$A^{\dagger}A$ is hermitian, and so it has real spectrum.  Because it
is essentially a square, it has positive real spectrum.
Consequently $\mbox{spec}(\mathcal{L}) = \mbox{spec}(-A^{\dagger}A)$ is
real and negative.  So the solutions to the equation
\[ \fddot = \mathcal{L}f\] 
consists of sines and cosines in the time variable multiplying the
eigenfunctions of $\mathcal{L}$, and hence has no growing mode and is
strictly stable.

So what gives?

The problem stems from the Von Neuman stability analysis which sets
$f(r,t) = e^{i\kk r} e^{i\ww t}$ instead of using the more general
stability analysis where $f(r,t) = g(r) e^{i\ww t}$.  The Von Neumann
form is used because it results in algebraic equations instead of
differential equations, which are much easier to solve.
The Von Neumann analysis yielded
\[\ww = \pm \sqrt{\kk^2  - \frac{5i\kk}{r}} = 
\pm \kk\sqrt{1 - \frac{5i}{\kk r}}\]
Since in this analysis $\kk \gg 1/r$, approximate using
Taylor's theorem to obtain
\[\ww \approx \pm \kk \mp \frac{5i}{2 r}.\]
Since velocity $v$ is 
\[ v = -\frac{\partial \ww}{\partial \kk} \approx \mp 1.\]
then $r \approx r_0 \mp t$, so
\begin{eqnarray*}
 \displaystyle \exp\left({\int i\ww {\rm d}t}\right) &\approx&
e^{\pm i\kk t}\exp\left({\pm\int\frac{5}{2(r_0 \mp t)}{\rm d}t}\right)\\ &=& 
e^{\pm i \kk t} \exp\left({-\frac{5}{2}\ln(r_0 \mp t)}\right)\\ &=& r^{-5/2}e^{\pm i\kk t}.\end{eqnarray*}
What happened was that  we let
$g(r) = e^{i\kk r}$ and so the $r^{-5/2}$ behavior had no choice but to
come out in the $e^{i\ww t}$ portion of the equation.

If, however, we forcibly put the factor of $r^{-5/2}$ into the Von Neumann
stability analysis, via
\[ f(r,t) = r^{-5/2} e^{i\kk r}e^{i\ww t} \]
plugging $f$ into [\ref{gPDE}] one obtains the equation:
\begin{eqnarray*}
 \omega^{2}
 &=& {\displaystyle  \kappa ^{2}  + \frac {15}{4r^2}}
\end{eqnarray*}
which has solutions,
\[
\ww = \pm {\displaystyle \frac {1}{2}} \,{\displaystyle \frac {\sqrt{
15 + 4\,r^{2}\,\kappa ^{2}}}{r}}  
\]

\medskip
\noindent So now, $\ww$ has only real roots, because the
``correct'' form for $g(r)$ was used. 

\section{The whole equation}

\medskip

Similarly, any analysis of the general equation~[\ref{gPDE}] will have
some ``growing modes'' analogous to the factor of $r^{-5/2}$ found in
the linear equation that do not affect the general stability.  One can
put this factor in explicitly to remove these modes.

Set 
\[ f(r,t) = f_0(r,t) + \ee r^{(-5/2)} e^{i\kk r} e^{i \ww t} \]
and linearize [\ref{gPDE}] in $\ee$ to obtain the equation:

\begin{eqnarray*}
\omega^{2} &=&    {\displaystyle \kappa ^{2}  - \frac{ 4i\,\ww\fOdot}{f_0 + r^2}}
+ {\displaystyle \frac {15}{4r^2}}  
{\displaystyle  +\frac {8r^2}{(f_0 + r^{2})^{2}}- \frac{ 10 (f_0' + 2r)}{r(f_0 +r^2)}} \\
& &{\displaystyle  + 
\frac {4i\,\kappa(f_0' + 2r)}{f_0+r^2} + 
\frac{2(f_0'+ 2\,r)^{2}}{(f_0 + r^{2})^2}}+ {\displaystyle 
\frac{2\fOdot^{2}}{(f_0 + r^{2})^2}}\\
\end{eqnarray*}

Solutions to this equation are found to be

\begin{eqnarray*}
\ww &=& {\displaystyle \frac {1}{2}} \biggl( - 4ir\fOdot \pm \Bigl[ 
- 8r^{2}\fOdot^{2} + 4\kappa^{2}r^{6} \Bigr. \biggl.\\
 & &  + 4\kappa^{2}r^{2}f_0^{2}  -8r^{2}f_0'^{2} + 
15f_0^{2}  - 40rf_0'f_0 + 16ir^{4}f_0'\kappa \\
 & & + 16ir^{2}\kappa f_0' f_0 - 72r^{3}f_0' + 32ir^{3}\kappa f_0 + 
32i\kappa r^{5} + 8\kappa^{2}r^{4}f_0 \\
 & & \biggl.\Bigl. - 50f_0r^{2} - 65r^{4}\Bigr]^{1/2}\biggr)
 \biggl/ {\vrule height0.37em width0em depth0.37em} \biggr. \! 
 \! (r(f_0 + r^{2})),
%
%
\end{eqnarray*}

Now address the realm where $\fOdot$ and $f'_0$ are small, and
$\kk$ is large.  Use a first order Taylor approximation to the
square root to obtain, 
\begin{eqnarray}
\ww &\approx& {\displaystyle \frac{-2i\fOdot}{\left(f_0 + r^2\right)}} \pm \kk \biggl[1 + {\displaystyle 
\frac{2i\left(f_0'+ 2r\right)}{\kk\left(f_0+r^2\right)} - \frac{5_0' f_0}{\kk^2r(f_0 + r^2)^2} - \frac{\fOdot^2 - f_0'^2}{\kk^2(f_0 + r^2)^2}}\nonumber\\
& & - \frac{72rf_0' + 50f_0 + 65r^2}{8\kk^2(f_0+r^2)^2}
+ O\left(\frac{1}{\kk^2}\right) + O\left(\frac{1}{\kk^4r^2}\right)\biggr]\label{wcontapprox}  
\end{eqnarray}
There are clearly some negative imaginary parts of $\ww$, but since
$\kk \gg {\displaystyle \frac{1}{r}}$ they 
are all bounded.

\bigskip

\section{Discretization scheme}

\medskip

After much preliminary work on stability with a standard finite
differencing scheme, it was discovered that the main stability
problems were generated by the linear part of the equation, namely

\begin{eqnarray*}
\fddot = f'' + {\displaystyle \frac{5f'}{r}}
\end{eqnarray*}

\noindent as $r \rightarrow 0$.  Using a naive centered difference
scheme for this part of the equation, one has irreconcilable problems
with an instability the origin which is not present in the continuum
equation.  The ``general rule of thumb'' to use is when you have a
problem, discretize the problem in the natural way for its
differential operator.  Proceeding along these lines:

\begin{eqnarray*}
{\cal{D}} f = f'' + {\displaystyle \frac{5f'}{r}}
\end{eqnarray*}

\noindent and it was  shown in section 2.1 that ${\cal{D}} =
r^{-5}\partial_r r^5 \partial_r$.  Discretize with $q$ indexing the
space variable $r$ so $q\delr = r$ and $n$ indexing the time variable
$t$ so $n\delt = t$. The discretization for ${\cal{D}}f$ is

\begin{eqnarray*}
& & (q\delr)^{-5} \Biggl[\left[\left(q + \frac{1}{2}\right)\delr
\right]^{5} \left({\displaystyle \frac{f((q+1)\delr,n\delt) 
- f(q\delr,n \delt)}{\delr}}\right)\Biggr.\\
& &
\Biggl. - \left[\left(q - \frac{1}{2}\right)\delr\right]^{5}\left({\displaystyle 
\frac{f(q\delr, n\delt) - f((q-1)\delr, n\delt)}{\delr}}\right)\Biggr]\Bigg/ \delr
\end{eqnarray*}

This differencing scheme removed the problems at the origin
completely.  Other than this, centered differences are used to discretize
the equation [\ref{gPDE}].

\bigskip
\section{Outline of the stability analysis}

\medskip

The stability analysis for this discretization scheme
will be approached in the following way.  

First, analyze the stability as $r \rightarrow 0$ by linearizing
the difference scheme as a matrix ${\cal{L}}$ and finding its eigenvalues
and eigenvectors.

Second, use Von Neumann analysis on the difference scheme
in the realm where $r$ is bounded away from zero and $f(r,t) =
f_0(r,t) + \ee e^{i\kk q\delr}e^{i\ww n\delt}$ where $f_0(r,t) \equiv
c$, a constant; $f_0(r,t) \equiv c$ does solve the equation exactly.
Find $\ww$ in terms of $\kk$ and show that the growing modes are all
bounded.

Third, analyze what happens when $f_0$ is not presumed to be
constant, but the derivatives are presumed to be close to zero.  The
equation for $\ww$ in terms of $\kk$ under this new circumstance gains
additional dependencies on $\kk$ and $\ww$.  Given $\kk,$ one can show
that inverting this map on $\ww$ yields contraction maps for $\ww$ in
balls of radius $|\kk|/2$ about $\ww =\pm\kk$.  So by the Contraction
Mapping Principle, there are solutions for $\ww$ in the balls about
$\pm\kk$.  From the form of the equation it can be seen that there are two
solutions for $\ww$, therefore these are they.  Now let $\ww_0$
indicate the solution for $\ww$ with $f_0$ presumed to be constant,
and $\ww_1$ indicate the solution for $\ww$ with no such assumption, one
can use the conclusions of the Contraction Mapping Principle to show
that the difference $|\ww_1 - \ww_0|$ is bounded.  Hence the negative
imaginary part of $\ww_1$, i.e. the growing mode, is bounded.

\bigskip
\section{Stability near zero}

\medskip

The Von Neumann stability analysis is not valid for $r \rightarrow 0$.
To analyze the stability here,  let $f = f_0 + \ee \df$ and linearize
the equation [\ref{gPDE}] in $\ee$.  Represent this as
\[ \ddot{\df} = {\mathcal{L}}_1(\df) + {\mathcal{L}}_2(\dot{\df}),\]
where ${\mathcal{L}}_1$ and ${\mathcal{L}}_2$ are linear operators,
and $\mathcal{L}_2$ is close to a multiple of the identity matrix, and
so one can replace it with $cI$.  Then discretize in space, and find
the eigenvalues and eigenvectors of $\mathcal{L}_1$, and relate these
to the eigenvalues and eigenvectors of
\begin{equation}{\label{bigmat}}\left( \begin{array}{cc} 0 & I \\ \mathcal{L}_1 & cI  \end{array} \right)
\left[ \begin{array}{c} \df \\ \dot{\df} \end{array} \right] =
\left[ \begin{array}{c} \dot{\df} \\ \ddot{\df} \end{array} \right]. 
\end{equation}

This equation is 
\[ M\vec{v} = \dot{\vec{v}}.\]
if $\lambda$ is an eigenvalue then one has 
\[ \dot{\vec{v}} = \lambda v.\]
We know the solutions to this, for the $i^{\rm th}$  component of
$\vec{v}$ we get
\[ v_i = A_i e^{\lambda t}. \]
Hence $v_i$ does not grow with time if $\lambda$ does not have a positive
real component.

Now, to relate the eigenvalues and eigenvectors of $\mathcal{L}_1$ to
the eigenvalues of the matrix in [\ref{bigmat}], presume that the
vector
\[ \left[\begin{array}{c} a \\ b \end{array}\right]\] 
is an eigenvector of the matrix in equation [\ref{bigmat}] with eigenvalue
$\lambda$.  Therefore
\[\left( \begin{array}{cc} 0 & I \\ \mathcal{L}_1 & cI \end{array}
\right)\left[\begin{array}{c} a \\ b \end{array}\right] = 
\lambda\left[\begin{array}{c} a \\ b \end{array}\right]. 
\] 
The top part of this equation states that $b = \lambda a$.  Now one can
use this in the bottom part of the equation in the following manner:
\begin{eqnarray*} \mathcal{L}_1 a + c b &=& \lambda b \\
\mathcal{L}_1 a + c \lambda a &=& \lambda^2 a 
\end{eqnarray*}
In particular one sees that $a$ must be an eigenvector of 
$\mathcal{L}_1$, and that if the eigenvalues of $\mathcal{L}_1$ are
equal to $\alpha$ then one can find the eigenvalues of the whole matrix via
the equation 
\[ \alpha a + c \lambda a = \lambda^2 a.\]
The solutions are
\[ \lambda = \frac{c \pm \sqrt{c^2 + 4\alpha}}{2}.\]
So if, as we are about to find, $\alpha < 0$, we have two cases:
\[ \left\{ \begin{array}{c@{\quad:\quad}c} c < 0 & \Re(\lambda) < 0 \\
c > 0 & \Re(\lambda) < c \end{array}\right. \] So any solution of
[\ref{bigmat}] can be written as a linear combination of solutions
with at worst bounded growth rates.

Now let's turn to showing that $\alpha < 0$ if $\alpha$ is an
eigenvalue of $\mathcal{L}_1$.  Allow $\df(q,n)$ to represent $\delta
f(q\delr, n\delt)$ and likewise with $f_0(q,n) \equiv f_0(q\delr,
n\delt)$, compute the following linearized equation, which is
discretized in space only:

\begin{eqnarray*} \lefteqn{\ddot{\df}(q,n)  = } \\
& & -2\left(\frac{\displaystyle 2q \left(\df(q+1, n) - 
\df(q-1,n)\right) - \frac{4q^2(\delr)^2\df(q,n)}{f_0(q,n) + 
(q\delr)^2}}{f_0(q,n) + (q\delr)^2}\right)  \\
& & 
+ (q\delr)^{-5} \Biggl[\left[\left(q + \frac{1}{2}\right)\delr
\right]^{5} \left({\displaystyle \frac{\df(q+1,n) 
- \df(q,n)}{\delr}}\right)\Biggr.\\
& &
\Biggl. - \left[\left(q - \frac{1}{2}\right)\delr\right]^{5}\left({\displaystyle 
\frac{\df(q, n) - \df(q-1, n)}{\delr}}\right)\Biggr]\Bigg/\delr \\ 
& &- 8\left( \frac{(q\delr)^2 \df(q,n)}{\left(f_0(q,n) + 
(q\delr)^2\right)^2}\right) \nonumber  -2 \left(\frac{(\fOdot(q,n))^2\df(q,n)}
{ \left(f_0(q,n) + (q\delr)^2\right)^2}\right) \\
& & +
4\left(\frac{\fOdot(q,n)\dot{\df}(q,n)}{f_0(q,n) + (q\delr)^2}\right) 
\end{eqnarray*}

One is concerned about behavior as $r \rightarrow 0$, and so the
functions $f_0(q\delr,n\delt)$ and $\fOdot(q\delr, n\delt)$ can be
approximated by $f_0(0,n\delt)$ and $\fOdot(0, n\delt)$ which shall
be called $f_0$ and $\fOdot$.  Likewise, $f_0 + (q\delr)^2$ will be
replaced with $f_0$, and this yields the following equation:

\begin{eqnarray} \lefteqn{\ddot{\df}(q,n)  = }\nonumber \\
& & -2\left(\frac{\displaystyle 2q \left(\df(q+1, n) - 
\df(q-1,n)\right)- \frac{4q^2(\delr)^2\df(q,n)}{f_0}}{f_0}\right) \nonumber \\
& & 
+ (q\delr)^{-5} \Biggl[\left[\left(q + \frac{1}{2}\right)\delr
\right]^{5} \left({\displaystyle \frac{\df(q+1,n) 
- \df(q,n)}{\delr}}\right)\Biggr.\nonumber\\
& &
\Biggl. - \left[\left(q - \frac{1}{2}\right)\delr\right]^{5}\left({\displaystyle 
\frac{\df(q, n) - \df(q-1, n)}{\delr}}\right)\Biggr]\Bigg/\delr {\label{premat}}\\ 
& &- 8\left( \frac{(q\delr)^2 \df(q,n)}{\left(f_0\right)^2}\right) \nonumber  -2 \left(\frac{(\fOdot)^2\df(q,n)}
{ \left(f_0\right)^2}\right) \nonumber\\
& & +
4\left(\frac{\fOdot\dot{\df}(q,n)}{f_0}\right) \nonumber
\end{eqnarray}

Since this analysis is for $r \rightarrow 0$,
\begin{eqnarray*} \mathcal{L}_2 &=&  4\left(\frac{\fOdot}{f_0}\right) I \\
&\approx& c I \\
\end{eqnarray*}
is a good approximation.

To find $\mathcal{L}_1$ use [\ref{premat}] and the quadratic fit
boundary condition at the origin, i.~e.
\[ \df(0,t) = \frac{4}{3} \df(\delr,t) - \frac{1}{3}\df(2\delr, t). \]
If one lets 
\[ \mathcal{L}_1 = [a_{i,j}],\]
and $f_0(k,t) = f_0(k\delr,t)$, 
then one obtains the following tridiagonal matrix:
\begin{eqnarray*} a_{1,1} &=& \frac{4}{3}\Biggl( \left(\frac{4}{f_0 }
\right) + (\delr)^{-2}\left[\left(\frac{1}{2}\right)^{5}
\right]\Biggr)\\
& & +(\delr)^{-2}\left[-\left(\frac{3}{2}\right)^5 - \left(\frac{1}{2}\right)^5
\right] - 2\left(\frac{\fOdot}{f_0}\right)^2\\
a_{1,2} &=& -\frac{1}{3}\left(\left(\frac{4}{f_0}
\right)+ (\delr)^{-2}\left[\left(\frac{1}{2}\right)^{5}\right]\right)\\ 
& & -\left(\frac{4}{f_0}\right)  +(\delr)^{-2}\left[\left(\frac{3}{2}\right)^5\right]\\
a_{k,k-1} &=&\frac{4k}{f_0}
+  k^{-5}(\delr)^{-2}\left[\left(k - \frac{1}{2}\right)^{5}
\right]\\ 
a_{k,k} &=& k^{-5}(\delr)^{-2}\left[-\left(k + \frac{1}{2}\right)^5 - \left(k - \frac{1}{2}\right)^5\right]  - 
2\left(\frac{\fOdot}{f_0}\right)^2\\
a_{k,k+1} &=& \left(\frac{-4k}{f_0}\right) + 
k^{-5}(\delr)^{-2}\left[\left(k + \frac{1}{2}\right)^5
\right]\\
\end{eqnarray*}

One can use Maple to compute the eigenvalues and eigenvectors of this
matrix while changing the size of the matrix $n$, and the values of
$f_0$, $\fOdot$, and $\delr$.

Since $\fOdot$ is always much less than $f_0$, the contributions from
terms with $\fOdot$ are negligible.  Further, reparametrizing by
multiplying a factor $p$ times $\fOdot$, $\delr$ and $p^2$ times $f_0$
yields eigenvalues $1/p^2$ times the originals.  In fact, the largest of
the terms in any of these matrix elements should be that multiplied by
$\delr^{-2}$, hence it is to be expected that rescaling $\delr$ itself
by a factor $p$ should result in the eigenvalues changing by
approximately $1/p^2$.

For example, when $n = 5$, $f_0 = 1$, $\fOdot = -0.01$, $\delr = 0.01$
the matrix is:
\[\left[ 
{\begin{array}{ccccc}
-75828. & 75828. & 0 & 0 & 0 \\
2381.1 & -32891. & 30510. & 0 & 0 \\
0 & 4030.7 & -25633. & 21602. & 0 \\
0 & 0 & 5145.1 & -23149. & 18004. \\
0 & 0 & 0 & 5925.0 & -22010.
\end{array}}
 \right]. 
\]
It has eigenvalues with multiplicity $m$ and eigenvectors given by:
\[ \begin{array}{*{2}{r}@{\;\;\; [}r*{4}{@{,\,}r}@{]}}
\multicolumn{1}{c}{\mbox{Eigenvalue}} & \multicolumn{1}{c}{m} & 
\multicolumn{5}{c}{\mbox{Eigenvector}} \\ 
  -79876.&1& 8.8171&-.47042&.036302&-.0034019&.00034922.\\
  -42162.&1&-2.0274&-.90138&.43183&-.16212&.047685 \\
  -31531.&1&6.3939&3.7401&-.33103&-.60680&.37688  \\
  -18654.&1&1.7172&1.2964&.47128&-.08927&-.15705  \\
  -7316.6&1&-1.6083&-1.4544&-1.0938&-.65654&-.26464 \\
\end{array}
\]

Now rescale by $p=5$ and use parameters $f_0 =25,\: \fOdot =
-0.05,\: \delr = 0.05, \: n=5$ to get the matrix:
\[ \left[{\begin{array}{ccccc}
-3033.1 & 3033.1 & 0 & 0 & 0 \\
95.243 & -1315.6 & 1220.4 & 0 & 0 \\
0 & 161.23 & -1025.3 & 864.08 & 0 \\
0 & 0 & 205.80 & -925.96 & 720.17 \\
0 & 0 & 0 & 237.00 & -880.40
\end{array}}
 \right]. 
\]
This has eigenvalues with multiplicity $m$ and eigenvectors:
\[\begin{array}{*{2}{r}@{\;\;\; [}r*{4}{@{,\,}r}@{]}}
\multicolumn{1}{c}{\mbox{Eigenvalue}} & \multicolumn{1}{c}{m} & 
\multicolumn{5}{c}{\mbox{Eigenvector}} \\ 
-3195.9 & 1 &  8.8181 & -.47043 & .036302 & -.0034009 & .00034900  \\
-1686.6 & 1 & -2.0282 & -.90108 & .43197 & -.16222 & .047711 \\
-1261.1 & 1 & 6.3888 & 3.7349 & -.33118 & -.60692 &.37736  \\
-745.84 & 1 & 1.7173 & 1.2958 & .47102 & -.08941 & -.15695  \\
-292.56 & 1 & -1.6088 & -1.4547 & -1.0937 & -.65607 & -.26455\\
\end{array}
\]
These are $1/25^{\mbox{th}}$ of those of the previous case, as expected.

Now,  vary $\fOdot$, using parameters $f_0 = 1,\;\fOdot = -0.1,\;
\delr = 0.01,\; n = 5$ and get the matrix:
\[
\left[ 
{\begin{array}{ccccc}
-75828. & 75828. & 0 & 0 & 0 \\
2381.1 & -32891. & 30510. & 0 & 0 \\
0 & 4030.7 & -25633. & 21602. & 0 \\
0 & 0 & 5145.1 & -23149. & 18004. \\
0 & 0 & 0 & 5925.0 & -22010.
\end{array}}
 \right] 
\]
With eigenvalues:
\[\begin{array}{*{2}{r}@{\;\;\; [}r*{4}{@{,\,}r}@{]}}
\multicolumn{1}{c}{\mbox{Eigenvalue}} & \multicolumn{1}{c}{m} & 
\multicolumn{5}{c}{\mbox{Eigenvector}} \\ 
-79876. & 1 & 8.8171 & -.47042 & .036302 & -.0034019 & .00034922\\
-42162. & 1 & -2.0274 & -.90138 & .43183 & -.16212 & .047685\\
-31531. & 1 & 6.3939 & 3.7401 & -.33103 & -.60680 & .37688  \\
-18654. & 1 & 1.7172 & 1.2964 & .47128 & -.08927 & -.15705  \\
-7316.6 & 1 & -1.6083 & -1.4544 & -1.0938 & -.65654 & -.26464 \\
\end{array}
\]
These are unchanged to this level of precision from those found with
$\fOdot = -0.01$.  The contribution of the terms with $\fOdot$ is
negligible, as expected.

Now, rescale $\delr = 0.1$ while leaving $f_0 = 1$ and $\fOdot = -0.01$
and get the matrix and eigenvalues:
\[  \left[ 
{\begin{array}{ccccc}
-753.00 & 753.00 & 0 & 0 & 0 \\
31.731 & -328.91 & 297.18 & 0 & 0 \\
0 & 52.187 & -256.33 & 204.14 & 0 \\
0 & 0 & 67.291 & -231.49 & 164.20 \\
0 & 0 & 0 & 79.050 & -220.10
\end{array}}
 \right] 
\]
\[ \begin{array}{*{2}{r}@{\;\;\; [}r*{4}{@{,\,}r}@{]}}
\multicolumn{1}{c}{\mbox{Eigenvalue}} & \multicolumn{1}{c}{m} & 
\multicolumn{5}{c}{\mbox{Eigenvector}} \\ 
-806.30 & 1 & 8.1450 & -.57701 & .057337 & -.0069800 & .00094304  \\
-436.27 & 1 & -2.3858 & -1.0045 & .61732 & -.28715 & .10502  \\
-317.81 & 1 & 4.0282 & 2.3296 & -.34264 & -.49247 &.39909 \\
-175.73 & 1 & 1.3007 & .99754 & .37537 & -.10673 & -.19021  \\
-53.872 & 1 & 1.2249 & 1.1373 & .92190 & .62377 & .29669  \\
\end{array}
\]
So rescaling $\delr$ by a factor of 10 rescales the eigenvalues by
approximately a factor of $1/100^{\mbox{th}}$, with the scaling best
closest to the origin.  This is consistent with the original
equations.

Lastly,  check to see what effect rescaling $n$ or the size of the
matrix has.  Ideally since we want to explore behavior near $r = 0$ we
would want to take $n \rightarrow \infty$ as $\delr \rightarrow 0$ and
$n\delr \rightarrow 0$, but this isn't possible, so instead, simply double
the size of $n$ to $n=10$.  The matrix and eigenvalues are:
\[ \left[
{\begin{array}{*{10}{c@{\!}}}\Sc{
-75828.} & \Sc{  75828. } & \Sc{  0 } & \Sc{  0 } & \Sc{  0 } & \Sc{  0 } & \Sc{  0 } & \Sc{  0 } & \Sc{  0 } & \Sc{  0} \\
\Sc{
2381.1 } & \Sc{  -32891. } & \Sc{  30510. } & \Sc{  0 } & \Sc{  0 } & \Sc{  0 } & \Sc{  0 } & \Sc{  0 } & \Sc{  0 } & \Sc{  0} \\
\Sc{
0 } & \Sc{  4030.7 } & \Sc{  -25633. } & \Sc{  21602. } & \Sc{  0 } & \Sc{  0 } & \Sc{  0 } & \Sc{  0 } & \Sc{  0 } & \Sc{  0} \\
\Sc{
0 } & \Sc{  0 } & \Sc{  5145.1 } & \Sc{  -23149. } & \Sc{  18004. } & \Sc{  0 } & \Sc{  0 } & \Sc{  0 } & \Sc{  0 } & \Sc{  0} \\
\Sc{
0 } & \Sc{  0 } & \Sc{  0 } & \Sc{  5925.0 } & \Sc{  -22010. } & \Sc{  16085. } & \Sc{  0 } & \Sc{  0 } & \Sc{  0 } & \Sc{  0} \\
\Sc{
0 } & \Sc{  0 } & \Sc{  0 } & \Sc{  0 } & \Sc{  6496.2 } & \Sc{  -21394. } & \Sc{  14897. } & \Sc{  0 } & \Sc{  0 } & \Sc{  0} \\
\Sc{
0 } & \Sc{  0 } & \Sc{  0 } & \Sc{  0 } & \Sc{  0 } & \Sc{  6931.7 } & \Sc{  -21023. } & \Sc{  14091. } & \Sc{  0 } & \Sc{  0} \\
\Sc{
0 } & \Sc{  0 } & \Sc{  0 } & \Sc{  0 } & \Sc{  0 } & \Sc{  0 } & \Sc{  7273.9 } & \Sc{  -20783. } & \Sc{13509. } & \Sc{  0} \\
\Sc{
0 } & \Sc{  0 } & \Sc{  0 } & \Sc{  0 } & \Sc{  0 } & \Sc{  0 } & \Sc{  0 } & \Sc{  7550.2 } & \Sc{  -20618. } & \Sc{  13068.} \\
\Sc{
0 } & \Sc{  0 } & \Sc{  0 } & \Sc{  0 } & \Sc{  0 } & \Sc{  0 } & \Sc{  0 } & \Sc{  0 } & \Sc{  7777.8 } & \Sc{ -20501.}
\end{array}}\right]
 \]
\[\begin{array}{*{2}{c}*{4}{r}}
\multicolumn{1}{c}{\mbox{Eigenvalue}} & \multicolumn{1}{c}{m} & 
\multicolumn{4}{c}{\mbox{Eigenvector}} \\ 
-79898. & 1 & [8.8170, & -.47043, & .036302, & -.0034177,  \\
& & .00035827, & -.000049446, &\!\!\!\! .35708\!\times\!\! 10^{-5}, &\!\!\!\! -.44969\!\times\!\! 10^{-5},  \\
-42854. & 1 & [2.0301, & .88204, & -.44662, &.19165,  \\
 & & -.082308, & .036070, & -.016080, & .0071721, \\
& & -.0030613,& .0010666] \\
& &\!\!\!\! -.6539\!\times\!\! 10^{-6}, &\!\!\!\!-.92087\!\times\!\! 10^{-6}]\\
-39076. & 1 & [7.1614, & 3.4681, & -1.2625, & .13923,\\
& & .23670, & -.30248, & .25528, & -.17811, \\
& & .10369, & -.043440]\\
-34796. & 1 & [9.2387, & 4.9953, & -1.0334, & -.49271, \\
& & .61276, & -.30607, & .00732, & .14346, \\
& & -.15312, & .08324] \\
-29191. & 1 & [-19.285, & -11.852, & .068578, & 2.1976, \\
& & -.75346, &-.47022, & .57537, & -.10236,\\
& & -.24492, & .21945]\\
-22916. & 1 & [9.1374, & 6.3726, & 1.3700, & -1.0152,\\
& & -.40642, & .39472, & .13659, & -.21252, \\
& & -.04067, & .13029] \\
-16554. & 1 & [-3.3235, & -2.5963, & -1.1309, & .00853, \\
& & .32669, & .10839, & -.10703, & -.08724, \\
& & .03047, & .059846] \\
-10673. & 1 & [8.5362, & 7.3303, & 4.6724, & 1.8693,\\
& & -.040111, & -.71838, & -.50033, & -.014049,  \\
& & .25853, & .20488] \\
-5754. & 1 & [1.7509, & 1.6167, & 1.3015, & .89606, \\ 
& & .49400, & .16895, & -.03811, & -.12426,\\
& & -.11777, & -.062081] \\
-2158. & 1 & [.57440, & .55752, & .51684, & .45759, \\
& & .38593, & .30774, & .22905, & .15526,\\ 
& & .090760, & .038476] \\
\end{array}
\]
As before, all eigenvalues are negative.  The modes localized near
zero don't change.  The eigenvalues increase towards zero as $n$
increases because there are growing modes of this equation away from
$r=0$.

One can conclude that the eigenvalues of this matrix under reasonable
initial conditions will always all be negative, as required.  This
differencing scheme does not have any instabilities generated at the
origin.


\bigskip
\section{Stability away from zero: simplified model}

\medskip

In this section the stability of difference equation derived from
[\ref{gPDE}] with the differencing scheme outlined in section 2.1
under the special condition that $f(r,t) =f_0(r,t) + \ee e^{i\kk
q\delr}e^{i\ww n \delt}$ where $f_0(r,t) \equiv c,$ a constant, will
be addressed.

An astute reader will note that the factor of $r^{-5/2}$ is missing in
this analysis.  Its explicit inclusion makes the resultant expression
for $\ww$ far uglier than even it is now.  A part of the growing mode
is analogous to the $r^{-5/2}$ term, but not exactly the same.

Plug into [\ref{gPDE}] and linearize in $\ee$ to obtain the following
equation:
\begin{eqnarray} \frac{e^{i\ww \delt} + e^{-i\ww \delt} -2}{(\delt)^2} &=&
 +\: \displaystyle{\frac{\left(q + \displaystyle{\frac{1}{2}}\right)^5
\left(e^{i\kk \delr} - 1\right) - 
\left(q - \displaystyle{\frac{1}{2}}\right)^{5}
\left(1 - e^{-i\kk \delr}\right)}{q^5(\delr)^2}}\nonumber\\
& &-\frac{4q(e^{-i\kk\delr}-e^{i\kk \delr})}{f_0 + (q\delr)^2}.\label{lin}
 \end{eqnarray}
Let  $x = e^{i\ww \delt}$ and 
\[ J = -\frac{4q(e^{-i\kk\delr}-e^{i\kk \delr})}{f_0 + (q\delr)^2} 
 +\: \displaystyle{\frac{\left(q + \displaystyle{\frac{1}{2}}\right)^5
\left(e^{i\kk \delr} - 1\right) - 
\left(q - \displaystyle{\frac{1}{2}}\right)^{5}
\left(1 - e^{-i\kk \delr}\right)}{q^5(\delr)^2}}.
\]
Hence $J$ represents the right hand side of the equation.  One can reduce equation [\ref{lin}] to
\[ x^2 - (2 + J\delt^2)x + 1 = 0\] and so it has two
solutions if we try to solve for $x =e^{i\ww t}$.  Hence it has two
(logarithmic) solutions for $\ww$.  These are obtained via the quadratic
formula
\[ x = e^{i\ww\delt} = 1 + \frac{J\delt^2}{2} \pm \sqrt{J}\delt\sqrt{1 + 
{\displaystyle{\frac{J\delt^2}{4}}}} \]
so
\[ \ww = \frac{-i\ln\left(1 + \displaystyle{\frac{J\delt^2}{2}} \pm \sqrt{J}\delt\sqrt{1 + 
{\displaystyle{\frac{J\delt^2}{4}}}}\right)}{\delt}\] 

Clearly the biggest concern is determining exactly how big $J\delt^2$
is. Let $\kk\delr = \te$, then reduce exponentials to sines and
cosines appropriately, and expand the factors of $q\pm 1/2$, and
simplify using such information as $q = r/\delr$, to obtain
\begin{eqnarray*} J &=& 
\frac{-8i\kk r}{f_0 + r^2}\left(\frac{\sin\te}{\te}\right) 
+ \kk^2\left(\frac{2\cos\te -2}{\te^2}\right) 
+ \frac{5i\kk}{r}\left(\frac{\sin\te}{\te}\right)\\
& & 
+ \frac{5\kk^2\delr^2}{2r^2}\left(\frac{2\cos\te -2}{\te^2}\right)
+ \frac{5i\kk\delr^2}{2r^3}\left(\frac{\sin\te}{\te}\right)\\
& & + \frac{5\kk^2\delr^4}{16r^4}\left(\frac{2\cos\te -2}{\te^2}\right) 
+ \frac{i\kk\delr^4}{16r^5}\left(\frac{\sin\te}{\te}\right)
\end{eqnarray*}
Analyze the behavior as $\delr \rightarrow 0$ and $\delt
\rightarrow 0$, under the assumption that $r > 0$ is fixed and that
$\delt \ll \delr$.  One also knows that $\kk\delr < \pi$.  Consequently,
all terms go to zero as $\delr \rightarrow 0$ except the following:
\[\frac{-8i\kk r}{f_0 + r^2}\left(\frac{\sin\te}{\te}\right) 
+ \kk^2\left(\frac{2\cos\te -2}{\te^2}\right) 
+ \frac{5i\kk}{r}\left(\frac{\sin\te}{\te}\right)
+ \frac{5\kk^2\delr^2}{2r^2}\left(\frac{2\cos\te - 2}{\te^2}\right).\]
The trigonometric part is strictly bounded, and the simple condition
that $\delt \ll c\delr$ with $c$ chosen sufficiently small, implies
that $\kk\delt^2$ or $(\kk\delt)^2 \ll (c\pi)^2$ can be made as
small as one likes.

In particular, one may use the Taylor approximation to the square root.
Then one may then use the Taylor approximation to the logarithm, and with
some cancellation between the first and second term, one obtains:
\begin{eqnarray*} \ww &\approx& \frac{-i\ln\left[\displaystyle{\frac{J\delt^2}{2} 
\pm \sqrt{J}\delt\left(1 + \displaystyle{\frac{J\delt^2}{8}} + O(J^2\delt^4)
\right)}\right]}{\delt}\\
&\approx& \frac{-i\left[\pm \sqrt{J}\delt + 
O\left((\sqrt{J}\delt)^3\right)\right]}{\delt} \\
&\approx& \left[\pm i\sqrt{J} + O\left(\sqrt{J}^{\,\,3}\delt^2\right)\right]
\end{eqnarray*} 
Recall, as $\delr \rightarrow 0$ all of the terms
in $J$  go to $0$ except for
\[\frac{-8i\kk r}{f_0 + r^2}\left(\frac{\sin\te}{\te}\right) 
+ \kk^2\left(\frac{2\cos\te -2}{\te^2}\right) 
+ \frac{5i\kk}{r}\left(\frac{\sin\te}{\te}\right)
+ \frac{5\kk^2\delr^2}{2r^2}\left(\frac{2\cos\te - 2}{\te^2}\right).\]
So one has
\begin{eqnarray*}\lefteqn{\sqrt{J} \approx }\\
& & \sqrt{\left[\frac{-8i\kk r}{f_0 + r^2}\left(\frac{\sin\te}{\te}\right) 
+ \kk^2\left(\frac{2\cos\te -2}{\te^2}\right) 
+ \frac{5i\kk}{r}\left(\frac{\sin\te}{\te}\right)\right.}\\
& &\overline{\left. + \frac{5\kk^2\delr^2}{2r^2}\left(\frac{2\cos\te - 2}{\te^2}\right) + O(\delr)\right]}.\\
\end{eqnarray*}

Since $\kk$ is large while $\delr$ is small, the largest term here
should be the $\kk^2$ unmultiplied by $\delr$, and so approximating
the square root this can be modified to:
\begin{eqnarray*}\lefteqn{\sqrt{J} \approx }\\
& & i\kk\sqrt{\frac{2-2\cos\te}{\te^2}}\left[1 +
\frac{4ir}{\kk(f_0 + r^2)}
\left(\frac{\sin\te}{\te}\right)\left(\frac{\te^2}{2 - 2\cos\te}
\right)\right.\\ 
& & \left. - \frac{5i}{2\kk
r}\left(\frac{\sin\te}{\te}\right) \left(\frac{\te^2}{2-2\cos\te}\right) - \frac{5\delr^2}{4r^2} +
O\left(\frac{1}{\kk^2}\right)\right].\\
\end{eqnarray*}
Therefore the leading order term of $\sqrt{J}$ is of size $O(\kk)$.
Recalling $\kk\delr < \pi$ one can make the correction
$O\left(\sqrt{J}^{\,\,3}\delt^2\right)$ as small as one likes by
requiring that $\delt < c\delr^{3/2}$ and choosing the factor $c$
appropriately.  And given this choice of $\delt$, then one has that
\begin{eqnarray*} \ww &\approx& \pm \sqrt{\frac{2-2\cos\te}{\te^2}}\biggl[\kk
\biggr.\\
& & \underbrace{+\frac{4ir}{(f_0 + r^2)}
\left(\frac{\sin\te}{\te}\right)\left(\frac{\te^2}{2 -2\cos\te}\right)
- \frac{5i}{2 r}\left(\frac{\sin\te}{\te}\right) 
\left(\frac{\te^2}{2-2\cos\te}\right)}_{\mbox{bounded terms}}\\
& &\biggl.\underbrace{-\frac{5\kk\delr^2}{4r^2} + O\left(\frac{1}{\kk}\right) + O\left(\kk\sqrt{J}^{\,\,3}\delt^2\right)}_{\mbox{terms that go to zero}}\biggr].\\
\end{eqnarray*}
Furthermore it is clear that if one permitted $r \rightarrow 0$ that
the dominant piece would be 
\[\ww\approx \pm \kk\sqrt{\frac{2-2\cos\te}{\te^2}} -\frac{5i}{2r}\left(\frac{\sin\te}{\te}\right)
\left(\frac{\te^2}{2 - 2\cos\te}\right)\]
and as $\te \rightarrow 0$ this goes to 
\[\ww \approx \pm \kk \mp\frac{5i}{2r}\]
or exactly what we obtained in appendix A.1 that was so often
``corrected'' by the addition of a factor of $r^{-5/2}$.

\section{The complications}

\bigskip
Now we must deal with the reality that we have thus far assumed $f_0
\equiv c$, and in general this is not so.  We need to know that when
$f_0$ is a function with small bounded derivatives, that the growing modes
remain close to those in the previous analysis.

In the previous section we let $f(r,t) = f(q\delr, n\delt) = f_0(r,t) + \ee
e^{i\kk q\delr}e^{i\ww n\delt}$ under the assumption that $f_0 \equiv c$.
Call the solution for $\ww$ in terms of $\kk$ in this
analysis $\ww_0$.  The equation solved to do this was
\begin{eqnarray*} \frac{e^{i\ww \delt} + e^{-i\ww \delt} -2}{(\delt)^2} &=&
 +\: \displaystyle{\frac{\left(q + \displaystyle{\frac{1}{2}}\right)^5
\left(e^{i\kk \delr} - 1\right) - 
\left(q - \displaystyle{\frac{1}{2}}\right)^{5}
\left(1 - e^{-i\kk \delr}\right)}{q^5(\delr)^2}}\\
& &-\frac{4q(e^{-i\kk\delr}-e^{i\kk \delr})}{f_0 + (q\delr)^2}.
 \end{eqnarray*}
Characterize this equation as
\[ g(\ww) = \old(\kk).\]

If one does not assume that $f_0 \equiv c$ then one gets the equation
\begin{eqnarray} \frac{e^{i\ww \delt} + e^{-i\ww \delt} -2}{(\delt)^2} &=&
 +\: \displaystyle{\frac{\left(q + \displaystyle{\frac{1}{2}}\right)^5
\left(e^{i\kk \delr} - 1\right) - 
\left(q - \displaystyle{\frac{1}{2}}\right)^{5}
\left(1 - e^{-i\kk \delr}\right)}{q^5(\delr)^2}}\nonumber\\
& &-\frac{4q(e^{-i\kk\delr}-e^{i\kk \delr})}{f_0 + (q\delr)^2} - \frac{4f_0'}{f_0 + r^2}\left(\frac{e^{i\kk \delr} - e^{-i\kk\delr}}{2\delr}\right)\nonumber \\
& & + \frac{2\left(f_0'\right)^2}{(f_0 + r^2)^2} + \frac{8rf_0'}{(f_0 + r^2)^2}
 - \frac{2\fOdot^2}{(f_0 + r^2)^2} \nonumber \\
& &+ \frac{4\fOdot}{f_0 + r^2}\left(\frac{e^{i\ww\delt} - e^{-i\ww\delt}}{2\delt}\right)\label{gwgen}
\end{eqnarray}
This equation is quadratic in $e^{i\ww\delt}$; hence it has two
solutions for $e^{i\ww\delt}$, and hence two solutions for $\ww$ for each
branch of the logarithm.  We wish to get a handle on these two
solutions for $\ww$.  Equation [\ref{gwgen}] can be characterized as
\begin{equation} g(\ww) = \old(\kk) + \newk(\kk) + \neww(\ww)\label{gwform}\end{equation}
with 
\begin{eqnarray}\newk(\kk) &=&  - \frac{4f_0'}{f_0 + r^2}\left(\frac{e^{i\kk \delr} - e^{-i\kk\delr}}{2\delr}\right)  + \frac{2\left(f_0'\right)^2}{(f_0 + r^2)^2}\nonumber\\
& & + \frac{8rf_0'}{(f_0 + r^2)^2}
 - \frac{2\fOdot^2}{(f_0 + r^2)^2}\label{newkform} \end{eqnarray}
and
\begin{equation} \neww(\ww) =  
\frac{4\fOdot}{f_0 + r^2}\left(\frac{e^{i\ww\delt} - 
e^{-i\ww\delt}}{2\delt}\right).\label{newwform}\end{equation}
Here $\fOdot$ is substituted for the finite difference
\[ \frac{f_0(q\delr, (n+1)\delt) - f_0(q\delr, (n-1)\delt)}{2\delt},\]
and likewise with $f_0'$ and the centered difference equivalent in the
$r$ variable.  

Now note that 
\begin{equation} g(\ww) = \frac{2\cos\ww\delt -2}{\delt^2},\label{gwcos}
\end{equation} and
\[\ww_0 = g^{-1}(\old(\kk))\]
and
\[\ww_1 = g^{-1}(\old(\kk) + \newk(\kk) + \neww(\ww_1)).\]
Further if one lets
\[\ww_s = g^{-1}(\old(\kk) + s(\newk(\kk) + \neww(\ww_s)))\]
then 
\begin{eqnarray}
| \ww_1 - \ww_0| &=& \left|\int_0^1 \frac{\partial \ww_s}{\partial s}ds\right| \nonumber \\
&\leq&  \int_0^1\left| \bigl[(g^{-1})'\bigr]\bigl(\newk(\kk) + 
\neww(\ww_s)\bigr)\right| ds\nonumber \\
&\leq& \max\left(|(g^{-1})')|\right)\,\max\left(|\newk(\kk) 
+ \neww(\ww_s)|\right) \label{wdiffest}
\end{eqnarray}
So finding a bound on $\ww$ in terms of $\kk$ will allow one to find a bound
for $| \ww_1 - \ww_0|$ in terms of $\kk$.

Now we will show that the map
\[ T(\ww) = g^{-1}(\old(\kk) + \newk(\kk) + \neww(\ww))\]
is a contraction mapping on the ball $|\ww - \kk| < |\kk|/2 = B(\kk, \kk/2)$.  Using [\ref{gwcos}],
one finds that 
\[ g'(\ww) = \frac{-2\sin(\ww\delt)}{\delt}. \]
We've already stated one needs to choose $\delt < c \delr^{3/2}$ so
specifically, one can make $\delt < \delr/150$.  Then since 
$|\kk \delr| < \pi$ and $|\ww| < 3|\kk|/2$ this implies that
$|\ww\delt| < \pi/100$ and one has
\[ |g'(\ww)| = |2\ww|\left|\frac{\sin(\ww\delt)}{\ww\delt}\right| > 
1.98|\ww|.\]

Now if $\ww_1$ and $\ww_2$ are in $B(\kk, \kk/2)$ then assign $y_1$ and $y_2$ as follows and  find that
\begin{eqnarray*}y_1  &=& \old(\kk) + \newk(\kk) + \neww(\ww_1)\\
y_2 &=& \old(\kk) + \newk(\kk) + \neww(\ww_2)\\
y_2 -y_1 &=& \neww(\ww_2) - \neww(\ww_1). 
\end{eqnarray*}
Letting,
\[ y(s) = y_1 + s(y_2-y_1),\]
calculate 
\begin{eqnarray} 
\left|T(\ww_2) - T(\ww_1)\right| &=& \nonumber\\ 
|g^{-1}\bigl(y_2\bigr) - g^{-1}\bigl(y_1\bigr)| &=& \left| \int_0^1 \bigl(g^{-1}\bigr)'\bigl(y(s)\bigr) \bigl(y_2 - y_1\bigr) ds\right| \nonumber\\
&\leq &  \int_0^1 \mbox{max}\left|\bigl(g^{-1}\bigr)'\bigl(y(s)\bigr)\right|
\left|\bigl(y_2 - y_1\bigr)\right| ds \nonumber\\
&\leq&  \mbox{max}\left|\bigl(g^{-1}\bigr)'\bigl(y(s)\bigr)\right|
\left|\bigl(y_2 - y_1\bigr)\right|.\label{est} 
\end{eqnarray}

Simple calculus yields:
\[ \bigl(g^{-1}\bigr)'\bigl(g(\ww)\bigr) = \frac{1}{g'(\ww)}.\]
By choice of $y_1 $ and $y_2$ and $\ww_1, \ww_2
\in B(\kk, \kk/2)$, 
\begin{equation}\mbox{max} \left| \bigl(g^{-1}\bigr)'\bigl(y(s)\bigr)\right| = \frac{1}{1.98|\ww|} \leq\frac{2}{1.98|\kk|}.\label{dgmax}\end{equation}
Since $|\ww_1|, |\ww_2| \leq 3|\kk|/2$ and $|\ww_1\delt|, |\ww_2\delt| < \pi/100$, we have
\begin{eqnarray} |\neww(\ww_2) - \neww(\ww_1)| &=& \left|\frac{4\fOdot}{f_0 +r^2}\left(\frac{i\sin(\ww_2\delt)}{\delt} - \frac{i\sin(\ww_1\delt)}{\delt}\right)\right|\nonumber\\
&=& \left|\frac{4\fOdot}{f_0 +r^2}\right|\left|\left(
\frac{\sin(\ww_2\delt)}{\delt} - 
\frac{\sin(\ww_1\delt)}{\delt}\right)\right|\nonumber\\
&=& \left|\frac{4\fOdot}{(f_0+r^2)}\right|\left|\int_{\ww_1}^{\ww_2} \cos(s\delt) ds\right|\nonumber\\
&\leq&\left|\frac{4\fOdot}{(f_0+r^2)}\right|\left|\ww_2 - \ww_1\right|.\label{newest}
\end{eqnarray}

Now plug [\ref{dgmax}] and [\ref{newest}] into [\ref{est}], to obtain:
\begin{eqnarray*} 
|T(\ww_2) - T(\ww_1)| =
|g^{-1}(y_2) - g^{-1}(y_1)| \leq \frac{2}{1.98|\kk|}\left|\frac{4\fOdot}{(f_0+r^2)}\right|\left|\ww_2 - \ww_1\right|.
\end{eqnarray*}
The quantity
\[ \left|\frac{\fOdot}{f_0+r^2}\right| \]
defines the time scale for the problem.  In order for this entire
analysis to make sense one expects 
\[ \kk \gg \mbox{time scale}\qquad \mbox{and}\qquad\kk\gg \mbox{length scale}.\]  
Therefore, this is sufficient to show that
$T(w)$ is a contraction map from $B(\kk, \kk/2)$ to itself.  So by the
Contraction Mapping Principle, one concludes that there exists a
fixed point of $T$ in $B(\kk, \kk/2)$.  Clearly this argument holds
just as well for $-T$ and $B(-\kk, \kk/2)$, so there is a fixed point
for $-T$ in $B(-\kk, \kk/2)$. Since the left hand side of [\ref{gwgen}] is
quadratic in $e^{i\ww\delt}$, there are two solutions for $\ww$ and
these are they.

Lastly, we finish estimating in equation [\ref{wdiffest}].  We already
have a perfectly good estimate for $\bigl(g^{-1}\bigr)'$ from [\ref{dgmax}].
We merely need to estimate 
\[ \left|\newk(\kk) + \neww(\ww)\right|.\]
Replacing the exponentials with their trigonometric forms in [\ref{newkform}]
and [\ref{newwform}] to get:
\begin{eqnarray}\newk(\kk) &=&  - \frac{4f_0'}{f_0 + r^2}\left(\frac{2i\sin(\kk \delr)}{2\delr}\right)  + \frac{2\left(f_0'\right)^2}{(f_0 + r^2)^2}\nonumber\\
& & + \frac{8rf_0'}{(f_0 + r^2)^2}
 - \frac{2\fOdot^2}{(f_0 + r^2)^2}\nonumber \end{eqnarray}
and
\begin{equation} \neww(\ww) =  
\frac{4\fOdot}{f_0 +
r^2}\left(\frac{2i\sin{\ww\delt}}{2\delt}\right).\nonumber\end{equation}
Recall also that $\kk \gg 1/r$ so $\kk r \gg 1$.  Now if $\te$ is real,
\[ \left| \sin \te\right| \leq \te,\]
and if $\te$ is complex with $|\te|$ sufficiently small, as it would be
if $\te = \ww\delt$,  then
\[ \left| \sin \te\right| \leq 1.01|\te|,\]
 one has
\begin{eqnarray*}\left|\newk(\kk) + \neww(\ww)\right| &\leq& \left| 
\frac{4f_0'}{f_0+r^2}\right||\kk| + \left| \frac{2\left(f_0'\right)^2}{(f_0 + r^2)^2}\right| + \left|\frac{8rf_0'}{(f_0 + r^2)^2}\right|\\
& & + \left| \frac{2\fOdot^2}{(f_0 + r^2)^2}\right| +
\left| \frac{4\fOdot}{f_0 +r^2}\right| |1.01\ww|
\end{eqnarray*}
By my previous arguments about $\kk$ and the length and time scales,
the three central terms are much much smaller than the others.
Plugging $|\ww| \leq 3|\kk|/2$ and this and [\ref{dgmax}] into
[\ref{wdiffest}] one concludes that under these assumptions
\begin{eqnarray*}
|w_1 - w_0| &\leq& \frac{2}{1.98|\kk|}\left(\left| 
\frac{4f_0'}{f_0+r^2}\right||\kk| + \left| \frac{2\left(f_0'\right)^2}{(f_0 + r^2)^2}\right| + \left|\frac{8rf_0'}{(f_0 + r^2)^2}\right|\right.\\
& &\left. + \left| \frac{2\fOdot^2}{(f_0 + r^2)^2}\right| +
\left| \frac{4\fOdot}{f_0 +r^2}\right| \frac{3.03|\kk|}{2}\right).
\end{eqnarray*}
Hence, $|\ww_1 - \ww_0|$ is bounded, which implies in turn that the
imaginary part of $\ww_1$ is bounded since the imaginary part of
$\ww_0$ was.  Once again the growing modes are bounded.

Further, if we compare with equation[\ref{wcontapprox}], we see
analogues between the terms in these two equations.  
\begin{eqnarray*}
\frac{2}{1.98}\left|\frac{4f_0'}{f_0 +r^2}\right| \leftrightarrow \frac{2i(f_0' + 2r)}{f_0 + r^2}\end{eqnarray*}
and 
\[ \frac{3.03}{1.98}\left|\frac{4\fOdot}{f_0+r^2}\right| \leftrightarrow \frac{-2i\fOdot}{f_0 + r^2}. \]

\section{Convergence estimates}

This section contains an analysis of the convergence of the
differencing scheme for the equation
\begin{equation} \fddot = f'' + \frac{5f'}{r} + \frac{2\fdot^2}{f + r^2} 
- \frac{2(f')^2}{f + r^2} - \frac{8rf'}{f + r^2}. \label{myPDE} 
\end{equation}
We will show that as $\delr \rightarrow 0$ and $\delt \rightarrow 0$
that the solution found for $f(r,t)$ converges to an actual solution
of the partial differential equation.

First consider
the scheme to forward integrate the equation and the error in it.
Substitute the appropriate differences in [\ref{myPDE}]
then solve for $f(r,t + \delt)$ in the difference for $\fddot$ i.e. solve:
\begin{eqnarray*} f_c(r,t+\delt) &=& 2f_c(r,t) - f_c(r,t-\delt) \\
& &+ \delt^2\left(f_c'' + \frac{5f_c'}{r} + \frac{2\fdot_c^2}{f_c +
r^2} - \frac{2(f_c')^2}{f_c + r^2} - \frac{8rf_c'}{f_c +
r^2}\right).\end{eqnarray*} Here $f_c$ is used instead of $f$ to
indicate that this is the numerically calculated value rather than the
one that is a solution to the PDE.  Also assume the derivatives are
represented by the appropriate differences.  Approximate $f_c(r, t +
\delt)$ for use in $\fdot_c^2$ by using first $f_c(r, t + \delt) =
2f_c(r,t) - f_c(r,t-\delt)$, and then one iterates 6 times, finding a
new $f_c(r,t+\delt)$ from solving the equation with the previous
value.  Using greater numbers of iterations on $f_c(r,t+\delt)$ does
not change the answer to the numerical precision on the computer, so
this will not be considered as a source of error.

The error comes from the discretizations.  Let $\delt^2 C_1$ be
the error in the difference used for $\fddot$, and $C_2$ be the
accumulated error in the differences for the left hand side of
[\ref{myPDE}], then one can see that the error is of the form
\begin{eqnarray}
|f(r,t+\delt) - f_c(r,t+\delt)| = |\delt^2 C_1 + \delt^2 C_2| \leq
\delt^2\left(|C_1| + |C_2|\right). \nonumber
\end{eqnarray}
The error in going 2 steps would be
\begin{eqnarray}
|f(r,t+2\delt) - f_c(r,t+2\delt)| &\leq&
2\delt^2\left(|C_1| + |C_2|\right) \nonumber \\ 
& & + e^{W\delt} \left[\delt^2\left(|C_1| + |C_2|\right)\right]. \nonumber
\end{eqnarray}
With the term involving $e^{W \delt}$ coming from the stability
analysis, with $W \geq |\Re(i\ww)|$.  One knows from the stability
analysis that the negative imaginary part of $\ww$ is bounded, hence
$W$ is.  The error in going $n$ steps would therefore be:
\begin{eqnarray}
|f(r,t+n\delt) - f_c(r,t+n\delt)| &\leq&
n\delt^2\left(|C_1| + |C_2|\right) \nonumber \\ 
& & + \sum_{j=1}^{n} e^{W j\delt} \left[\delt^2\left(|C_1| + |C_2|\right)\right]. \nonumber
\end{eqnarray}
And one also knows that
\begin{eqnarray}
\sum_{j=1}^{n} e^{Wj\delt} \leq \int_1^n e^{Wj\delt}dj =
\frac{e^{Wn\delt} - e^{W\delt}}{W\delt} \end{eqnarray} Since $W$ was
strictly bounded, one represents the above as $\frac{C_3}{\delt}$.
Now see that the error in going $n$ steps is
\begin{eqnarray}
|f(r,t+n\delt) - f_c(r,t+n\delt)| &\leq&
n\delt^2\left(|C_1| + |C_2|\right) \nonumber \\ 
& & + \delt C_3\left(|C_1| + |C_2|\right). \nonumber
\end{eqnarray}
Since $C_3$ is bounded, one merely needs to show that
$C_1$ and $C_2$ are bounded to know that as $\delt \to 0$ clearly this
error goes to zero.

Now we use Taylor's Theorem to find the errors in using the differences,
i.~e. to find $C_1$ and $C_2$.

We use a generic function $h(x)$ and approximate  to fourth order:
\begin{eqnarray*}
h(x + \delta) &=& h(x) +\delta h'(x) + \frac{\delta^2}{2}h''(x) + \frac{\delta^3}{6}h'''(x) + \frac{\delta^4}{24}h^{(4)}(\zeta_1)  \\
h(x - \delta) &=& h(x) -\delta h'(x) + \frac{\delta^2}{2}h''(x) - \frac{\delta^3}{6}h'''(x) + \frac{\delta^4}{24}h^{(4)}(\zeta_2).  
\end{eqnarray*}
Here $\zeta_1 \in [x, x+\delta]$, $\zeta_2 \in [x-\delta,x]$.  Use a
third order approximation, subtracting, plus a little algebra, to obtain:
\[
\frac{h(x +\delta) - h(x
-\delta)}{2\delta}-\frac{\delta^2}{6}h'''(\xi_1) = h'(x).\] where
$\xi_1 \in [x-\delta, x+\delta]$ and replaced $h'''(\zeta_1) +
h'''(\zeta_2) = 2h'''(\xi_1)$.  Likewise, using the fourth order
approximation, add the equations, and do some algebra to obtain:
\[ \frac{h(x+\delta) + h(x-\delta) -2h(x)}{\delta^2} - \frac{\delta^2}{12}h^{(4)}(\xi_2)= h''(x). \]

Applying this to equation [\ref{myPDE}], one has
\begin{eqnarray*} \fddot(r,t) &=& \frac{f(r, t +\delt) + f(r, t-\delt) - 2f(r,t)}{\delt^2} - \frac{\delt^2}{12}\frac{\partial^4f}{\partial t^4}(r,\eta_1)\\
\frac{2\fdot^2(r,t)}{f(r,t)+r^2} &=& \frac{\left(f(r,t +\delt) - f(r,t-\delt)\right)^2}{2\delt^2\left(f(r,t) + r^2\right)} \\
& &-2\left(\frac{f(r,t+\delt)-f(r,t-\delt)}{\delt\left(f(r,t) + r^2\right)}\right)\left(\frac{\delt^2}{6}\frac{\partial^3 f}{\partial t^3}(r,\eta_2)\right) \\
& & +\frac{2}{f(r,t) + r^2} \left(\frac{\delt^2}{6}\frac{\partial^3 f}{\partial t^3}(r,\eta_2)\right) \\
\frac{2\left(f'(r,t)\right)^2}{f(r,t) + r^2} &=& \frac{\left(f(r+\delr,t) - f(r-\delr,t)\right)^2}{2\delr^2\left(f(r,t) + r^2\right)} \\
& &-2\left(\frac{f(r+\delr,t)-f(r-\delr,t)}{\delr\left(f(r,t) + r^2\right)}\right)\left(\frac{\delr^2}{6}\frac{\partial^3 f}{\partial r^3}(\xi_1,t)\right) \\
& & +\frac{2}{f(r,t) + r^2} \left(\frac{\delr^2}{6}\frac{\partial^3 f}{\partial r^3}(\xi_1,t)\right) \\
\frac{8rf'(r,t)}{f(r,t) + r^2} &=& \frac{4\left(f(r+\delr,t) - f(r-\delr,t)\right)}{\delr\left(f(r,t) + r^2\right)} \\
& & + \frac{8r}{f(r,t) + r^2}\left(\frac{\delr^2}{6}\frac{\partial^3 f}{\partial r^3}(\xi_1,t)\right).
\end{eqnarray*}

The first of these equations says that if
\[ \frac{\partial^4 f}{\partial t^4}(r, \eta_1) \] is bounded then the error
$C_1$ goes to zero as $\delt^2 \rightarrow 0$.

If $f(r,t)$ is chosen in $C(4,4)$, the space of functions with four
bounded space/time derivatives, all of these differences have bounded
errors.

Now attending to the discretization for
\begin{eqnarray}
f''(r,t) + \frac{5f'(r,t)}{r} &\approx& \Biggl[\left[r + \frac{1}{2}\delr\right]^{5} \left({\displaystyle \frac{f((r +\delr,t) 
- f(r,t)}{\delr}}\right)\Biggr. \label{lindisc1}\\
& &
\Biggl. - \left[r- \frac{1}{2}\delr\right]^{5}\left({\displaystyle 
\frac{f(r, t) - f(r -\delr , t)}{\delr}}\right)\Biggr]\Bigg/ (r^5\delr ), \nonumber
\end{eqnarray}
one needs to find an approximation of the error on this.  Start
with  $h(r) = r^5 f'(r)$.  Consequently:
\begin{eqnarray*} h'(r) &=& 5r^4 f'(r) + r^5 f''(r) = r^5\left(f''(r) + \frac{5f'(r)}{r}\right) \\
h''(r) &=& 20r^3f'(r) + 10r^4f''(r) + r^5 f'''(r) \\
h'''(r) &=& 60r^2f'(r) + 60r^3f''(r) + 15r^4 f'''(r) + r^5 f^{(4)}(r).
\end{eqnarray*}
By taking a Taylor series for $h(r),$ obtain:
\begin{eqnarray*}
h\left(r + \frac{\delr}{2}\right) &=& h(r) + \frac{\delr}{2} h'(r) + \frac{\delr^2}{8}h''(r) + \frac{\delr^3}{48}h'''(\zeta_1) \\
h\left(r - \frac{\delr}{2}\right) &=& h(r) - \frac{\delr}{2} h'(r) + \frac{\delr^2}{8}h''(r) - \frac{\delr^3}{48}h'''(\zeta_2). 
\end{eqnarray*}
Subtracting and doing some algebra as before yields:
\begin{eqnarray*}
h'(r) &=& \frac{h\left(r +\frac{\delr}{2}\right) - h\left(r - \frac{\delr}{2}\right)}{\delr} - \frac{\delr^3}{24}h'''(\eta).
\end{eqnarray*}
Now start substituting in definitions for $h(r)$, $h'(r)$, and let
\[\rho_3 \in \left[r - \frac{\delr}{2}, r+\frac{\delr}{2}\right]\] 
etc. to obtain:
\begin{eqnarray}
\lefteqn{r^5\left(f''(r) + \frac{5f'(r)}{r}\right) =}\nonumber\\
& & \frac{\left(r + \frac{\delr}{2}\right)^5 
f'\left(r + \frac{\delr}{2}\right) - \left(r - \frac{\delr}{2}\right)^5 f'\left(r - \frac{\delr}{2}\right)}{\delr} \nonumber \\
& & - \frac{\delr^3}{24}\left(60r^2f'(\rho_3) + 60r^3f''(\rho_3) + 15r^4 f'''(\rho_3) + r^5 f^{(4)}(\rho_3)\right). \label{lindisc2}
\end{eqnarray}
A centered difference approximation gives:
\begin{equation}
f'\left(r + \frac{\delr}{2}\right) = \frac{f(r + \delr) - f(r)}{\delr} -
\frac{\delr^2}{24} f'''(\xi_1)\label{fpdisc1},
\end{equation}
and similarly
\begin{equation}
f'\left(r - \frac{\delr}{2}\right) = \frac{f(r) - f(r-\delr)}{\delr} - \frac{\delr^2}{24} f'''(\xi_2)\label{fpdisc2}
\end{equation}
So substituting [\ref{fpdisc1}] and [\ref{fpdisc2}] into
[\ref{lindisc2}], one has the difference equation and
approximation for [\ref{lindisc1}].  Letting $\rho_4 \in [r, r + \delr]$
and $\rho_5 \in [r-\delr, r]$
\begin{eqnarray}
\lefteqn{f''(r,t) + \frac{5f'(r,t)}{r} =}\nonumber\\
& & \Biggl[\left[r + \frac{\delr}{2}\right]^{5} 
\left({\displaystyle \frac{f(r +\delr,t) 
- f(r,t)}{\delr}} \right)\Biggr. \nonumber\\
& &
\Biggl. - \left[r- \frac{\delr}{2}\right]^{5}\left({\displaystyle 
\frac{f(r, t) - f(r -\delr , t)}{\delr}} \right)\Biggr]\Bigg/ (r^5\delr ) 
\nonumber\\
& &- \left[\frac{\left(r + 
\frac{\delr}{2}\right)^5}{r^5\delr}\right]
\left(\frac{\delr^2}{24} \frac{\partial^3 f}{\partial r^3}(\rho_4,t)\right) 
+ \left[\frac{\left(r - \frac{\delr}{2}\right)^5}{r^5\delr}\right]
\left(\frac{\delr^2}{24} \frac{\partial^3 f}{\partial r^3}(\rho_5,t)\right) 
\nonumber\\
& &- \frac{\delr^3}{24r^5}\left(60r^2\frac{\partial f}{\partial r}(\rho_3,t) 
+ 60r^3\frac{\partial^2 f}{\partial r^2}(\rho_3,t) 
+ 15r^4 \frac{\partial^3 f}{\partial r^3}(\rho_3, t) \nonumber \right.\\
& &\left. + r^5 \frac{\partial^4 f}{\partial r^4}(\rho_3,t)\right). \nonumber
\end{eqnarray}.

Once again, one sees that if $f(r,t)$ is chosen in $C(4,4)$, the error
in this difference is bounded.  Under this condition, putting all of this information together, we expect the error in
\[ |f(r,t+n\delt) -f_c(r,t+n\delt)| \leq K\delt,\]
where $K$ is bounded and $K \rightarrow 0$ only when both $\delt
\rightarrow 0$ and $\delr \rightarrow 0$, but not when only one of
$\delt$ or $\delr$ go to zero.

To confirm this, convergence tables have been run on this program.
$f(0,100)$ and $f(10,100)$ are used as indicators.  

First, the convergence as $\delt \rightarrow 0$ is investigated.  One
run is made with $\delr = 0.100$ and $\delt = 0.00125$, since we
cannot do a run with $\delr$ and $\delt$ infinitely small, and the
stability analysis tells us that we must keep $\delt \leq C
\delr^{3/2}.$ Call this $f_\infnty$.  Then $\delt$ is allowed to take
on a variety of larger values and each time we can find the errors $E0
= |f(0,100) - f_\infnty(0,100)|,$ $E10 = |f(10,100) -
f_\infnty(10,100)|,$ $h = \delt - 0.00125$.  Then with any two of
these one can calculate 
\begin{equation}  \ln(E0_a/E0_b)/\ln(h_a/h_b)\qquad\mbox{and}\qquad
 \ln(E10_a/E10_b)/\ln(h_a/h_b)\label{lnstuf}\end{equation} In this
problem, this quotient of natural logarithms should be close to $1$
since in theory the error is $K\delt$.

The complete set of initial conditions are $f(0,t) = 1.0$,
$R_{\mbox{max}} = 100$, $\dot{f}(0,t) = -0.01$.  The data is in Table \ref{convtaba}.

\begin{table}[H]
\begin{center}
\caption{4+1 dimensional model: Convergence data 1.}
\label{convtaba}
\[ \begin{array}{rrrr}
\multicolumn{1}{c}{\delr} & \multicolumn{1}{c}{\delt} &\multicolumn{1}{c}{f(0,100)} & \multicolumn{1}{c}{f(10,100)}\\
0.100& 0.00125& 0.247164210997& 0.245952971169\\
0.100& 0.05000& 0.247225131669& 0.246014184991\\
0.100& 0.04000& 0.247212637097& 0.246001630149\\
0.100& 0.02000& 0.247187644555& 0.245976517632\\
0.100& 0.01000& 0.247175146883& 0.245963959986\\
0.100& 0.00500& 0.247168897823& 0.245957680759\\
\end{array}
\]
\end{center}
\end{table}

In Table \ref{convtaba}, skipping the first row and proceeding
downward, calculate Table \ref{convtabb}, where the previous line is the
line associated with subscript ``a'' in equation [\ref{lnstuf}], and the 
current
line is associated with subscript ``b'' in equation [\ref{lnstuf}].
It is clear that the quotient of the natural logarithms as in
[\ref{lnstuf}] is close to one and gets closer as the size of $\delt$
decreases, as the theory predicts.

\begin{table}[H]
\begin{center}
\caption{4+1 dimensional model: Convergence data 2.}
\label{convtabb}
\[ \begin{array}{rrrrr}
\multicolumn{1}{c}{h} & \multicolumn{1}{c}{E0} &\multicolumn{1}{c}{E10} & \multicolumn{1}{c}{\mbox{ln quot. $E0$}}& \multicolumn{1}{c}{\mbox{ln quot. $E10$}} \\
0.04875& 0.000060920672& 0.000061213822& &\\
0.03875& 0.000048426100& 0.000048658980& 0.999822247666& 0.999835333093\\
0.01875& 0.000023433558& 0.000023546463& 0.999907462503& 0.999894968679\\
0.00875& 0.000010935886& 0.000010988817& 0.999972932084& 0.999944166093\\
0.00375& 0.000004686826& 0.000004709590& 0.999995539233& 0.999975691565\\
\end{array}
\]
\end{center}
\end{table}

Next, the convergence as $\delr \rightarrow 0$ is investigated.  One
run is made with $\delr = 0.00625$ and $\delt = 0.00125$, since we
cannot do a run with $\delr$ and $\delt$ infinitely small, and
consider this to be the value of $f_\infnty$.  Then $\delr$ is allowed
to take on a variety of larger values.  Likewise, we calculate the
values in [\ref{lnstuf}], but this time, our theory has no prediction
for the value, as the error is predicted to decrease as $K\delt$.

Again, the complete set of initial conditions are $f(0,t) = 1.0$,
$R_{\mbox{max}} = 100$, $\dot{f}(0,t) = -0.01$.  The data is in Table
\ref{convtabc}.

\begin{table}[H]
\begin{center}
\caption{4+1 dimensional model: Convergence data 3.}
\label{convtabc}
\[ \begin{array}{rrrr}
\multicolumn{1}{c}{\delr} & \multicolumn{1}{c}{\delt} &\multicolumn{1}{c}{f(0,100)} & \multicolumn{1}{c}{f(10,100)}\\
0.00250& 0.00125& 0.250239554784& 0.248987627424\\
0.10000& 0.00125& 0.247164210997& 0.245952971169\\
0.05000& 0.00125& 0.249463484942& 0.248227707633\\
0.02500& 0.00125& 0.250046460114& 0.248801253159\\
0.01250& 0.00125& 0.250192709322& 0.248943556603\\
0.00625& 0.00125& 0.250229305018& 0.248978377699\\
\end{array}
\]
\end{center}
\end{table}

In Table \ref{convtabc}, skipping the first row and proceeding
downward, calculate Tabel [\ref{convtabd}], where the previous line is the
line associated with subscript ``a'' in [\ref{lnstuf}], and the curent
line is associated with subscript ``b'' in [\ref{lnstuf}].
The quotient of natural logarithms is closest to being an integer when
the error caused by $\delr$ is much greater than the error caused by
$\delt$.  This is as this sort of analysis would predict when the
error is a sum of a piece that goes to zero as $\delr \rightarrow 0$
and another piece that goes to zero as $\delt \rightarrow 0$.

\begin{table}[H]
\begin{center}
\caption{4+1 dimensional model: Convergence data 4.}
\label{convtabd}
\[ \begin{array}{rrrrr}
\multicolumn{1}{c}{h} & \multicolumn{1}{c}{E0} &\multicolumn{1}{c}{E10} & \multicolumn{1}{c}{\mbox{ln quot. $E0$}}& \multicolumn{1}{c}{\mbox{ln quot. $E10$}} \\
0.09750& 0.003075343787& 0.003034656255&&\\
0.04750& 0.000776069842& 0.000759919791& 1.914735161591& 1.925458076682\\
0.02250& 0.000193094670& 0.000186374265& 1.861663705261& 1.880927424171\\
0.01000& 0.000046845462& 0.000044070821& 1.746545407713& 1.778154140081\\
0.00375& 0.000010249766& 0.000009249725& 1.549300513769& 1.591718519800\\
\end{array}
\]
\end{center}
\end{table}

\clearpage

 
\chapter{Stability and convergence of the $\IC P^1$  model, charge 1 sector}
\bigskip
In this chapter we analyze the stability of the equation
\begin{equation}  \fddot = f'' + \frac{3f'}{r} - \frac{4rf'}{f^2 + r^2} + \frac{2f}{f^2 + r^2}\left(\fdot^2 - f'^2 \right) \label{gPDE2}
\end{equation}
and the associated differencing scheme used in finding numerical
solutions.  This analysis will proceed analogously to that of the 4+1
dimensional model outlined in appendix A.  Once again $f = f(r,t)$, and $r$
is a radial variable, hence $r > 0$. It will be shown that the
stability behavior of the differential equation is qualitatively the
same as that of the difference equation, and that the difference
equation converges to the differential equation as $\Delta r
\rightarrow 0$ and $\Delta t \rightarrow 0$.

\bigskip
\section{Continuum stability: simplified model}

As in section A.1, the first thing we will address is the stability of the linear part of the partial differential equation [\ref{gPDE2}].  This is:
\begin{equation} \fddot = f'' + \frac{3f'}{r} \label{lPDE2}\end{equation}
Setting
\[ f(r,t) = e^{i\kk r} e^{i \ww t}, \]
and plugging this into [\ref{lPDE2}] yields the equation:
\[ \ww^2 = \kk^2 - \frac{3i\kk}{r}.\]
Solve this for $\ww$ and find that
\[ \ww = \pm \sqrt{\frac{\kk^2 r -3i\kk}{r}}.\]
Hence $\ww$ always has a negative imaginary part, therefore this equation has a growing mode and is not stable, as we saw before with equation [\ref{gPDE}]
in section A.1.

Once again, we can abstractly consider the equation
\[ \ddot f = f'' + \frac{3f'}{r} = \mathcal{L} f\]
where $\mathcal{L} f $ is a linear operator with
\[ \mathcal{L} = r^{-3}\partial_r r^{3} \partial_r\]
and hence that
\[ \mathcal{L} = r^{-3/2}(r^{-3/2}\partial_r r^{3/2})(r^{3/2}\partial_r r^{-3/2}) r^{3/2} = -B^{-1}A^\dagger A B \]
where $B = r^{-3/2}$ and $A = r^{-3/2}\partial_r r^{3/2}.$
And, just as before $A^\dagger A$ is hermitian so it has real spectrum, and since it is essentially a square, it has positive real spectrum.  Consequently
$\mbox{spec}(\mathcal{L}) = \mbox{spec}(-A^\dagger A)$ is real and negative.  The solutions of the equation 
\[ \fddot = \mathcal{L}f\]
consists of sines and cosines in the time variable multiplying the eigenfunctions of $\mathcal{L},$ and hence has no growing mode and is strictly stable.

This is resolved in exactly the same manner as in section A.1.  In
this case we find that under the normal operating conditions of $\kk
\gg 1/r$ we have a factor of $r^{-3/2}$ that is not accounted for in
the Von Neumann stability analysis that had no choice but to appear in
the $e^{i\ww t}$ portion of the equation.

When one forcibly puts this factor of $r^{-3/2}$ into the Von Neumann stability analysis via 
\[ f(r,t) = r^{-3/2} e^{i\kk r}e^{i\ww t}\]
and plugs into [\ref{lPDE2}], one calculates:
\[ \ww^2 = \kk^2 + \frac{3}{4r^2}.  \]
Hence 
\[ \ww = \pm \frac{\sqrt{3 + 4r^2\kk^2}}{2r},\]
and $\ww$ has no negative imaginary part.

\bigskip
\section{Continuum stability}

\medskip

The first thing to address is the stability of partial differential
equation [\ref{gPDE2}].

We set \[ f(r,t) = f_0(r,t) + \ee r^{-3/2}
e^{i\kk r}e^{i\ww t},\] and plug into [\ref{gPDE2}] then linearize in
$\ee$ to obtain:

\begin{eqnarray*} \omega^2 &=& \kappa^2 + \frac{3}{4r^2} 
+ \frac{4i\kk r - 6}{r^2 + f_0^2} - \frac{8f_0'f_0r}{(r^2 + f_0^2)^2} - \frac{4if_0\fOdot\ww}{r^2 + f_0^2}\\
& & - \frac{6f_0f_0'}{r(r^2 + f_0^2)} + \frac{4i\kk f_0 f_0'}{r^2 + f_0^2}
- \frac{\left(2 - \displaystyle{\frac{4f_0^2}{r^2 + f_0^2}}\right)\left(\fOdot^2 - f_0'^2\right)}{r^2 + f_0^2}
\end{eqnarray*}
 
Solving this for $\ww$ yields:
\begin{eqnarray}\ww &=& -\frac{2if_0\fOdot}{(r^2 + f_0^2)} \pm \kappa\left[ 1 + \frac{4i f_0^2}{\kk(r^2 + f_0^2)^2} + \frac{4i f_0 f_0'}{\kk(r^2 + f_0^2)}\right.\nonumber\\
& &
+ \frac{3f_0^4}{4\kk^2r^2(r^2 + f_0^2)^2} - \frac{6f_0^3 f_0'}{\kk^2 r (f_0+r^2)^2} - \frac{2\fOdot^2}{\kk^2(r^2 + f_0^2)}\nonumber\\
& &\left. + \frac{2f_0'^2(r^2 - f_0^2)}{\kk^2(r^2 + f_0^2)^2}
- \frac{14f_0f_0'r}{\kk^2 (r^2 + f_0^2)^2} 
- \frac{21r^2 + 18 f_0^2 }{4\kk^2 (r^2 + f_0^2)^2}\right]^{1/2} \label{wcontapprox2}\end{eqnarray}

The imaginary parts are bounded.  If we restrict our concern to the
realm where $\kk$ is large and $\kk \gg \displaystyle{\frac{1}{r}}$,
we actually have that all of the rest of the terms in the square root
are bounded and small.

\bigskip
\section{Discretization scheme}
\medskip

As in section A.3, it was found that the main stability problems were generated by the linear part of the equation, namely:
\[ \fddot = f'' + \frac{3f'}{r}\]
as $r \rightarrow 0$.  As before we use the natural differential operator
\[ \mathcal{D}f = r^{-3}\partial_r (r^{3} f) = f'' + \frac{3f'}{r},\]
and we discretize it in the natural way.  Letting $q\delr = r$ and
$n\delt = t$, this discretization is
\begin{eqnarray*}& & (q\delr)^{-3}\left[\left[\left(q + \frac{1}{2}\right)\delr\right]^3 \frac{f((q+1)\delr, n\delt) - f(q\delr, n\delt)}{\delr} \right.\\
& &\left.- \left[\left(q - \frac{1}{2}\right)\delr\right]^3\frac{f(q\delr, n\delt) - f((q-1)\delr, n\delt)}{\delr}\right]\Bigg/ \delr. \end{eqnarray*}

\bigskip
\section{Outline of stability analysis}
\medskip

The stability analysis for this equation will be approached in exactly
the same way as in section A.4 for the previous model.

\bigskip
\section{Stability near zero}
\medskip

Once again we follow our previous methodologies.  A review of section A.5  might be in order.  One has
\[ \ddot{\df} = \mathcal{L}_1(\df) + \mathcal{L}_2(\dot{\df}),\]
with $\mathcal{L}_1$ and $\mathcal{L}_2$ linear operators.  As before
$\mathcal{L}_1$ and $\mathcal{L}_2$ are linear operators with $\mathcal{L}_2$ close to a multiple of the identity matrix.  The analysis of the eigenvalues proceeds as before, and once again all hinges on showing that if $\alpha$ is an eigenvalue of $\mathcal{L}_1$ then $\alpha < 0$.  

Allow $\df(q,n)$ to represent $\df(q\delr, n\delt)$ and likewise with
$f_0(q,n) \equiv f_0(q\delr, n\delt)$, and compute the following
linearized equation, discretized in space only:

\begin{eqnarray}\lefteqn{\ddot{\df}(q,n) = }\nonumber\\
& & (q\delr)^{-3}\left[\left(q + \displaystyle{\frac{1}{2}}\right)^3\delr^3\left(\frac{\df(q+1,t) - \df(q,t)}{\delr}\right)  \right.\nonumber\\
& &\left. - \left(q - \displaystyle{\frac{1}{2}}\right)^3\delr^3\left(\frac{\df(q,t) - \df(q-1,t)}{\delr}\right)\right]\bigg/\delr\label{premat2}\\
& & + \frac{4f_0(q,t) \fOdot(q,t)\dot{\df}(q,t)}{q^2\delr^2 + f_0^2(q,t)} + \frac{2f_0(q,t)\left(\df(q+1,t) - \df(q-1,t)\right)}{q\delr^2\left(q^2\delr^2 + f_0^2(q,t)\right)}\nonumber\\
& & + \frac{2\fOdot^2(q,t)\df(q,t)}{q^2\delr^2 + f_0^2(q,t)} - \frac{4f_0^2(q,t)\fOdot^2(q,t)\df(q,t)}{\left(q^2\delr^2 + f_0^2(q,t)\right)^2}\nonumber
\end{eqnarray}

One sees immediately from this that
\begin{eqnarray*} \mathcal{L}_2 &=& \left(\frac{4f_0\fOdot}{r^2 + f_0^2}\right)I\\
&\approx& cI\end{eqnarray*}
is a good approximation.

$\mathcal{L}_1$ is determined by [\ref{premat2}] and the quadratic fit boundary condition at the origin, i.e.
\[ \df(0,t) = \frac{4}{3}\df(\delr,t) - \frac{1}{3}\df(2\delr,t).\]
Let $\mathcal{L}_1 = [a_{i,j}],$ and $f_0 = f_0(r,t) = f_0(k\delr,t)$
and obtain the following tridiagonal matrix:
\begin{eqnarray*}
a_{1,1} &=& \frac{4}{3}\left(\frac{(1/2)^3}{\delr^2} +
\frac{2\delr}{\delr^2 + f_0^2}\right)- \frac{(3/2)^3}{\delr^2}
- \frac{(1/2)^3}{\delr^2} \\
& &+ \frac{2\fOdot^2}{\delr^2 + f_0^2} -
\frac{4f_0^2\fOdot^2}{\delr^2 + f_0^2}\\
a_{1,2} &=& -\frac{1}{3}\left(\frac{(1/2)^3}{\delr^2} +
\frac{2\delr}{\delr^2 + f_0^2}\right) + \frac{(3/2)^3}{\delr^2}
- \frac{2\delr}{\delr^2 + f_0^2}\\
a_{k,k-1} &=& \frac{(k-1/2)^3}{k^3\delr^2} + \frac{2k^2\delr}{k^2\delr^2 + f_0^2}\\
a_{k,k} &=& -\frac{(k+1/2)^3}{k^3\delr^2} - \frac{(k-1/2)^3}{k^3\delr^2} +\frac{2\fOdot^2}{ k^2\delr^2+f_0^2} - \frac{4f_0^2\fOdot^2}{(k^2\delr^2+ f_0^2)^2}\\
a_{k, k+1} &=& \frac{(k+1/2)^3}{k^3\delr^2} - \frac{2k^2\delr}{k^2\delr^2 + f_0^2.}\end{eqnarray*}
One can now use Maple or another program to compute the eigenvalues
and eigenvectors of this matrix while changing the size of the matrix
$n$ and the values of $f_0$ $\fOdot$ and $\delr$.

Since $\fOdot$ is always much less than $f_0$, the contributions from
the terms with $\fOdot$ are negligible.  One expects that rescaling
$\delr$ by a factor $p$ should result in the eigenvalues changing by
approximately $1/p^2$, since the largest terms are multiplied by $\delr^2$.

Let's see this explicitly. When $n=5, f_0 = 1, \fOdot = -0.01, \delr =
0.01$ the matrix is: 
\[\left[\begin{array}{ccccc}
-33333.3 & 33333.3 & 0 & 0 & 0 \\
4218.8 & -23750.0 & 19531.2 & 0 & 0 \\
0 & 5787.2 & -21666.7 & 15879.4 & 0 \\
0 & 0 & 6699.5 & -20937.5 & 14238.0 \\
0 & 0 & 0 & 7290.5 & -20600.0\\
\end{array}\right].\]
This matrix has eigenvalues with multiplicity $m$ and eigenvectors given by
\[\begin{array}{*{2}{r}@{\;\;\; [}r*{4}{@{,\,}r}@{\,]}}
\multicolumn{1}{c}{\mbox{Eigenvalue}} & \multicolumn{1}{c}{m} & 
\multicolumn{5}{c}{\mbox{Eigenvector}} \\ 
-43905.3 & 1 & -0.947 & 0.300 & -0.105 & 0.038 & -0.012\\
-24319.3 & 1 & -1.784 & -0.482 & 0.399 & 0.109 & -0.214\\
-35118.8 & 1& 3.303 & -0.177 & -0.610 & 0.582 & -0.292 \\
-12891.4 & 1 & 2.504 & 1.535 & 0.313 & -0.387 & -0.366\\
-4052.7 & 1 & 0.728 & 0.640 & 0.488 & 0.308 & 0.136
\end{array}
\]

Now rescale $\fOdot = -0.1$, which is multiplying by a factor of $10$.
We expect this to have little or no effect. 
\[\left[\begin{array}{ccccc}
-33333.3 & 33333.3 & 0 & 0 & 0 \\
4218.8 & -23750.0 & 19531.2 & 0 & 0 \\
0 & 5787.2 & -21666.7 & 15879.4 & 0 \\
0 & 0 & 6699.5 & -20937.5 & 14238.0 \\
0 & 0 & 0 & 7290.5 & -20600.0
\end{array}\right]\]
This matrix has eigenvalues with multiplicity $m$ and eigenvectors given by
\[\begin{array}{*{2}{r}@{\;\;\; [}r*{4}{@{,\,}r}@{\,]}}
\multicolumn{1}{c}{\mbox{Eigenvalue}} & \multicolumn{1}{c}{m} & 
\multicolumn{5}{c}{\mbox{Eigenvector}} \\ 
-43905.4 & 1 & -0.947 & 0.300 & -0.105 & 0.038 & -0.012 \\
-35118.8 & 1 & -3.303 & 0.177 & 0.610 & -0.582 & 0.292 \\
-24319.3 & 1 & -1.784 & -0.482 & 0.399 & 0.109 & -0.214 \\
-12891.4 & 1 & 2.504 & 1.535 & 0.313 & -0.387 & -0.366 \\
-4052.7 & 1 & 0.728 & 0.640 & 0.488 & 0.308 & 0.136
\end{array}
\]

Now rescale $\delr =0.1$ so that $\delr$ is $10$ times the original
with $f_0 = 1, \fOdot = -0.01$.  This we expect to scale the
eigenvalues by a factor of 100.  The matrix is:
\[\left[\begin{array}{ccccc}
-333.1 & 333.1 & 0 & 0 & 0 \\
43.0 & -237.5 & 194.5 & 0 & 0\\
0 & 59.5 & -216.7 & 157.1 & 0 \\
0 & 0 & 69.8 & -209.4 & 139.6 \\
0 & 0 & 0 & 76.9 & -206.0
\end{array}\right].\]

The eigenvalues with multiplicity $m$ and eigenvectors are:
\[\begin{array}{*{2}{r}@{\;\;\; [}r*{4}{@{,\,}r}@{\,]}}
\multicolumn{1}{c}{\mbox{Eigenvalue}} & \multicolumn{1}{c}{m} & 
\multicolumn{5}{c}{\mbox{Eigenvector}} \\ 
-440.7 & 1 & 0.945 & -0.305 & 0.110 &-.041 & .0136\\
-353.0 & 1 & -3.145 & 0.188 & 0.583 & -0.576 & 0.302 \\
-243.4 & 1 & 1.745 & 0.470 & -0.399 & -0.110 & 0.226 \\
-127.6 & 1 & 2.333 & 1.439 & 0.298 & -0.376 & -0.369 \\
-38.0 & 1 & -0.714 & -0.633 & -0.491 & -0.319 & -0.146
\end{array}.\]

These are indeed scaled by a factor of 100 from the original.

Now rescale so $f_0 = 10$ is $10$ times as large with $\delr = 0.01, \fOdot = -0.01$.  The matrix is:
\[\left[\begin{array}{*{5}{c}}
-33333.3 & 33333.3 & 0 & 0 & 0 \\
4218.8 & -23750.0 & 19531.2 & 0 & 0 \\
0 & 5787.0 & -21666.7 & 15879.6 & 0 \\
0 & 0 & 6699.2 & -20937.5 & 14238.3 \\
0 & 0 & 0 & 7290.0 & -20600.0
\end{array}\right]\]

The eigenvalues with multiplicity $m$ and eigenvectors are:
\[\begin{array}{*{2}{r}@{\;\;\; [}r*{4}{@{,\,}r}@{\,]}}
\multicolumn{1}{c}{\mbox{Eigenvalue}} & \multicolumn{1}{c}{m} & 
\multicolumn{5}{c}{\mbox{Eigenvector}} \\ 
-43905.2 & 1 & -0.9471 & 0.3004 & -0.1054 & 0.0381 & -0.0119\\
-35118.5 & 1 & 3.3030 & -0.1769 & -0.6105 & 0.5816 & -0.2920\\
-24319.3 & 1 & -1.7836 & -0.4823 & 0.3993 & 0.1091 & -0.2138\\
-12891.5 & 1 & 2.5037 & 1.5354 & 0.3128 & -0.3867 & -0.3657\\
-4053.0 & 1 & 0.7281 & 0.6396 & 0.4877 & 0.3079 & 0.1357
\end{array} \]
These are virtually the same as in the original with $f_0 = 1.0$.

The final rescaling is to reset the matrix size $n=10$ with $\delr = .01$,
$f_0 = 1, \fOdot = -0.01$.  The matrix is: 
\[\!
\left[\begin{array}{*{10}{c@{}}}
\Sc{-33333.} & \Sc{33333.} & \Sc{0} & \Sc{0} & \Sc{0} & \Sc{0} & \Sc{0} & \Sc{0} & \Sc{0} & \Sc{0} \\
\Sc{4218.8} & \Sc{-23750.} & \Sc{19531.} & \Sc{0} & \Sc{0} & \Sc{0} & \Sc{0 } & \Sc{ 0} & \Sc{0} & \Sc{0} \\
\Sc{0} & \Sc{5787.2} & \Sc{-21667.} & \Sc{15879.} & \Sc{0} & \Sc{0} & \Sc{0 } & \Sc{0} & \Sc{0} & \Sc{0} \\
\Sc{0} & \Sc{0} & \Sc{6699.5} & \Sc{-20938.} & \Sc{14238.} & \Sc{0} & \Sc{0} & \Sc{0} & \Sc{0} & \Sc{0} \\
\Sc{0} & \Sc{0} & \Sc{0} & \Sc{7290.5} & \Sc{-20600.} & \Sc{13310.} & \Sc{0} & \Sc{ 0} & \Sc{0} & \Sc{0} \\
\Sc{0} & \Sc{0} & \Sc{0} & \Sc{0} & \Sc{7703.3} & \Sc{-20417.} & \Sc{12713.} & \Sc{ 0} & \Sc{0} & \Sc{0} \\
\Sc{0} & \Sc{0} & \Sc{0} & \Sc{0} & \Sc{0} & \Sc{8007.5} & \Sc{-20306.} & \Sc{ 12299.} & \Sc{0} & \Sc{0} \\
\Sc{0} & \Sc{0} & \Sc{0} & \Sc{0} & \Sc{0} & \Sc{0} & \Sc{8241.0} & \Sc{-20234. }  & \Sc{ 11993.} & \Sc{0} \\
\Sc{0} & \Sc{0} & \Sc{0} & \Sc{0} & \Sc{0} & \Sc{0} & \Sc{0} & \Sc{8425.8} & \Sc{-20185.} & \Sc{ 11759.} \\
\Sc{0} & \Sc{0} & \Sc{0} & \Sc{0} & \Sc{0} & \Sc{0} & \Sc{0} & \Sc{0} & \Sc{8576.} & \Sc{ -20150.}
\end{array}\right]
\]
The eigenvalues with multiplicity $m$ and eigenvectors are:
\[\begin{array}{*{2}{c}*{4}{r}}
\multicolumn{1}{c}{\mbox{Eigenvalue}} & \multicolumn{1}{c}{m} & 
\multicolumn{4}{c}{\mbox{Eigenvector}} \\ 
-43977.1 & 1 & [\;\;\,0.9460, & -0.3021, & 0.1085, & -0.0423,  \\
& & 0.0175, & -0.0075, & 0.0033, & -0.0015,\\
& & 0.0006, & -0.0002\,]\\
-39218.2 & 1 & [-2.5044, & 0.4421, & 0.1908, & -0.3720,\\
& & 0.3879, & -0.3388, & 0.2660, & -0.1885, \\
& & 0.1156, & -0.0520\,]\\
-35858.2 & 1 & [-3.5682, & 0.2703, &  0.6032, & -0.6376, \\
& & 0.3843, & -0.0914, & -0.1219, & 0.2136, \\
& & -0.1945, & 0.1062\,] \\
-31051.4 & 1 & [\;\;\,6.2021, & 0.4246,  & -1.4984, & 0.7308,\\
& & 0.1859, & -0.5463, & 0.3443, & 0.0549, \\
& & -0.2861, & 0.2250\,] \\
-25358.7 & 1 & [-2.3840, & -0.5703, & 0.5619, & 0.0772,\\
& & -0.2884, & 0.0608, & 0.1511, & -0.1017,\\
& & -0.0604, & 0.0994\,] \\
-19307.4 & 1 & [-3.8375, & -1.6147, & 0.4616, & 0.6571,\\
& & -0.1420, & -0.3737, & 0.0534, & 0.2477,\\
& & -0.01757, & -0.1788\,]\\
-13418.0 & 1 & [-4.6228, & -2.7619, & -0.4625, & 0.7663, \\
& & 0.6223, & -0.0839, & -0.4233, & -0.1824,\\
& & 0.1872, & 0.2384\,]\\
-8177.3 & 1 & [-2.2447, & -1.6940, & -0.8658, & -0.1181, \\
& & 0.3015, & 0.3462, & 0.1505, & -0.0769, \\
& & -0.1808, & -0.1295\,]\\
-4002.9 & 1 & [-1.2609, & -1.1095, & -0.8494, & -0.5405,\\
& & -0.2432, & -0.0072,& 0.1381, & 0.1877,\\
& & 0.1592, & 0.0845\,]\} \\
-1210.9 & 1 & [\;\;\,0.5060, & 0.4876, & 0.4534, & 0.4063,\\
& & 0.3497, & 0.2868, & 0.2214, & 0.1570, \\
& & 0.0969, & 0.0439\,].
\end{array}\]
Once again, all eigenvalues are negative.  The modes localized near
zero don't change.  The eigenvalues increase towards zero as $n$
increases because there are growing modes of this equation away from
$r=0$.

One can conclude that the eigenvalues of this matrix under reasonable
initial conditions will always be negative, as required.  This
differencing scheme does not have any instabilities generated at the
origin.  

\bigskip
\section{Stability away from zero: simplified model}
\medskip

The analysis in this section shall follow that of section A.6.

In this section the stability of the difference equation derived from
[\ref{gPDE2}] with the differencing scheme outlined in section 3.1
under the special condition that $f(r,t) = f_0(r,t) + \epsilon e^{i\kk
q\delr}e^{i\ww n\delt}$ where $f_0(r,t) \equiv c,$ a constant, will be
addressed.

Explicit inclusion to the factor $r^{-3/2}$ is omitted, because it
does not impact this analysis to omit it.

Plug the previous expression for $f(r,t)$ into  [\ref{gPDE2}] and linearize in $\ee$ to obtain the following:

\begin{eqnarray}
\frac{e^{i\ww \delt} + e^{-i\ww\delt} - 2}{\delt^2} &=&
+ \frac{\left(q + \displaystyle{\frac{1}{2}}\right)^3\left(e^{i\kk\delr} - 1\right) - \left(q - \displaystyle{\frac{1}{2}}\right)^3\left(1 - e^{-i\kk\delr}\right)}{q^3\delr^2} \nonumber\\
& & - \frac{2q^2\delr(e^{i\kk\delr} - e^{-i\kk\delr})}{q^2\delr^2 + f_0^2}.\label{lin2}
\end{eqnarray}

As before let $x = e^{i\ww\delt}$ and 
\[ J =  \frac{\left(q + \displaystyle{\frac{1}{2}}\right)^3\left(e^{i\kk\delr} - 1\right) - \left(q - \displaystyle{\frac{1}{2}}\right)^3\left(1 - e^{-i\kk\delr}\right)}{q^3\delr^2}  - \frac{2q^2\delr(e^{i\kk\delr} - e^{-i\kk\delr})}{q^2\delr^2 + f_0^2}.\]
Reduce equation [\ref{lin2}] to
\[ x^2 - (2 + J\delt^2)x + 1 = 0.\]
This is exactly as in section A.6.  As before there are two
logarithmic solutions for $\ww$ given by
\[ \ww = \frac{-i\ln\left(1 + \displaystyle{\frac{J\delt^2}{2}} \pm \sqrt{J}\delt\sqrt{1 + 
{\displaystyle{\frac{J\delt^2}{4}}}}\right)}{\delt}.\]

As in section A.6, the biggest concern is determining exactly how big
$J\delt^2$ is. Let $\kk\delr = \te$, then reduce exponentials to sines
and cosines appropriately, and expand the factors of $q\pm 1/2$, and
simplify using such information as $q = r/\delr$, to obtain
\begin{eqnarray*} J &=& \kk^2\left(\frac{2\cos\te -2}{\te^2}\right)
+ \frac{3i\kk}{r}\left(\frac{\sin\te}{\te}\right) + 
\frac{3\kk^2\delr^2}{4r^2}\left(\frac{2\cos\te -2}{\te^2}\right)\\
& & - \frac{i\kk\delr^2}{4r^3}\left(\frac{\sin\te}{\te}\right) - \frac{4ir^2\kk}{r^2 + f_0^2}\left(\frac{\sin\te}{\te}\right).
\end{eqnarray*}
As $\delr \rightarrow 0$ one of these terms go to zero but the
following do not:
\[ \kk^2\left(\frac{2\cos\te -2}{\te^2}\right)
+ \frac{3i\kk}{r}\left(\frac{\sin\te}{\te}\right) + 
\frac{3\kk^2\delr^2}{4r^2}\left(\frac{2\cos\te -2}{\te^2}\right)-
\frac{4ir^2\kk}{r^2 + f_0^2}\left(\frac{\sin\te}{\te}\right).\] 
As before, the trigonometric parts of these terms is strictly bounded,
and we are working in the case where $r$ is bounded away from zero.
Our only concern is with the size of $\kk$, but we know that $\kk\delr
< \pi$ and $\delt \ll \delr$, hence we can make $\delt < c \delr$ and
with c chosen sufficiently small, we can make $\kk\delt$ as small as
we like.  In particular we may use the Taylor approximation to the
square root and then apply the Taylor approximation to the logarithm.
Just as in section A.6, with some cancellations, one obtains:
\begin{eqnarray*} \ww &\approx& \frac{-i\ln\left[\displaystyle{\frac{J\delt^2}{2} 
\pm \sqrt{J}\delt\left(1 + \displaystyle{\frac{J\delt^2}{8}} + O(J^2\delt^4)
\right)}\right]}{\delt}\\
&\approx& \frac{-i\left[\pm \sqrt{J}\delt + 
O\left((\sqrt{J}\delt)^3\right)\right]}{\delt} \\
&\approx& \left[\pm i\sqrt{J} + O\left(\sqrt{J}^{\,\,3}\delt^2\right)\right]
\end{eqnarray*} 
Recall, as $\delr \rightarrow 0$ all of the terms
in $J$  go to $0$ except for
\[ \kk^2\left(\frac{2\cos\te -2}{\te^2}\right)
+ \frac{3i\kk}{r}\left(\frac{\sin\te}{\te}\right) + 
\frac{3\kk^2\delr^2}{4r^2}\left(\frac{2\cos\te -2}{\te^2}\right)-
\frac{4ir^2\kk}{r^2 + f_0^2}\left(\frac{\sin\te}{\te}\right).\] 
So one has
\begin{eqnarray*}
\sqrt{J} &\approx& \sqrt{\left[\kk^2\left(\frac{2\cos\te -2}{\te^2}\right)
+ \frac{3i\kk}{r}\left(\frac{\sin\te}{\te}\right) + 
\frac{3\kk^2\delr^2}{4r^2}\left(\frac{2\cos\te -2}{\te^2}\right)\right.} \\
& &\overline{\left. -
\frac{4ir^2\kk}{r^2 + f_0^2}\left(\frac{\sin\te}{\te}\right) + O(\delr)\right]}
\end{eqnarray*}
Since $\kk$ is large while $\delr$ is small, approximating this square root yields: \nopagebreak
\begin{eqnarray*}
\sqrt{J} &\approx& i\kk\sqrt{\frac{2-2\cos\te}{\te^2}}\left[1 + 
\frac{3i}{2\kk r}\left(\frac{\sin\te}{\te}\right)\left(\frac{\te^2}{2-2\cos\te}\right) 
+ \frac{3\delr^2}{8r^2}\right.\nopagebreak \\ 
& & - \frac{i\delr^2}{8\kk r^2}\left(\frac{\sin\te}{\te}\right) \left(\frac{\te^2}{2-2\cos\te}\right) - 
\frac{2ir^2}{\kk(r^2 + f_0^2)}\left(\frac{\sin\te}{\te}\right)\left(\frac{\te^2}{2-2\cos\te}\right) \\
& &\left. + O\left(\frac{1}{\kk^2}\right)\right].
\end{eqnarray*}
As before, the leading order term of $\sqrt{J}$ is $O(\kk)$.  We can
make the correction $O(\sqrt{J}^3\delt^2)$ as small as we like by
requiring that $\delt < c\delr^{3/2}$ and choosing the factor $c$
appropriately.  Given these choices, one has:
\begin{eqnarray*}
\ww &\approx& \pm\sqrt{\frac{2-2\cos\te}{\te^2}}\Biggl[\kk\Biggr. \\
& & + \frac{3i}{2 r}\left(\frac{\sin\te}{\te}\right)\left(\frac{\te^2}{2-2\cos\te}\right) 
+ \frac{3\kk\delr^2}{8r^2}\nopagebreak \\ 
& & - \frac{i\delr^2}{8 r^2}\left(\frac{\sin\te}{\te}\right) \left(\frac{\te^2}{2-2\cos\te}\right) - 
\frac{2ir^2}{(r^2 + f_0^2)}\left(\frac{\sin\te}{\te}\right)\left(\frac{\te^2}{2-2\cos\te}\right) \\
& &\Biggl. + O\left(\frac{1}{\kk}\right) + O\left(\sqrt{J}^3\delt^2\right)\Biggr].
\end{eqnarray*}
It is also clear as in section A.6 that if one permitted $r \rightarrow 0$ the dominant piece would be:
\[ \ww \approx \pm\kk\sqrt{\frac{2-2\cos\te}{\te^2}} \mp 
\frac{3i}{2r}\left(\frac{\sin\te}{\te}\right)\left(\frac{\te^2}{2-2\cos\te}\right) \]
and as $\te \rightarrow 0$ this goes to 
\[ \ww \approx \pm \kk \mp \frac{3i}{2r} \]
or exactly what we obtained in section B.1 that was so often ``corrected'' by the addition of the factor $r^{-3/2}$.

\bigskip
\section{The complications}
\medskip
We must now deal with the reality that the assumption $f_0 \equiv c$
is not in general true.  We shall, as in section A.7, analyze the
difference between the solution for $\ww$ when $f_0\equiv c$, called
$\ww_0$, and the solution for $\ww$ when $f_0 \not\equiv c$, called
$\ww_1$.   

To find $\ww_0$ we solved [\ref{lin2}] an equation which can be
characterized as 
\[ g(\ww) = \old(\kk).\]
It was:
\begin{eqnarray*}
\frac{e^{i\ww \delt} + e^{-i\ww\delt} - 2}{\delt^2} &=&
+ \frac{\left(q + \displaystyle{\frac{1}{2}}\right)^3\left(e^{i\kk\delr} - 1\right) - \left(q - \displaystyle{\frac{1}{2}}\right)^3\left(1 - e^{-i\kk\delr}\right)}{q^3\delr^2} \\
& & - \frac{2q^2\delr(e^{i\kk\delr} - e^{-i\kk\delr})}{q^2\delr^2 + f_0^2}.
\end{eqnarray*}
If we do not assume that $f_0 \equiv c$ then the same linearization in
$\ee$ of [\ref{gPDE2}] with $f(q\delr,n\delt) = f_0(q\delr,n\delt) +
\ee e^{i\kk q \delr} e^{i\ww n \delt}$ yields the following equation:
\begin{eqnarray}
\frac{e^{i\ww \delt} + e^{-i\ww\delt} - 2}{\delt^2} &=&
+ \frac{\left(q + \displaystyle{\frac{1}{2}}\right)^3\left(e^{i\kk\delr} - 1\right) - \left(q - \displaystyle{\frac{1}{2}}\right)^3\left(1 - e^{-i\kk\delr}\right)}{q^3\delr^2} \nonumber \\
& & - \frac{2q^2\delr(e^{i\kk\delr} - e^{-i\kk\delr})}{q^2\delr^2 + f_0^2} - \frac{8q^2\delr^2 f_0'f_0}{(r^2 + f_0^2)^2} \nonumber \\
& & + \frac{\fOdot f_0(e^{i\ww \delt} - e^{-i\ww\delt})}{2\delt(r^2 + f_0^2)} 
- \frac{f_0' f_0(e^{i\kk \delr} - e^{-i\kk\delr})}{2\delr(r^2 + f_0^2)}\nonumber \\
& & + \frac{2(\fOdot^2 - f_0'^2)}{r^2 + f_0^2} 
- \frac{4f_0^2(\fOdot^2 - f_0'^2)}{(r^2 + f_0^2)^2}. \label{gwgen2} 
\end{eqnarray}
Characterize this equation as
\[ g(w) = \old(\kk) + \newk(\kk) + \neww(\ww)\]
with
\begin{eqnarray} \newk(\kk) &=& 
- \frac{f_0' f_0(e^{i\kk \delr} - e^{-i\kk\delr})}{2\delr(r^2 + f_0^2)} 
- \frac{8q^2\delr^2 f_0'f_0}{(r^2 + f_0^2)^2} \nonumber\\
& &
+ \frac{2(\fOdot^2 - f_0'^2)}{r^2 + f_0^2} 
- \frac{4f_0^2(\fOdot^2- f_0'^2)}{(r^2 + f_0^2)^2} \label{newkform2}
\end{eqnarray}
and \begin{eqnarray} 
\neww(\ww) &=&+ \frac{\fOdot f_0(e^{i\ww \delt}
- e^{-i\ww\delt})}{2\delt(r^2 + f_0^2)} \label{newwform2}\end{eqnarray} 
Here $\fOdot$ is substituted for the finite difference
\[ \frac{f_0(q\delr, (n+1)\delt) - f_0(q\delr, (n-1)\delt)}{2\delt},\]
and likewise with $f_0'$.  

From here we proceed as in section A.7.  We have
\begin{equation} g(\ww) = \frac{2\cos\ww\delt -2}{\delt^2},\label{gwcos2}
\end{equation}
\[\ww_0 = g^{-1}(\old(\kk)),\]
and
\[\ww_1 = g^{-1}(\old(\kk) + \newk(\kk) + \neww(\ww_1)).\]
Let
\[\ww_s = g^{-1}(\old(\kk) + s(\newk(\kk) + \neww(\ww_s)))\]
then 
\begin{eqnarray}
| \ww_1 - \ww_0| &=& \left|\int_0^1 \frac{\partial \ww_s}{\partial s}ds\right| \nonumber \\
&\leq&  \int_0^1\left| \bigl[(g^{-1})'\bigr]\bigl(\newk(\kk) + 
\neww(\ww_s)\bigr)\right| ds\nonumber \\
&\leq& \max\left(|(g^{-1})')|\right)\,\max\left(|\newk(\kk) 
+ \neww(\ww_s)|\right) \label{wdiffest2}
\end{eqnarray}
So finding a bound on $\ww$ in terms of $\kk$ will allow one to find a bound
for $| \ww_1 - \ww_0|$ in terms of $\kk$.

We start by showing that 
\[ T(\ww) = g^{-1}(\old(\kk) + \newk(\kk) + \neww(\ww))\]
is a contraction mapping on the ball $|\ww - \kk| < |\kk |/2 = B(\kk,
|\kk|/2)$.  Then we will know that there exists a solution for $\ww
\in B(\kk, |\kk|/2)$ and hence $|\ww| \leq 3|\kk|/2$.  

As before using [\ref{gwcos2}]
\[ g'(\ww) = \frac{-2\sin(\ww\delt)}{\delt}. \]
We've already stated one needs to choose $\delt < c \delr^{3/2}$ in
section B.6, so specifically, one can make $\delt < \delr/150$.  Then
since $|\kk \delr| < \pi$ and $|\ww| < 3|\kk|/2$ this implies that
$|\ww\delt| < \pi/100$ and one has
\[ |g'(\ww)| = |2\ww|\left|\frac{\sin(\ww\delt)}{\ww\delt}\right| > 
1.98|\ww|.\]

If $\ww_1$ and $\ww_2$ are in $B(\kk, \kk/2)$,  assign $y_1$ and $y_2$ as follows and  find that
\begin{eqnarray*}y_1  &=& \old(\kk) + \newk(\kk) + \neww(\ww_1)\\
y_2 &=& \old(\kk) + \newk(\kk) + \neww(\ww_2)\\
y_2 -y_1 &=& \neww(\ww_2) - \neww(\ww_1). 
\end{eqnarray*}
Letting,
\[ y(s) = y_1 + s(y_2-y_1),\]
calculate 
\begin{eqnarray} 
\left|T(\ww_2) - T(\ww_1)\right| &=& \nonumber\\ 
|g^{-1}\bigl(y_2\bigr) - g^{-1}\bigl(y_1\bigr)| &=& \left| \int_0^1 \bigl(g^{-1}\bigr)'\bigl(y(s)\bigr) \bigl(y_2 - y_1\bigr) ds\right| \nonumber\\
&\leq &  \int_0^1 \mbox{max}\left|\bigl(g^{-1}\bigr)'\bigl(y(s)\bigr)\right|
\left|\bigl(y_2 - y_1\bigr)\right| ds \nonumber\\
&\leq&  \mbox{max}\left|\bigl(g^{-1}\bigr)'\bigl(y(s)\bigr)\right|
\left|\bigl(y_2 - y_1\bigr)\right|.\label{est2} 
\end{eqnarray}

Simple calculus yields:
\[ \bigl(g^{-1}\bigr)'\bigl(g(\ww)\bigr) = \frac{1}{g'(\ww)}.\]
By choice of $y_1 $ and $y_2$ and $\ww_1, \ww_2
\in B(\kk, \kk/2)$, 
\begin{equation}\mbox{max} \left| \bigl(g^{-1}\bigr)'\bigl(y(s)\bigr)\right| = \frac{1}{1.98|\ww|} \leq\frac{2}{1.98|\kk|}.\label{dgmax2}\end{equation}
Analyzing since $|\ww_1|, |\ww_2| \leq 3|\kk|/2$ and $|\ww_1\delt|, |\ww_2\delt| < \pi/100$,
\begin{eqnarray} |\neww(\ww_2) - \neww(\ww_1)| &=& \left|\frac{f_0\fOdot}{r^2 + f_0^2}\left(\frac{i\sin(\ww_2\delt)}{\delt} - \frac{i\sin(\ww_1\delt)}{\delt}\right)\right|\nonumber\\
&=& \left|\frac{f_0\fOdot}{r^2 + f_0^2}\right|\left|\left(
\frac{\sin(\ww_2\delt)}{\delt} - 
\frac{\sin(\ww_1\delt)}{\delt}\right)\right|\nonumber\\
&=& \left|\frac{f_0\fOdot}{r^2 + f_0^2}\right|\left|\int_{\ww_1}^{\ww_2} \cos(s\delt) ds\right|\nonumber\\
&\leq&\left|\frac{f_0\fOdot}{r^2 + f_0^2}\right|\left|\ww_2 - \ww_1\right|.\label{newest2}
\end{eqnarray}
Now plugging [\ref{dgmax2}] and [\ref{newest2}] into [\ref{est2}], obtain:
\begin{eqnarray*} 
|T(\ww_2) - T(\ww_1)| =
|g^{-1}(y_2) - g^{-1}(y_1)| \leq \frac{2}{1.98|\kk|}\left|\frac{f_0\fOdot}{r^2+ f_0^2}\right|\left|\ww_2 - \ww_1\right|.
\end{eqnarray*}
The quantity
\[ \left|\frac{f_0\fOdot}{f_0+r^2}\right| \]
defines the time scale for the problem.  In order for this entire
analysis to make sense one expects 
\[ \kk \gg \mbox{time scale}\qquad \mbox{and}\qquad\kk\gg \mbox{length scale}.\]  
Therefore, this is sufficient to show that
$T(w)$ is a contraction map from $B(\kk, \kk/2)$ to itself.  So by the
Contraction Mapping Principle, one concludes that there exists a
fixed point of $T$ in $B(\kk, \kk/2)$.  Clearly this argument holds
just as well for $-T$ and $B(-\kk, \kk/2)$, so there is a fixed point
for $-T$ in $B(-\kk, \kk/2)$. Since the left hand side of [\ref{gwgen}] is
quadratic in $e^{i\ww\delt}$, there are two solutions for $\ww$ and
these are they.
Lastly, we finish estimating in equation [\ref{wdiffest2}].  We already
have a perfectly good estimate for $\bigl(g^{-1}\bigr)'$ from [\ref{dgmax2}].
We merely need to estimate 
\[ \left|\newk(\kk) + \neww(\ww)\right|.\]
Replacing the exponentials with their trigonometric forms in [\ref{newkform2}]
and [\ref{newwform2}] to get:
\begin{eqnarray}\newk(\kk) &=&  - \frac{if_0' f_0}{r^2 + f_0^2}\left(\frac{\sin(\kk \delr)}{\delr}\right) 
- \frac{8q^2\delr^2 f_0'f_0}{(r^2 + f_0^2)^2} \nonumber\\
& &
+ \frac{2(\fOdot^2 - f_0'^2)}{r^2 + f_0^2} 
- \frac{4f_0^2(\fOdot^2- f_0'^2)}{(r^2 + f_0^2)^2} \nonumber
\end{eqnarray}
and \begin{eqnarray} 
\neww(\ww) &=&+ \frac{\fOdot f_0}{r^2 + f_0^2}\left(\frac{i\sin(\ww\delt)}{\delt}\right) \nonumber\end{eqnarray} 
Recall also that $\kk \gg 1/r$ so $\kk r \gg 1$.  Now if $\te$ is real,
\[ \left| \sin \te\right| \leq \te,\]
and if $\te$ is complex with $|\te|$ sufficiently small, as it would be
if $\te = \ww\delt$,  then
\[ \left| \sin \te\right| \leq 1.01|\te|,\]
 one has
\begin{eqnarray*}\left|\newk(\kk) + \neww(\ww)\right| &\leq& 
 \left| \frac{f_0' f_0}{r^2 + f_0^2}\right| |\kk|
+ \left| \frac{8r^2 f_0'f_0}{(r^2 + f_0^2)^2}\right| 
+\left| \frac{2(\fOdot^2 - f_0'^2)}{r^2 + f_0^2}\right| \\
& &+ \left| \frac{4f_0^2(\fOdot^2- f_0'^2)}{(r^2 + f_0^2)^2}\right| 
+ \left|\frac{\fOdot f_0}{r^2 + f_0^2}\right| |1.01 \ww|\end{eqnarray*} 

By my previous arguments about $\kk$ and the length and time scales,
the three central terms are much much smaller than the others.
Plugging $|\ww| \leq 3|\kk|/2$ and this and [\ref{dgmax2}] into
[\ref{wdiffest2}] one concludes that under these assumptions
\begin{eqnarray*}
|w_1 - w_0| &\leq& \frac{2}{1.98|\kk|}\left( \left| \frac{f_0' f_0}{r^2 + f_0^2}\right| |\kk|
+ \left| \frac{8r^2 f_0'f_0}{(r^2 + f_0^2)^2}\right| 
+\left| \frac{2(\fOdot^2 - f_0'^2)}{r^2 + f_0^2}\right| \right.\\
& &\left.+ \left| \frac{4f_0^2(\fOdot^2- f_0'^2)}{(r^2 + f_0^2)^2}\right| 
+ \left|\frac{\fOdot f_0}{r^2 + f_0^2}\right| \frac{3.03|\kk|}{2}\right)
\end{eqnarray*} 
Hence, $|\ww_1 - \ww_0|$ is bounded, which implies in turn that the
imaginary part of $\ww_1$ is bounded since the imaginary part of
$\ww_0$ was.  Once again the growing modes are bounded.

Further, if we compare with equation[\ref{wcontapprox2}], we see
analogues between the terms in these two equations.  
\begin{eqnarray*}
\frac{2}{1.98}\left|\frac{f_0'f_0}{f_0^2 +r^2}\right| \leftrightarrow \frac{4if_0 f_0'}{r^2 + f_0^2} + \frac{4if_0^2}{(r^2 + f_0^2)^2}\end{eqnarray*}
and 
\[ \frac{3.03}{1.98}\left|\frac{\fOdot f_0}{r^2 + f_0^2}\right| \leftrightarrow \frac{-2if_0\fOdot}{r^2 + f_0^2}. \]

\section{Convergence estimates}
This section contains an analysis of the convergence of the differencing scheme for the equation 
\begin{equation}
\fddot = f'' + \frac{3f'}{r} - \frac{4rf'}{f^2 + r^2} + \frac{2f}{f^2 + r^2}\left(\fdot^2 - f'^2\right).\label{myPDE2} \end{equation}
We will show that as $\delr \rightarrow 0$ and $\delt \rightarrow 0$ that the solution found for $f(r,t)$ converges to an actual solution of the partial differential equation.  As in this entire chapter, the arguments will essentially mimic those given before in section A.7.

Substituting the appropriate differences into [\ref{myPDE2}] and
forward integrating the equation one can solve for $f_c(r,t+\delt)$,
where $f_c$ is used instead of $f(r,t)$ to indicate it is a calculated
value (at $(r,t)$).  One obtains
\begin{eqnarray*} f_c(r,t+\delt) &=& 2f_c(r,t) - f_c(r,t-\delt)\\
& & + \delt^2\left[ f_c'' + \frac{3f_c'}{r} - \frac{4rf_c'}{f_c^2 + r^2} + \frac{2f_c}{f_c^2 + r^2}\left(\fdot_c^2 - f_c'^2\right)\right].
\end{eqnarray*}
Here we also assume the derivatives are represented by the appropriate
differencs as outlined in previous sections.  Approximate
$f_c(r,t+\delt)$ for use in $\fdot_c^2$ by using first $f_c(r,t+\delt)
= 2f_c(r,t) - f_c(r,t-\delt)$ and then iterating the solution found
for $f_c(r,t+\delt)$ six times, finding a new one from solving with
the previous value.  Using a greater number of iterations on
$f_c(r,t+\delt)$ does not change the answer to the numerical precision
on the computer, so this is not considered a source of error.

The error comes from the discretizations.  If we let $\delt^2C_1$ be
the error in the difference for $\fddot$ and $C_2$ be the accumulated
error in the differences for the left hand side of [\ref{myPDE2}],
then one may complete the exact same computations as in section A.8
to find the error in n steps is bounded by
\begin{eqnarray*} |f(r,t+n\delt) - f_c(r,t+n\delt)| &\leq& n\delt^2(|C_1| + |C_2|)\\
& & + \delt C_3(|C_1| + |C_2|)\end{eqnarray*} where $C_3$ is a
constant.  

Taylor's theorem will be applied extensively to the terms of equation
[\ref{myPDE2}] to show that $C_1$ and $C_2$ are constants.  We see
\begin{eqnarray*} 
\fddot(r,t) &=& \frac{f(r,t+\delt) + f(r,t-\delt) - 2f(r,t)}{\delt^2}
-\frac{\delt^2}{12}\frac{\partial^4f}{\partial t^4}(r,\eta_1)\\
\frac{4rf'(r,t)}{f^2(r,t) + r^2} &=& \frac{2(f(r+\delr,t) - f(r-\delr,t))}{\delr(f^2(r,t) + r^2)} \\
& & + \frac{4r}{f^2(r,t) + r^2}\left(\frac{\delr^2}{6}\frac{\partial^3 f}{\partial r^3}(\xi_1,t)\right)\\
\frac{2f(r,t)\fdot^2(r,t)}{f^2(r,t) + r^2} &=& \frac{f(r,t)(f(r,t+\delt)-f(r,t-\delt))^2}{2\delt^2(f^2(r,t) + r^2)} \\
& & -\left(\frac{2f(r,t)(f(r,t+\delt) - f(r,t-\delt))}{\delt(f^2(r,t) + r^2)}\right)\times \\
& &
\left(\frac{\delt^2}{6}\frac{\partial^3 f}{\partial t^3}(r,\eta_2)\right)\\
& & + \frac{2f(r,t)}{f^2(r,t) + r^2}\left(\frac{\delt^2}{6}\frac{\partial^3 f}{\partial t^3}(r,\eta_2)\right)\\
\frac{2f(r,t)f''^2(r,t)}{f^2 + r^2} &=& \frac{f(r,t)(f(r+\delr,t)-f(r-\delr,t))^2}{2\delr^2(f^2(r,t) + r^2)} \\
& & -\left(\frac{2f(r,t)(f(r+\delr,t) - f(r-\delr,t))}{\delr(f^2(r,t) + r^2)}\right)\times \\
& &\left(\frac{\delr^2}{6}\frac{\partial^3 f}{\partial r^3}(\xi_2,t)\right)\\
& & + \frac{2f(r,t)}{f^2(r,t) + r^2}\left(\frac{\delr^2}{6}\frac{\partial^3 f}{\partial r^3}(\xi_2,t)\right)\\
\end{eqnarray*}
The first of these equations, for example, says that if 
\[\frac{\partial^4 f}{\partial t^4}(r,\eta_1)\]
is bounded then the error $C_1\rightarrow 0$ as $\delt^2 \rightarrow
0$.  If $f(r,t)$ is chosen in $C(4,4)$, the space of functions with
five bounded space/time derivatives, all of these differences have
bounded errors.

Now we must attend to the discretization error for
\begin{eqnarray} f''(r,t) + \frac{3f'(r,t)}{r} &\approx&
\left[\left(r + \frac{\delr}{2}\right)^3\left(\frac{f(r+\delr,t) - f(r,t)}{\delr}\right) \label{r3p}\right. \\
& & \left. - \left(r - \frac{\delr}{2}\right)^3\left( \frac{f(r,t) - f(r-\delr,t)}{\delr}\right)\right]\Bigg/(r^3\delr).\nonumber\end{eqnarray}
If we let $h(r) = r^3 f'(r)$ we have:
\begin{eqnarray*}
h'(r) &=& 3r^2f'(r) + r^3 f''(r) = r^3\left(f''(r) + \frac{3f'(r)}{r}\right)\\
h''(r) &=& 6rf'(r) + 6r^2 f''(r) + r^3f'''(r)\\
h'''(r) &=& 6f'(r) + 18rf''(r) + 9r^2f'''(r) + r^3 f^{(4)}(r).
\end{eqnarray*}
Using the appropriate Taylor Series, one can calculate that
\[ h'(r) = \frac{h(r + \frac{\delr}{2}) - h(r -\frac{\delr}{2})}{\delr} - \frac{\delr^3}{24} h'''(\eta).\] 
Now if we allow 
\[ \rho_1 \in \left[ r - \frac{\delr}{2}, r + \frac{\delr}{2}\right]\]
and substitute, to obtain
\begin{eqnarray}\lefteqn{ r^3 \left(f''(r) + \frac{3f'(r)}{r}\right) =}\nonumber\\
& &\frac{\left(r + \frac{\delr}{2}\right)^3 f'\left(r +
\frac{\delr}{2}\right) -\left(r - \frac{\delr}{2}\right)^3 f'\left(r - \frac{\delr}{2}\right)}{\delr} \nonumber\\ &
& - \frac{\delr^3}{24}\left(6f'(\rho_1) + 18rf''(\rho_1) +
9r^2f'''(\rho_1) + r^3f^{(4)}(\rho_1)\right)\label{r3p2}
\end{eqnarray}

Now calculate central difference approximations for $f'(r + \delr/2)$
and $f'(r-\delr/2)$ and substitute into [\ref{r3p2}].  Let $\rho_2 \in [r, r + \delr]$ and $\rho_3 \in [r -\delr, r]$.  The resultant equation is
\begin{eqnarray*}
\lefteqn{f''(r,t) + \frac{3f'(r,t)}{r} = }\\
& &\left[ \left(r + \frac{\delr}{2}\right)^3\left(\frac{f(r + \delr, t) - f(r,t)}{\delr}\right)\right. \\
& &\left. - \left(r - \frac{\delr}{2}\right)^3\left(\frac{f(r,t) - f(r-\delr,t)}{\delr}\right)\right]\Bigg/(r^3\delr)\\
& & - \left[\frac{\left(r + \frac{\delr}{2}\right)^3}{r^3\delr}\right]\left(\frac{\delr^2}{24}\frac{\partial^3 f}{\partial r^3}(\rho_2,t)\right) +
\left[\frac{\left(r - \frac{\delr}{2}\right)^3}{r^3\delr}\right]\left(\frac{\delr^2}{24}\frac{\partial^3 f}{\partial r^3}(\rho_3,t)\right)\\
& & -\frac{\delr^3}{24r^3}\left(6f'(\rho_1) + 18rf''(\rho_1) +
9r^2f'''(\rho_1) + r^3f^{(4)}(\rho_1)\right).
\end{eqnarray*}
Once again one sees that if $f(r,t)$ is chosen in $C(4,4)$, the error in this difference is bounded.  Under this condition, putting all of this information together, we expect the error in 
\[ |f(r,t+n\delt) - f_c(r,t+n\delt)| \leq K\delt\]
where $K$ is bounded and $K\rightarrow 0$ only when both $\delt
\rightarrow 0$ and $\delr \rightarrow 0$, but not when only one of
$\delt$ or $\delr$ go to zero.

To confirm this, convergence tables have been created for this program.
$f(0,60)$ and $f(10,60)$ are used as indicators.

First, the convergence as $\delt \rightarrow 0$ is investigated.  One
run is made with $\delr = 0.100$ and $\delt = 0.001$, since we
cannot do a run with $\delr$ and $\delt$ infinitely small, and the
stability analysis tells us that we must keep $\delt \leq C
\delr^{3/2}.$ Call this $f_\infnty$.  Then $\delt$ is allowed to take
on a variety of larger values and each time we can find the errors $E0
= |f(0,60) - f_\infnty(0,60)|,$ $E10 = |f(10,60) -
f_\infnty(10,60)|,$ $h = \delt - 0.001$.  Then with any two of
these one can calculate 
\begin{equation}  \ln(E0_a/E0_b)/\ln(h_a/h_b)\qquad\mbox{and}\qquad
 \ln(E10_a/E10_b)/\ln(h_a/h_b)\label{lnstuf2}\end{equation} In this
problem, this quotient of natural logarithms should be close to $1$
since in theory the error is $K\delt$.

The complete set of initial conditions are $f(0,t) = 1.0$, $R_{\mbox{max}} = 100$, $\dot{f}(0,t) = -0.01$.  The data is in Table \ref{ctaba}. 

\begin{table}[H]
\begin{center}
\caption{$\IC P^1$ model, charge 1 sector: Convergence data 1.}
\label{ctaba}
\[\begin{array}{rrrr}
\multicolumn{1}{c}{\delr} & \multicolumn{1}{c}{\delt} &\multicolumn{1}{c}{f(0,60)} & \multicolumn{1}{c}{f(10,60)}\\
0.1&       0.001&   0.433979230592 &  0.431548378650 \\ 
0.1&       0.0025&  0.434046902945 &  0.431616258041\\
0.1&       0.005&   0.434159692320 &  0.431729392396\\
0.1&       0.0075&  0.434272484358 &  0.431842529337\\
0.1&       0.0125&  0.434498076252 &  0.432068810954\\
\end{array}\]
\end{center}
\end{table}

In Table \ref{ctaba}, skipping the first row and proceeding downward,
calculate Table \ref{ctabb}, where the previous line is the line
associated with subscript ``a'' in equation [\ref{lnstuf2}], and the
curent line is associated with subscript ``b'' in equation
[\ref{lnstuf2}].  It is clear that the quotient of the natural
logarithms as in [\ref{lnstuf2}] is close to one and gets closer as the
size of $\delt$ decreases, as the theory predicts.

\begin{table}[H]
\begin{center}
\caption{$\IC P^1$ model, charge 1 sector: Convergence data 2.}
\label{ctabb}
\[ \begin{array}{rrrrr}
\multicolumn{1}{c}{h} & \multicolumn{1}{c}{E0} &\multicolumn{1}{c}{E10} & \multicolumn{1}{c}{\mbox{ln quot. $E0$}}& \multicolumn{1}{c}{\mbox{ln quot. $E10$}} \\
0.0015&   0.0000676724&  0.0000678794& &\\
0.0040&    0.0001804617&  0.0001810137&   1.000011111 &  1.000011077\\
0.0065&   0.0002932538&  0.0002941507&   1.000024846 &  1.000024070\\   
0.0115&   0.0005188457&  0.0005204323&   1.000040720 &  0.994688335\\
\end{array}\]
\end{center}
\end{table}

Next, the convergence as $\delr \rightarrow 0$ is investigated.  One
run is made with $\delr = 0.01$ and $\delt = 0.001$, since we
cannot do a run with $\delr$ and $\delt$ infinitely small, and
consider this to be the value of $f_\infnty$.  Then $\delr$ is allowed
to take on a variety of larger values.  Likewise, we calculate the
values in [\ref{lnstuf2}], but this time, our theory has no prediction
for the value, as the error is predicted to decrease as $K\delt$.

Again, the complete set of initial conditions are $f(0,t) = 1.0$,
$R_{\mbox{max}} = 100$, $\dot{f}(0,t) = -0.01$.  The data is in Table \ref{ctabc}.

\begin{table}[H]
\begin{center}
\caption{$\IC P^1$ model, charge 1 sector: Convergence data 3.}
\label{ctabc}
\[ \begin{array}{rrrr}
\multicolumn{1}{c}{\delr} & \multicolumn{1}{c}{\delt} &\multicolumn{1}{c}{f(0,60)} & \multicolumn{1}{c}{f(10,60)}\\
0.010&      0.001&   0.434213647771&   0.431709360602 \\
0.020&      0.001&   0.434206452051&   0.431707448176 \\
0.025&      0.001&   0.434201061431&   0.431704981289 \\
0.040&      0.001&   0.434177743230&   0.431691546698 \\
0.050&      0.001&   0.434156269734&   0.431677574831 \\
0.100&      0.001&   0.433979230592&   0.431548378650 \\ 
0.200&      0.001&   0.433312345450&   0.431017145989 \\
\end{array}\]
\end{center}
\end{table}  

In Table \ref{ctabc}, skipping the first row and proceeding
downward, calculate Table \ref{ctabd}, where the previous line is
the line associated with subscript ``a'' in equation [\ref{lnstuf2}],
and the curent line is associated with subscript ``b'' in equation
[\ref{lnstuf2}].  The quotient of natural logarithms is closest to
being an integer when the error caused by $\delr$ is much greater than
the error caused by $\delt$.  This is as this sort of analysis would
predict when the error is a sum of a piece that goes to zero as $\delr
\rightarrow 0$ and another piece that goes to zero as $\delt
\rightarrow 0$.

\begin{table}[H]
\begin{center}
\caption{$\IC P^1$ model, charge 1 sector: Convergence data 4.}
\label{ctabd}
\[ \begin{array}{rrrrr}
\multicolumn{1}{c}{h} & \multicolumn{1}{c}{E0} &\multicolumn{1}{c}{E10} & \multicolumn{1}{c}{\mbox{ln quot. $E0$}}& \multicolumn{1}{c}{\mbox{ln quot. $E10$}} \\
0.010&    0.0000071957&  0.0000019124& & \\
0.015&    0.0000125863&  0.0000043793&   1.37897263&   2.043406152\\
0.030&    0.0000359045&  0.0000178139&   1.512310436&  2.024231211\\ 
0.040&    0.0000573780&  0.0000317858&   1.629570850&  2.012779570\\
0.090&    0.0002344172&  0.0001609820&   1.735588896&  2.000508645\\
0.190&    0.0009013023&  0.0006922146&   1.802345190&  1.952054808\\
\end{array}\]
\end{center}
\end{table}

\clearpage

\bibliography{diss2}
\begin{vita}
\index{Vita@\emph{Vita}}%

Jean-Marie Linhart \index{Jean-Marie Linhart}%
was born in Melrose Park, Illinois on June 2, 1969, the daughter of
Harriet Roeters Linhart and Carl George Linhart.  After completing her
work at Oak Park and River Forest High School, Oak Park, Illinois, in
1987, she entered The University of Chicago in Chicago, Illinois.  She
received the degree of Bachelor of Science in Mathematics from the
University of Chicago in June 1990.  She entered the Graduate School
of The University of Texas at Austin in August of 1990.  During the
summer of 1991 she was employed at Lawrence Livermore National
Laboratories in the Global Climate Research group.  While working on
her Master's degree, she worked as a Teaching Assistant in the
Department of Mathematics at the University of Texas at Austin.  She
received the degree of Master of Arts in Mathematics from the
University of Texas in May 1993.  She then worked at Lawrence
Livermore National Laboratory in the Global Climate Research group
until June of 1994.  From July of 1994 to October of 1995, she was
employed as a Mathematics Instructor at ITT Technical Institute in
Austin Texas.  In August of 1995, she returned to the Graduate School
of The University of Texas.  From August 1995 to August 1997, she was
employed in the Department of Mathematics at The University of Texas
at Austin as a Teaching Assistant and Assistant Instructor.  During
the summer of 1996 she was employed by Schlumberger Well Services in
the Formation Evaluation department.  She returned to employment by
Schlumberger Well Services in August of 1997.

\end{vita}

\end{document}